\documentclass[12pt]{article}
\pdfoutput=1
\usepackage{jheppub}
\usepackage{amsmath}
\usepackage{amsfonts}
\usepackage{amssymb}
\usepackage{latexsym}
\usepackage{tkz-graph}
\usepackage{tkz-euclide}
\usepackage{simplewick}
\RequirePackage{amssymb,amsmath,amsthm,amsfonts}
\RequirePackage{bbm}
\numberwithin{equation}{section}

\usepackage{mathtools}
\usepackage{simpler-wick}

\usepackage{epsf}
\usepackage{color}
\usepackage{graphicx}
\usepackage{subfig}
\usepackage{dsfont}

\usepackage[export]{adjustbox}
\usepackage{hyperref}
\definecolor{darkred}{rgb}{0.8,0.1,0.1}
\hypersetup{colorlinks=true, linkcolor=darkred, citecolor=blue, linktoc=page}

\graphicspath{{figures/}}
\usepackage[utf8]{inputenc} 

\AtBeginDocument{}

\def\tr{{\rm tr}}

\newcommand{\ZZ}{\ensuremath{\mathbb Z}}

\newcommand{\ra}{\ensuremath{\rightarrow}}
\newcommand{\bpm}{\ensuremath{\begin{pmatrix}}}
\newcommand{\epm}{\ensuremath{\end{pmatrix}}}

\newcommand{\expt}[1]{\left< #1 \right> }
\newcommand{\innn}[3]{\left< #1 \left| #2 \right| #3 \right>}

\newcommand{\dd}[2]{\frac{\partial #1}{\partial #2} }

\DeclareMathSymbol{:}{\mathord}{operators}{"3A}

\usepackage{comment}

\title{Multi-trace Correlators in the SYK Model and Non-geometric Wormholes}

\author[a]{Micha Berkooz,}
\author[a]{Nadav Brukner,}
\author[b]{Vladimir Narovlansky,}
\author[a,c]{and Amir Raz}
\emailAdd{micha.berkooz@weizmann.ac.il}
\emailAdd{nadav.brukner@weizmann.ac.il}
\emailAdd{narovlansky@princeton.edu}
\emailAdd{araz@utexas.edu}
\affiliation[a]{Department of Particle Physics and Astrophysics, \\ Weizmann Institute of Science, Rehovot 7610001, Israel}
\affiliation[b]{Princeton Center for Theoretical Science, \\  Princeton University, Princeton, NJ 08544, USA}
\affiliation[c]{Department of Physics, \\  University of Texas at Austin, Austin, TX 78712, USA}

\preprint{UTTG-02-21}

\abstract{
We consider multi-energy level distributions in the SYK model, and in particular, the role of global fluctuations in the density of states of the SYK model. The connected contributions to the moments of the density of states go to zero as $N \to \infty$, however, they are much larger than the standard RMT correlations. We provide a diagrammatic description of the leading behavior of these connected moments, showing that the dominant diagrams are given by 1PI cactus graphs, and derive a vector model of the couplings which reproduces these results. We generalize these results to the first subleading corrections, and to fluctuations of correlation functions. In either case, the new set of correlations between traces (i.e. between boundaries) are not associated with, and are much larger than, the ones given by topological wormholes.  The connected contributions that we discuss are the beginning of an infinite series of terms, associated with more and more information about the ensemble of couplings, which hints towards the dual of a single realization. In particular, we suggest that incorporating them in the gravity description requires the introduction of new, lighter and lighter, fields in the bulk with fluctuating boundary couplings.

}

\begin{document}

\setcounter{footnote}{0}

\maketitle

\setcounter{equation}{0}
\setcounter{footnote}{0}

\section{Introduction}

The Sachdev-Ye-Kitaev (SYK) model \cite{Sachdev_1993,Kitaev_talk} is a simple tractable model of quantum chaos and holography. The model consists of  $N$ Majorana fermions, and a Hamiltonian with all-to-all random interactions of $p$ fermions (the full definition is given in section \ref{sec:ModelDef}). It has been extensively analyzed in the large $N$ limit, both at finite $p$ \cite{MaldecenaStanford,Polchinski_2016,Kitaev_2018} and when $p\propto\sqrt{N}$, AKA the double scaled limit \cite{Berkooz_2019,Micha2018,Garc_a_Garc_a_2016,feng2018spectrum}.\footnote{For an introduction and review of the SYK literature, we refer the reader to \cite{Rosenhaus_2019} and references therein.} In the fixed $p$ large $N$ limit, the theory at low energies is governed by an effective Schwarzian action, which, via its relation to JT gravity, is a simple toy model of holography \cite{NADSbreaking,Almheiri2015,Jensen_2016}. In the double scaled limit $p \propto \sqrt{N}$, the model also has a Schwarzian regime but it is also related to a  $q$-deformation of random matrix theory \cite{Speicher2019}, and the full large $N$ density of states is given by the $q$-Gaussian distribution  \cite{Erdos14,Cotler:2016fpe,feng2018spectrum,Berkooz_2019}.

As we are dealing with a random Hamiltonian in the large $N$ limit, the natural objects to consider are disorder averaged. The simplest of these are expectation values of a single trace; for example we may consider the averaged moments of the Hamiltonian $m_k = \expt{\text{tr}(H^k)}_{J}$, where $\left<\cdot\right>_J$ represents an ensemble average over different random couplings in the Hamiltonian, or the averaged thermal partition function $Z(\beta) =  \expt{\text{tr}(e^{-\beta H})}_{J} $. These teach us about the expected spectral density of the Hamiltonian $\rho(E) = \expt{\sum_i \delta(E - E_i)}_J$.

We may also consider the correlations between the eigenvalues, which are computed through expectation values of higher trace objects. For example the spectral form factor $\expt{\tr\left(e^{(-\beta+it)H} \right) \tr\left(e^{(-\beta-it)H} \right)}_J$ at long times gives us insight into the nearest neighbor eigenvalue spacing \cite{Cotler:2016fpe}. This spacing follows a distribution given by a random matrix theory (RMT) universality class which only depends on the symmetries of the model \cite{Kanazawa_2017,Garc_a_Garc_a_2016,You_2017,Cotler:2016fpe,sun2019periodic,stanford2019jt}. The eigenvalues also exhibit long range correlations \cite{Altland_2018,verbaarschot2019}, which dominate at shorter time scales, and are the focus of this paper. More precisely, our main objects of focus are the multi-trace thermal expectation values
\begin{equation} \label{eq:def_themal_partition}
\mathcal{Z}(\beta_1,\beta_2,\ldots,\beta_n) = \expt{\tr\left[e^{-\beta_1 H}\right]
      \tr\left[e^{-\beta_2 H}\right] \ldots 
      \tr\left[e^{-\beta_n H}\right] }_J
\end{equation}
(and their behavior at relatively short time scales). 
Specifically we focus on the connected component, and to a few leading orders in the large $N$ expansion. 

The motivation for focusing on the connected correlators is to understand them in the gravitational dual theory. In a gravitational theory these objects usually come about from a path integral over a space-time with multiple boundaries, with the connected components representing a geometry which is connected, in some sort of way, in the bulk \cite{marolf_maxfield}. Such a picture arises in JT gravity, and its proposed duality to the $\beta$-ensemble RMT models with a Schwarzian envelope density of states. There, the connected components come from topological connections between boundaries, and replicate exactly the topological recursion results of RMT \cite{Saad:2019lba}. 

In contradistinction, we are discussing the full SYK model. There at shorter time scales, the gravitational interpretation is different - the SYK model contains the RMT results, but it also has modifications to the connected components that come from ``global fluctuations'' of the spectrum \cite{Altland_2018,verbaarschot2019}. These large scale fluctuations are suppressed as some power of $1/N$, rather than the exponential $2^{-N/2}$ suppression of the RMT terms\footnote{which is $1/dim({\cal H})$}, and so are dominant at least at early times. We will focus on these global fluctuations at leading orders in $1/N$, and try to fit them into the broader gravitational picture.
We note that in the double scaling limit these corrections go like $e^{-c \sqrt{N}\log(N)}$, which is exponentially suppressed but still much larger than $2^{-N/2}$. So at short time scales these are effects which are much larger than geometric wormholes. 

Studying the full SYK model is necessary because the JT gravity-RMT correspondence restricts to extremely low energies and very long times, and throws out many interesting effects at finite times. In particular we will see that, properly interpreted, these ``global fluctuations'' imply the existence of specific ultra light particles in the bulk (or low dimension operators). Furthermore, when one attempts to discuss the dual model at a given fixed realization, the relevant information resides in these additional fields.\footnote{As we were completing this work, \cite{Saad:2021rcu} appeared, which explores the issue of a single realization in a related but different setting. The setting is different enough that a direct comparison is difficult.}

\subsection{Definition of the model, and some known results} \label{sec:ModelDef}

The SYK model consists of  $N$ Majorana fermions $\psi_i$, $i=1,\cdots,N$, which satisfy the canonical commutation relations $\{\psi_i,\psi_j\} = 2\delta_{ij}$. 
Throughout the paper we will denote by $I$ the ordered index sets $I = \{i_1,\ldots,i_p\}$ ($1\leq i_1<i_2<..<i_p\leq N$) of length $p$, and  $\Psi_I$ as the $p$ fermion interaction $\Psi_I \equiv \psi_{i_1}\psi_{i_2}\ldots \psi_{i_p}$. The SYK Hamiltonian consists of random all-to-all interactions of $p$ fermions:
\begin{equation} \label{eq:hamiltonian}
H = \mathcal{J}~ i^{p/2} {\binom{N}{p}}^{-1/2}\sum_{|I|=p} J_{I}\Psi_{I} ,
\end{equation}
 where the sum runs over all index sets $I$ of length $p$, and
 $\cal{J}$ sets an energy scale which we will normalize to 1. We take $J_I$ to be i.i.d random couplings with zero mean and normalized variance \begin{align}
     \expt{J_I J_{I'}}_J = \delta_{I,I'},
 \end{align}  
 where expectation value over the random variables $J_I$ is denoted by $\expt{\cdot}_J$. Furthermore, we consider $p$ to be even and $J_I$ to have a Gaussian distribution in what follows, though this is not strictly necessary. (We will discuss non-Gaussian distributions in section \ref{sec:vector}.)

We define the multi-trace thermal expectation to be 
\begin{equation}
\mathcal{Z}(\beta_1,\beta_2,\ldots,\beta_n) \equiv \expt{\tr\left[e^{-\beta_1 H}\right]
\tr\left[e^{-\beta_2 H}\right] \ldots \tr\left[e^{-\beta_n H}\right]}_J,\ \ \ \ \tr(1)=1
\end{equation}
where we have redefined the trace in (\ref{eq:def_themal_partition}) by a factor of $2^{-N/2}$.
The calculation proceeds by expanding in multi-trace moments (and then evaluating them, and resumming) 
\begin{equation}
M(k_1,k_2,\ldots,k_n) \equiv \expt{\tr\left[H^{k_1}\right]
\tr\left[H^{k_2}\right] \ldots \tr\left[H^{k_n}\right] }_J.
\end{equation}
We will focus on the leading order (in $1/N$) term in the connected part of these moments. These are defined in the usual way by subtracting all lower disconnected moments. Explicitly the recursive definition for the connected part is 
\begin{equation} \label{eq:def_connected_moments}
\begin{split}
M_c(k_1) &= M(k_1),\\
M_c(k_1,k_2) &= M(k_1,k_2) - M_c(k_1) M_c(k_2),\\
M_c(k_1,k_2,\ldots,k_n) &= M(k_1,k_2,\ldots,k_n) ~ - \sum_{\substack{\text{partitions $p$} \\ \text{ of $\{1,\ldots,n\}$}}}~
\prod_{\{i_1,\ldots i_m\}\in p}M_c(k_{i_1},\ldots,k_{i_m}).
\end{split}
\end{equation}

We can similarly define the connected thermal partition functions in the same recursive manner, or as the exponentiation of the connected moments
\begin{align}
\mathcal{Z}_c(\beta_1,\beta_2,\ldots,\beta_n) &\equiv \sum_{k_1,\ldots,k_n = 0}^\infty
M_c(k_1,k_2,\ldots,k_n) \prod_{j=1}^n \frac{(-\beta_j)^{k_j}}{k_j!} .
\end{align}


To put this work in context, we present a brief reference to known results with regards to the SYK spectrum and spectral fluctuations. An additional more detailed comparison is carried out in subsection \ref{sec:TimeScl} where we use our explicit formulas to compare to existing analytic and numerical results.

Single trace correlators give us the overall shape of the spectrum. For the finite $p$, $N\rightarrow \infty$ they were computed in \cite{Sachdev_1993,MaldecenaStanford,Kitaev_talk,Polchinski_2016}.  In the double scaled limit, $N,p \ra \infty$ while keeping $N/p^2$ fixed, they were computed in \cite{Erdos14,Cotler:2016fpe,feng2018spectrum,Micha2018}, and there the SYK model has an asymptotic $q$-Gaussian spectrum. This $q$-Gaussian spectrum is a great approximation of the actual spectrum at finite $N$ and $p$ \cite{Garc_a_Garc_a_2017,Garc_a_Garc_a_2018_2,Jia_2018}. The center of this spectrum resembles a Gaussian distribution \cite{Garc_a_Garc_a_2016}, while the edge of the spectrum resembles the Schwarzian density of states $\rho(E) \sim \sinh(\sqrt{E-E_0})$ \cite{Garc_a_Garc_a_2017,Cotler:2016fpe,Stanford_Witten2017,Mertens_2017}. 

The nearest neighbor (or a few apart) level spacing in the SYK model follows that of a random matrix ensemble.\footnote{See \cite{anderson2010introduction} for an introduction to RMT ensembles, and  \cite{Guhr_1998} for a review of RMT applications in physics.} The precise RMT ensemble (GOE, GUE, or GSE) depends on the values $N \mod 8$ and $p\mod 4$, which give rise to different particle-hole symmetries, and have been completely classified in \cite{You_2017,Garc_a_Garc_a_2016,Kanazawa_2017,Cotler:2016fpe,sun2019periodic,stanford2019jt}. This level spacing statistics results in a universal RMT contribution to the spectral form factor which is dominant at exponentially late times \cite{Cotler:2016fpe}. Apart from these universal contributions, there are also global fluctuations of the spectrum which are less suppressed in the large $N$ limit \cite{feng2018spectrum,Cotler:2016fpe,verbaarschot2019}. It has been an active area of research in past few years to analytically show the SYK model has RMT spectral fluctuations, as well as to characterize the global fluctuations. 

The most straightforward approach to computing the spectral fluctuations of the SYK model is to start with its path integral formulation \cite{MaldecenaStanford,Polchinski_2016,Kitaev_2018}, and to study the path integral of two replicas of the SYK model \cite{saad2018}. The two replicas path integral describes the double trace thermal partition function $\expt{\tr\left(e^{-\beta_1 H}\right)\tr\left(e^{-\beta_2 H}\right)}$. It has been argued that the universal RMT contributions should arise from replica non-diagonal saddle points of this action \cite{saad2018,Saad:2019lba,Cotler:2016fpe}, thus by finding these saddle points one can try to understand the connected thermal partition function and the spectral form factor. Finding these saddle points both analytically and numerically has been an ongoing area of research \cite{Saad:2019lba,saad2018,Aref_eva_2019,Wang_2019,khramtsov2020spectral}.

A different method to study the spectral fluctuations is to derive a non-linear $\sigma$-model for the SYK model using random matrix theory techniques,\footnote{See \cite{Guhr_1998} and references therein for an overview of the super-symmetry technique and the non-linear $\sigma$-model in RMT} as was first done in \cite{Altland_2018}.\footnote{This is the first derivation for the Majorana SYK model, a $\sigma$ model for the complex SYK model was previously derived in \cite{VERBAARSCHOT198478}.} This method produces the universal RMT level spacing statistics, as well as corrections around it, and was also studied in \cite{verbaarschot2019}. This $\sigma$ model is also related to the study of spectral fluctuations using supersymmetry techniques  \cite{Sedrakyan_2020,Behrends_2020}. 

We note that this problem can also be mapped to the study of the cummulants of $q$-Brownian motion \cite{Speicher2019,PlumaRodriguez_2019}.

\section{Framework and Main Results}

The goal of this paper is two fold. The first is to calculate explicit expressions for multi-trace correlators in the SYK model at early times, and develop a general perturbative framework for calculating corrections to these observables. The second goal is to start exploring how to interpret such correlators in a gravitational theory. In particular we identify the leading deviations that an observer sees at a given realization of the couplings. Two notions that we will discuss are \textit{global modes}\footnote{The term is borrowed from a discussion with A.~Altland, D.~Bagrets, S.~Shenker and D.~Stanford.} and \textit{fluctuations parameters}. Both have to do with specific ways in which the spectrum of the theory can fluctuate over specific energy separations or time scales, although we will focus more on their implications to the gravitational picture. 

Of course, if we start with any fixed Hamiltonian then the full set of eigenvalues $\{\lambda_i\}_{i=1}^{\dim \mathcal{H}}$ is tantamount to the full model. But if we are interested in expectation values of simple operators on typical states, or some other coarse grained observables at shorter time scales, we can do with less information which is captured in a coarse grained description of the spectrum.\footnote{This is true so long as the Hamiltonian is chaotic \cite{Deutsch1991,Srednicki1994,Pollack_2020}, which is the case for the SYK model \cite{Jensen_2016}, and should also be the case for gravitational systems in general \cite{Shenker_2014,Chaos}}

To understand what is meant by a coarse grained description of the spectrum, consider the Hamiltonian in \eqref{eq:hamiltonian} for some arbitrary choice of couplings $J$, and the quantity
\begin{equation}
    \rho(E,\epsilon)=\frac{2^{-N/2} }{\epsilon} N_{E-\epsilon/2,E+\epsilon/2},
\end{equation}
where $N_{E-\epsilon/2,E+\epsilon/2}$ is the number of eigenvalues in the small interval $[E-\epsilon/2,E+\epsilon/2]$. Then the coarse grained spectrum is given by the limits \footnote{Note that the order of limits is important: we first take $N\rightarrow \infty$ and only then $\epsilon\rightarrow 0$. This  is precisely what is meant by coarse graining: we first take a coarse description of the spectrum using bins of size $\epsilon$, then we take an appropriate large $N$ limit, and only at the end we look at the fine grained structure by taking $\epsilon \ra 0$.}
\begin{equation}
    \rho_0(E)  \equiv \lim_{\epsilon\rightarrow 0} \lim_{N\rightarrow \infty} \rho(E,\epsilon)  .
\end{equation}
If this limit exists then it implies that the coarse grained spectrum coincides with the ensemble average spectrum
\begin{equation}
    \rho_0(E) = \lim_{N\ra \infty} \expt{\rho(E)}_J = \lim_{N\ra \infty} 2^{-N/2}\expt{\sum_{i=1}^{ \dim \mathcal{H}} \delta(E-\lambda_i)}_J,
\end{equation}
and that fluctuations around this average spectrum are small for any typical Hamiltonian in the ensemble. For a fixed Hamiltonian we can still use this averaged spectrum to compute observables like the thermal partition function, and it will give the correct result up to small differences which depend on the exact realizations of the couplings.

We can ask, however, about fluctuations of $\rho(E,\epsilon)$ at finite $N$, and then at infinite $N$. We expect that in some cases
\begin{equation} \label{eq:coarse_couplings}
    \delta\rho(E)=\lim_{\epsilon\rightarrow 0} \lim_{N\rightarrow \infty} F(N) \biggl( \rho(E,\epsilon)-\rho_0(E) \biggr) 
\end{equation}
will become a finite random function with some fixed distribution, if we choose the right normalization factor $F(N)$. We will find that
\begin{equation}\label{FlucSclng}
    F(N)\sim N^{\kappa}
\end{equation}
for some fixed positive $\kappa$, which depends on the precise scaling of the model at large $N$.

In other words, as we go over the various realizations with a measure dictated by their distribution, we can write
\begin{equation} \label{eq:coarse_spec}
    \rho(E)=\rho_0(E) + N^{-\kappa} \delta\rho(E)
\end{equation}
where $\delta\rho(E)$ is a random function drawn from some ensemble\footnote{Note that since a measure on a space of functions is the same as a path integral, we might need to keep $\epsilon$ as a small cutoff.}. All the information on the specific realization is projected into the distribution of this function. Thus we have coarse grained the information in all the $J$'s, but not to the full extent of averaging over all of them. 

The scaling \eqref{FlucSclng} is already a striking statement since in ordinary $\beta$-ensemble RMT models, and in the gravitational paraphernalia that goes along with it, the leading coarse grained information is of order $(\dim \mathcal{H})^{-1} \sim e^{-\text{const} \cdot N}$. So for SYK like models the fluctuations of the spectrum are much much larger. As we discuss below, $\delta\rho(E)$ is the leading {\it global mode}.

From the coarse grained spectrum, \eqref{eq:coarse_spec}, we can compute the leading fluctuations of the multi-energy distribution, $\rho(E_1,\ldots,E_n) \equiv \rho(E_1)\ldots \rho(E_n)$. These leading order fluctuations are encoded in the connected part of the multi-energy distribution with respect to the average over the random fluctuations $\delta \rho$. Denoting the ensemble from which $\delta\rho(E)$ is drawn as ${\cal P}(\{ \delta\rho\})$ the connected part of this multi-energy distribution is given by
\begin{equation}
    \rho_c(E_1,\cdots,E_n)= N^{-n\kappa} \int {\cal DP}(\delta\rho)\ \delta\rho(E_1)\cdots\delta\rho(E_n).
\end{equation}

Next we can proceed in two directions:

\begin{enumerate}
    \item  We may extend these fluctuations to an entire expansion 
\begin{equation} \label{eq:expansion_of_everything}
    \rho(E)= \rho_0(E) + \sum_i N^{-\kappa_i} \delta\rho_k (E) 
\end{equation}
for a series of increasing positive numbers $\kappa$. We expect this sum to be asymptotic. However, if we terminate the expansion at some fixed order, say $i=1,\cdots,L$ we can hope that there is some joint distribution of the vector of functions 
\begin{equation}
{\vec \delta\rho(E)} = \bigl(\delta\rho_1(E),\delta\rho_2(E),\cdots,\delta\rho_L(E)\bigr) ,
\end{equation}  
which encodes the leading and some subleading orders coarse grained information on the exact realization of the couplings. We will derive such an expansion for the SYK model in sections \ref{sec:NewFluc} and \ref{sec:gravitystuff}. 

\item In some cases the distribution is highly degenerate, which is the situation in the SYK model. In this case, the measure on the $\rho$'s is not on a space of functions with an infinite number of parameters. Rather, if we truncate the series at some finite order, the vector $\vec\delta\rho$ is a vector of functions that depend on a finite number of parameters, which we will denote as $h_i$'s. Since these parameters control changes to the entire spectrum we will refer to them as {\it fluctuation parameters}, and we will refer to how they deform the spectrum as {\it global modes}. We call these deformations \textit{global} as they encode spectral correlations of eigenvalues with a finite energy separation; this is in contrast to the local nearest neighbor eigenvalue statistics which follows that of a $\beta$-ensemble RMT universality class.

\end{enumerate}

For example, we will see in section \ref{H2DualAct} that in the SYK model the leading order fluctuation behaves as an overall rescaling of the spectrum, i.e., we can write
\begin{equation}
    \rho(E,h)=\exp \left({ \binom{N}{p}^{-1/2} h \partial_E E}\right) \rho_0(E)
\end{equation}
with some distribution on $h$ which we denote as $P(h)$. In this case the leading order fluctuation also captures some contributions to smaller terms (larger $\kappa_i$'s) in the expansion \eqref{eq:expansion_of_everything}. The connected leading contribution to the multi-trace correlator will then be
\begin{equation}
    \rho_c(E_1,\cdots,E_n) = \int dh \, P(h) \rho(E_1,h)\cdots\rho(E_n,h) ~-~~ \text{disconnected},
\end{equation}
where we are subtracting disconnected contribution with respect to the integral over $h$. In this case the {\it fluctuation parameter} is $h$ and the {\it global mode} is a breathing mode which scales the distribution. 

We can now start asking about an observer in a given realization of the couplings, and how they may measure a deviation from the ensemble averaged theory. Actually this whole construction was built to allow us to easily isolate this leading coarse grained information. Indeed equations \eqref{eq:coarse_couplings} and \eqref{eq:coarse_spec} tell us that the leading deviation from the averaged theory is given by measuring a particular deviation $\delta \rho(E)$, or equivalently measuring the particular values of the fluctuation parameters $\{h_i\}$. The question is whether we can identify this set of fluctuation parameters with partial information about the set of random couplings $J$. 

The answer turns out to be yes. Consider the set of couplings $J_I$, where $I$ is an index set of length $p$ (there are  $\binom{N}{p}$ distinct couplings), and notice that the SYK Hamiltonian along with the Gaussian ensemble of couplings is invariant under $SO(N)$ rotations. Thus different realizations of the couplings are parametrized by the orbits of $SO(N)$, i.e., by the cosets
\begin{equation}
    \{ J_I\}/SO(N) .
\end{equation}
Another way of parametrizing these cosets is by the $SO(N)$ invariant combinations of the $J$'s. For example the invariants containing the least number of $J$'s are
\begin{equation}
    (h_2)^2 = \binom{N}{p}^{-1} \sum_{|I|=p} J_I^2 , \qquad  (h_3)^3= \binom{N}{p/2}^{-3}\sum_{|I_i| = p, |I_1\cap I_2| = p/2 } J_{I_1} J_{I_2} J_{I_1 \oplus I_2},
\end{equation}
where $\oplus$ stands for the XOR operation. Already at the level of 4 $J$'s there are many different invariants. It turns out that the fluctuations parameters that we discussed before are just these $h$'s, hence the leading order fluctuations of the SYK model are characterized by $SO(N)$ invariant combinations of the random couplings which contain the least number of $J$'s. A more detailed explanation and discussion of this general result happens in section \ref{sec:gravitystuff}.

The main implication of these fluctuation parameters is that one needs to introduce corresponding light fields into the gravitational path integral, which participate in generating correlations between disconnected bulk geometries. Such correlations should be distinguished from the standard picture of gravitational wormholes, in which the wormhole connects two disconnected boundaries through a connected bulk geometry. Rather they seem to be related to multi-trace deformations involving these light fields \cite{Aharony:2001pa,Witten:2001ua,Berkooz:2002ug}. This allows for the description of these non-geometric wormholes as a perturbative expansion around the disconnected geometry saddle. This will also be shown using the path integral formulation in section \ref{sec:path_integral}. It is puzzling that these non-geometric wormholes can be the dominant term in the multi-boundary dynamics, say, for concreteness, by being the dominant contribution to the spectral form factor at intermediate times, as we will show in section \ref{sec:TimeScl}. This shows that these additional fields must be included in any consistent gravitational dual to the SYK model.

Suppose that indeed all the correlations are multi-trace deformations that live on the boundary of $AdS_2$ and consider the situation that $AdS_2$ (perhaps times an additional compact manifold) is obtained as a near horizon limit of some higher dimensional black hole. In this case, the boundary of $AdS_2$ should be thought of as the transition region surrounding the near horizon region. The suggestion above then implies that more light fields need to be introduced in the near horizon limit, with new interactions at the transition region. An outside observer might see them as new degrees of freedom, beyond standard fields, living close to the horizon.

\subsection{The outline of the paper}

In section \ref{sec:multi_trace} we compute the leading correction to the multi-trace moments. We start by considering the moments as a sum over \textit{multi-trace chord diagram}, which are generated by summing over all Wick contractions of the random couplings. Then we focus on the double trace moments in section \ref{sec:double_trace}, and show that the leading order correction to the double trace moments comes from the minimal number of contractions between the two traces. This minimal connectivity leads to a general structure for the leading order multi-trace contribution coming from a sum over \textit{cactus diagrams}, which is developed in section \ref{sec:cactus}, and proven in sections \ref{sec:gen_1pi} and \ref{sec:dominance}. In section \ref{sec:pf_dos_cor} we re-sum these contributions, allowing us to give explicit formulae for the leading order contribution to the connected parts of the multi-trace thermal partition function and density of states. 

We end section \ref{sec:multi_trace} by comparing this leading order correction to previous results of the spectral fluctuations. Specifically in section \ref{sec:TimeScl} we compare the leading order connected contribution of the spectral form factor to the RMT contribution and to numerical simulations from \cite{Cotler:2016fpe}. We show that this leading order contribution is the dominant connected contribution to the spectral form factor at early times, and even becomes the dominant contribution overall at intermediate times for large enough $N$.

In section \ref{sec:vector} we show that the leading order contribution calculated in section \ref{sec:multi_trace} can be recast as a dual $0$-dimensional large $M = \binom{N}{p}$ vector model. We then show, in section \ref{sec:vec_as_coup}, that this vector model is actually a theory of the random couplings in the SYK model. This allows us to extend our analysis to couplings that have a non-Gaussian distribution. Finally the gravitational interpretation of this leading connected contribution is discussed is section \ref{H2DualAct}, which follows the framework laid out in this introduction.

Section \ref{sec:NewFluc} focuses on the first sub-leading contribution to the multi-trace correlators, which is the leading contribution involving odd moments. Similar to the leading contribution, we first compute the double trace moments in section \ref{sec:double_trace_3_lines}, followed by a derivation of the general structure of these contributions in section \ref{sec:subleading_theory_J}. We conclude this portion with section \ref{sec:h3_efctv}, in which we recast this sub-leading contribution as a new fluctuation parameter which is embedded in a fluctuating effective Hamiltonian.

The general structure of the multi-trace correlations, and their gravitational description, is discussed in section \ref{sec:gravitystuff}. We start by systematically describing the high order corrections to the connected multi-trace correlators as a perturbative collection of fluctuation parameters in section \ref{sec:GenFluc}. This description follows the formalism described is this section. We then go on to interpret the dual gravitational picture by considering life in a single realization of the couplings in \ref{sec:SnglRlz}, and the connection of these connected contributions to geometric wormholes in \ref{sec:wormholes}. 

 In section \ref{sec:operators} we use the multi-trace chord diagram description to calculate the leading order contributions to the connected multi-trace expectations of random operators. Then in section \ref{sec:path_integral} we replicate the leading order connected contributions to the double trace correlators in the replica path integral formalism of the SYK model. In particular we show that the leading order contributions to the double trace correlators can be computed as a perturbative expansion around the disconnected saddle point of the 2-replica action for the collective fields $G$ and $\Sigma$. We end the paper with a brief discussion on future directions in section \ref{sec:discussion}.

\section{Multi-Trace Thermal Partition Function} \label{sec:multi_trace}

Our goal is to calculate the leading contribution to the connected part of the thermal partition function \eqref{eq:def_themal_partition}, by expanding in multi-trace moments.
%
%
%
For example, the connected part of double-trace correlator is given explicitly by
\begin{equation} \label{eq:double_tr_explicit}
    \begin{split}
        M_c(k_1,k_2)=\frac{\mathcal{J}^k i^{kp/2}}{ \binom{N}{p}^{k/2}} \sum_{I_1,\cdots,I_k}\expt{ \tr \left[ J_{I_1} \Psi_{I_1} \cdots J_{I_{k_1}} \Psi_{I_{k_1}}\right] \tr \left[ J_{I_{k_1+1}} \Psi_{I_{k_1+1} } \cdots J_{I_k} \Psi_{I_k} \right] }_{J,c}
    \end{split}
\end{equation}
where $k=k_1+k_2$. The final expression in this case takes a very simple form and is given in equations  \eqref{eq:two_pairings}, \eqref{eq:double_thermal_partition} and  \eqref{eq:ConnRho}.

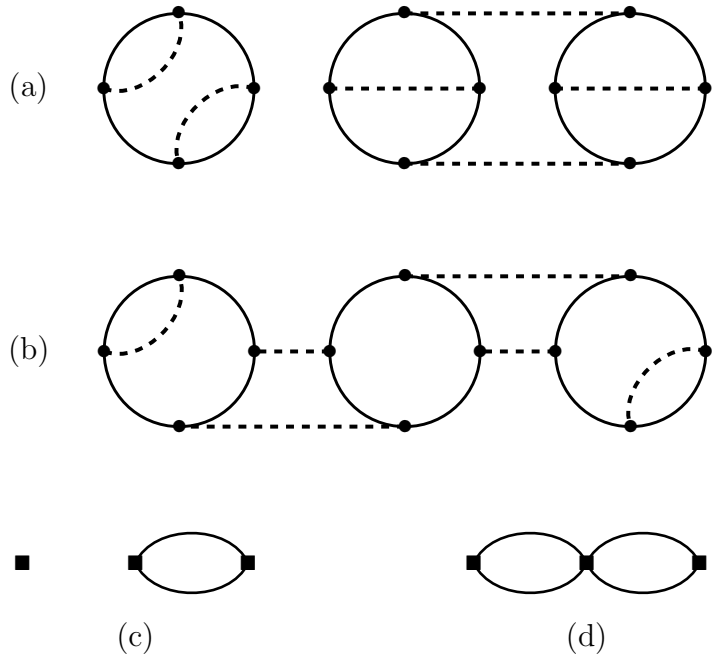
\begin{figure}
    \centering
    \begin{tikzpicture}[VertexStyle/.style={shape=coordinate}]
\SetGraphUnit{1.5}
 \tikzstyle{EdgeStyle}=[bend right = 65,line width=1.5pt,dashed]
 \node at (-1, 0) {(a)};
 \Vertex[x=0, y=0]{x1}
 \Vertex[x=1, y=1]{x2}
 \Vertex[x=2, y=0]{x3}
  \Vertex[x=1, y=-1]{x4}
 \Vertex[x=3, y=0]{y1}
 \Vertex[x=4, y=1]{y2}
 \Vertex[x=5, y=0]{y3}
  \Vertex[x=4, y=-1]{y4}
 \Vertex[x=6, y=0]{z1}
 \Vertex[x=7, y=1]{z2}
 \Vertex[x=8, y=0]{z3}
  \Vertex[x=7, y=-1]{z4}
 \Edges(x1,x2) 
 \Edges(x3,x4) 
 \tikzstyle{EdgeStyle}=[line width=1.5pt,dashed]
 \Edges(y1,y3) 
 \Edges(y2,z2)
 \Edges(z1,z3) 
 \Edges(y4,z4)
 \tikzstyle{EdgeStyle}=[bend left = 45,line width=1.1pt]
 \Edges(x1,x2,x3,x4,x1) 
  \Edges(y1,y2,y3,y4,y1) 
   \Edges(z1,z2,z3,z4,z1) 
 \node at (0,0) {$\bullet$};
 \node at (1,1) {$\bullet$};
 \node at (1,-1) {$\bullet$};
 \node at (2,0) {$\bullet$};
 \node at (3,0) {$\bullet$};
 \node at (4,1) {$\bullet$};
 \node at (4,-1) {$\bullet$};
 \node at (5,0) {$\bullet$};
 \node at (6,0) {$\bullet$};
 \node at (7,1) {$\bullet$};
 \node at (7,-1) {$\bullet$};
 \node at (8,0) {$\bullet$};
 \end{tikzpicture}
 
 \vspace{1cm}
 
 \begin{tikzpicture}[VertexStyle/.style={shape=coordinate}]
\SetGraphUnit{1.5}
 \tikzstyle{EdgeStyle}=[bend right = 65,line width=1.5pt,dashed]
 \node at (-1, 0) {(b)};
 \Vertex[x=0, y=0]{x1}
 \Vertex[x=1, y=1]{x2}
 \Vertex[x=2, y=0]{x3}
  \Vertex[x=1, y=-1]{x4}
 \Vertex[x=3, y=0]{y1}
 \Vertex[x=4, y=1]{y2}
 \Vertex[x=5, y=0]{y3}
  \Vertex[x=4, y=-1]{y4}
 \Vertex[x=6, y=0]{z1}
 \Vertex[x=7, y=1]{z2}
 \Vertex[x=8, y=0]{z3}
  \Vertex[x=7, y=-1]{z4}
 \Edges(x1,x2) 
 \Edges(z3,z4) 
 \tikzstyle{EdgeStyle}=[line width=1.5pt,dashed]
 \Edges(y1,x3) 
 \Edges(y2,z2)
 \Edges(z1,y3) 
 \Edges(y4,x4)
 \tikzstyle{EdgeStyle}=[bend left = 45,line width=1.1pt]
 \Edges(x1,x2,x3,x4,x1) 
  \Edges(y1,y2,y3,y4,y1) 
   \Edges(z1,z2,z3,z4,z1) 
 \node at (0,0) {$\bullet$};
 \node at (1,1) {$\bullet$};
 \node at (1,-1) {$\bullet$};
 \node at (2,0) {$\bullet$};
 \node at (3,0) {$\bullet$};
 \node at (4,1) {$\bullet$};
 \node at (4,-1) {$\bullet$};
 \node at (5,0) {$\bullet$};
 \node at (6,0) {$\bullet$};
 \node at (7,1) {$\bullet$};
 \node at (7,-1) {$\bullet$};
 \node at (8,0) {$\bullet$};
 \end{tikzpicture}
 
 \vspace{1cm}
 
 \begin{tikzpicture}[VertexStyle/.style={shape=coordinate}]
\SetGraphUnit{1.5}
 \tikzstyle{EdgeStyle}=[bend right = 65,line width=1pt]
 \node at (1.5, -1) {(c)};
 \node at (7.5, -1) {(d)};
 \Vertex[x=0, y=0]{x1}
 \Vertex[x=1.5, y=0]{x2}
 \Vertex[x=3, y=0]{x3}
 \Vertex[x=6, y=0]{y1}
 \Vertex[x=7.5, y=0]{y2}
 \Vertex[x=9, y=0]{y3}
 \Edges(x2,x3,x2) 
 \Edges(y1,y2,y3,y2,y1) 
 \node at (0,0) {{\scriptsize $\blacksquare$}};
 \node at (1.5,0) {{\scriptsize $\blacksquare$}};
 \node at (3,0) {{\scriptsize $\blacksquare$}};
 \node at (6,0) {{\scriptsize $\blacksquare$}};
 \node at (7.5,0) {{\scriptsize $\blacksquare$}};
 \node at (9,0) {{\scriptsize $\blacksquare$}};
 \end{tikzpicture}
 
    \caption{Two different chord diagrams that contribute to the moment $M(4,4,4)$. (a) and (b) are the full chord diagrams, while (c) and (d) are their contracted diagrams respectively. Diagram (b) (or it's contraction (d)) is connected while diagram (a) (or its contraction (c)) is not.
    }
    \label{fig:chord_diag}
\end{figure}

{\it Diagrammatic representation:} A convenient diagrammatic representation of this quantity is the following. As the distribution of the random variables $J_I$ is Gaussian, we use Wick's theorem to pair the $J_I$'s. We then represent this graphically as `chord diagrams' (this is similar to the construction in \cite{Berkooz_2019}, though we stress the following results are valid both in the double scaled limit and in the finite $p$ large $N$ limit). We then proceed as follows:
\begin{itemize}
    \item Each trace is drawn as a circle with $k_j$ nodes marked on it, one for each Hamiltonian insertion.\footnote{Morally speaking, but technically not quite correct, this circle corresponds to the thermal circle. It is exactly the thermal circle if we look at expressions like $\langle \tr((1+(\beta/L)H)^L) \rangle$ when $L\rightarrow\infty$.}
    \item Wick contraction of the coefficients $J_I$ is shown by drawing chords connecting the appropriate nodes. These chords can go between nodes in the same circle, or between different circles.
\item The connected moments only contain contributions from the pairings that fully connect the different traces.\footnote{By fully connected, we mean that for any two traces, there is a sequence of chords connecting them.}
\end{itemize}
An example of such a `multi-trace chord diagram' can be seen in figure \ref{fig:chord_diag}\textcolor{darkred}{(a,b)}.

Next, we can represent these multi-trace chord diagrams in a more concise way as follows: we represent each trace by a single vertex (instead of a circle), and preserve the external chords connecting the different traces, omitting the internal chords. Only if the resulting diagram is fully connected will it contribute to the connected moment.  An example of this is presented in figure \ref{fig:chord_diag}\textcolor{darkred}{(c,d)}. In order to avoid confusion with the vertices in chord diagrams, we represent each trace by a square-like vertex, and refer to these diagrams as {\it `multi-trace diagrams'}.

We will use this compact diagrammatic description to compute the leading order connected contributions to the moments and the thermal partition function. We shall start by focusing on the double trace expectations as an illustrative example in section \ref{sec:double_trace}. Then, we move on to the general multi-trace expectations in section \ref{sec:cactus}.

\subsection{Double trace expectation value} \label{sec:double_trace}

We start by computing the leading order contribution to the double trace connected moments $M_c(k_1,k_2)$ (see \eqref{eq:double_tr_explicit}). We shall do this by considering the number of index sets paired between the two traces, and show that the minimal number of pairings is 2 and that it dominates at leading order. 

Consider first a single pairing of $H$'s between the two traces. All other multi-fermion operators $\Psi_I$ are contracted within themselves in each trace, and the fermion species in these contractions appear in pairs. This leaves the fermion species, in a single $H$, unpaired within the same trace, and the result of the trace in the Hilbert space is zero (because $\tr(\psi_i)=0$). So a single pairing between the two traces gives an exactly vanishing contribution.

\begin{figure}[h]
\centering
\includegraphics[width=0.9\textwidth]{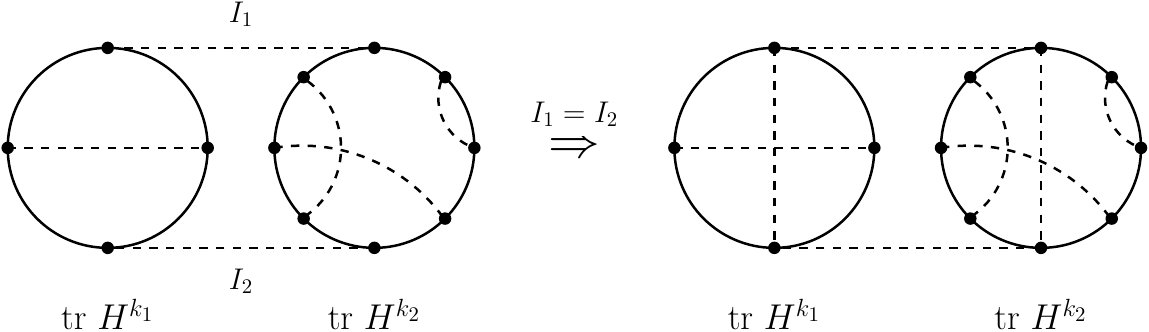}
\caption{An example of a diagram contributing to the double trace with $k_1=4$ and $k_2=8$.}
\label{fig:double_trace_CD}
\end{figure}

Next we may have two pairings with index sets $I_1$, $I_2$, each of them corresponding to a pairing between the two traces. An example of this is shown in the left hand side of figure \ref{fig:double_trace_CD}. In this case the traces do not vanish only if $I_1=I_2$ for the same reason as above.
Since the same index set $I_1=I_2$ appears twice in each trace, the corresponding nodes within each trace are effectively contracted, as shown in the right hand side of figure \ref{fig:double_trace_CD}. A priori the effective contractions within each trace are not independent (they are the same index set in the two traces), so naively we cannot do the sum over the index sets in each trace separately. However, since the trace is independent of permutations of the $i=1,\cdots,N$ sites, and we sum over all the other index sets, we have in \eqref{eq:double_tr_explicit} the exact relation
\begin{equation} \label{eq:double_tr_ind_indices}
\begin{split}
& \binom{N}{p} ^{-k/2} \sum _{I,J,K,\cdots } \tr \left( \Psi_I \Psi_J \cdots \Psi_I \cdots \right)  \tr \left( \Psi_I \Psi_K \cdots \Psi_I \cdots \right)  \\
& = \binom{N}{p} ^{-k/2-1} \sum _{I,I',J,K,\cdots } \tr \left( \Psi_I \Psi_J \cdots \Psi_I \cdots \right)  \tr \left( \Psi_{I'} \Psi_K \cdots \Psi_{I'} \cdots \right).
\end{split}
\end{equation}
We see that this factorizes into the product of the appropriate single-trace chord diagrams that appear in the expectation value of a single trace, down by a factor of $\binom{N}{p}^{-1}$ (originating from the constraint $I_1=I_2$).

Each choice of two internal diagrams for the two traces comes from $2 \frac{k_1}{2} \frac{k_2}{2} $ choices of which two lines go from one trace to the other,\footnote{The factor of 2 comes, when doing the original Wick contraction, from exchanging the endpoints of the two lines in one of the sides only.} so we get
\begin{equation} \label{eq:two_pairings}
    \left. M_c(k_1,k_2) \right|_{\text{2 pairings}}
     = \frac{k_1 k_2}{2} {\binom{N}{p}}^{-1} ~ M(k_1)  M(k_2) .
\end{equation}

In \cite{Cotler:2016fpe} and \cite{verbaarschot2019}, it was found that having $k$ pairings between the traces results in a suppression by a factor which is $\binom{N}{p}^{-k/2}$ for small $k$, and becomes a flat exponential suppression of $2^{-N}$ for large enough $k$. Thus the dominant contributions come from minimizing the number of pairings between the traces, at least for early times.

There are actually different ways in which more than two pairings can appear.
For example, we could have four index sets $I_1,\cdots ,I_4$ connecting the two traces, with one possibility for a non-vanishing contribution being $I_1=I_2$ and $I_3=I_4$.\footnote{In general, for four contractions, the constraint on the $I_j$'s is less restricting, as we will discuss in the general case later.}
Clearly this diagram is subleading because the $I_3=I_4$ imposes another constraint on the index sets. But there could also be more complicated situations. A particular case, which we deal with in section \ref{sec:NewFluc}, is the case in which three sets $I_1,I_2,I_3$ are identified between the two traces. By the non-vanishing trace condition, as before, $I_3$ is fixed to be those indices in $I_1$ and $I_2$ that are not in their intersection. However, since the size of $I_3$ should be $p$, we must have in addition that $|I_1 \cap I_2|=p/2$, giving a further suppression. This effect is subleading in the partition function, but it is the leading effect for traces with odd numbers of $H$'s, and hence can be isolated. But generally, the minimal number of pairings between the traces dominates the connected component, and hence the leading contribution is equation \eqref{eq:two_pairings}.  We note that this is not a new result, and was derived several times before \cite{feng2018spectrum,verbaarschot2019,Speicher2019}.

It will be convenient to organize the corrections that we calculate in terms of the small parameter
\begin{equation}
\epsilon \equiv {\binom{N}{p}}^{-1}    .
\end{equation}
We can now exponentiate \eqref{eq:two_pairings} to arrive at the leading order contribution to the connected double trace thermal partition function
\begin{equation} \label{eq:double_thermal_partition}
    \mathcal{Z}_c(\beta_1,\beta_2) = \frac{\epsilon}{2} \beta_1 \dd{\mathcal{Z}(\beta_1)}{\beta_1} \beta_2 \dd{\mathcal{Z}(\beta_2)}{\beta_2} + \text{higher order terms}. 
\end{equation}
Here $\mathcal{Z}(\beta)$ is the expected thermal partition function of the SYK model. If we consider the SYK model in the Schwarzian regime then $\mathcal{Z}(\beta)$ is the Schwarzian thermal partition function \cite{MaldecenaStanford,Stanford_Witten2017},
\begin{equation} \label{eq:Z_Schwarzian}
    \mathcal{Z}(\beta) = \beta^{-3/2} e^{-\beta E_0 + S_0+\frac{C}{2\beta}},
\end{equation}
while if we consider the SYK model in the double scaled limit then $\mathcal{Z}(\beta)$ was calculated in \cite{Micha2018,Berkooz_2019}. But, of course, equation \eqref{eq:double_thermal_partition} holds throughout all energy ranges, and for any length $p$ of the Hamiltonian (as long as the couplings are Gaussian).

We can also compute the connected spectral density function directly from \eqref{eq:two_pairings} to be
\begin{equation} \label{eq:ConnRho}
\rho_c(E,E') =  \frac{\epsilon}{2}~\frac{d}{dE}\left(E\rho_0(E)\right)~\frac{d}{dE'}\left({E'}\rho_0(E')\right),
\end{equation}
where $\rho_0(E)$ is the averaged spectral density function of the model.

\subsection{Multi-trace expectation values and cactus diagrams} \label{sec:cactus}

Next we would like to see what are the leading contributions for a general number of traces when considering the connected part after disorder average. We will see that these corrections can be described concisely via sums over 1PI `cactus' diagrams (to be defined below).

\begin{figure}[h]
\centering
\includegraphics[width=0.2\textwidth]{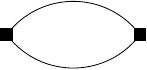}
\caption{The leading diagram contributing to the double trace, where the edges represent the chords connecting the different traces.}
\label{fig:traces2}
\end{figure}

\begin{figure}[h]
\centering
\includegraphics[width=0.2\textwidth]{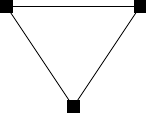}
\caption{A cyclic diagram for three traces.}
\label{fig:traces3}
\end{figure}

We again represent the different possibilities for chords connecting different traces by diagrams, as described above. In these diagrams, every trace is represented by a vertex, and the edges are the chords that go between the different traces. The computation in the previous subsection, for the leading order contribution to the connected double trace, just corresponds to the diagram in figure \ref{fig:traces2}. Another example, which contributes to the three traces connected correlator is shown in figure \ref{fig:traces3}. Because of the further suppression when having more than two external contractions in a trace, as was the case in the double trace, one could expect that only nodes of degree two will be present at leading order (like the diagram in figure \ref{fig:traces3}), but this is not the case as we will soon see.

To explain this, we will show that we can consider each diagram as a ``Feynman diagram'' in the following way. Each index set $I$ represents a momentum, $r$, valued in $\mathbb{Z} _2^N$, so that $r_i=1$ if $i \in I$ and otherwise $r_i=0$ (where $i=1,\cdots,N$). The condition that at a given vertex the $I$'s are such that each index $i$ appears an even number of times (i.e., the non-vanishing of the trace) is exactly $\sum r^{(j)}=0$ for $r^{(j)}$ being the incoming momenta to this vertex, and the equations are valued in $\mathbb{Z} _2^N$, which is precisely momentum conservation at the vertices. We see that this is a direct analogy to Feynman diagrams.

The first claim we would like to present is that \emph{in the connected average of $V$ traces, the leading behavior is $\epsilon ^{V-1} $}. To show this, recall that for every chord we have one explicit factor of $\epsilon $ coming just from the normalization of the Hamiltonian,  so overall this gives a scaling of $\epsilon ^E$ where $E$ is the number of edges.\footnote{We can also ignore all the insertions of $H$ which are contracted within the same trace since after summing over their index sets they do not contribute factors of $\epsilon$.} Similarly, each free index set $I$ (that we sum over) gives something of order at most $\sum _I 1 = \binom{N}{p} =\epsilon ^{-1} $. The question then is how many free momenta, i.e., index sets, we have.

Each vertex gives one momentum conservation equation, and of course as usual one vertex gives just overall momentum conservation of the external lines, that we do not have here; so it just gives $\delta (0)$ and no additional constraints.
So we have $V-1$ constraints on the momenta, so long as the graph is connected.
\footnote{It is important in this argument that each such constraint really fixes one momentum, or one index set $I$, and for this it is crucial that we consider connected diagrams, as otherwise we get less than $V-1$ such constraints. For a general graph the number of constraints is $V-N_c$, where $N_c$ is the number of connected components.}


Now, without further restrictions, all that those constraints would do is to eliminate precisely $V-1$ of the free momenta. However, an important point is that our momenta are in addition restricted to satisfy $|I|=p$. If a particular momentum conservation boils down to be setting the sum of two momenta to be vanishing, this $|I|=p$ restriction does not change the counting. However, this restriction can give further constraints if more momenta are involved. For instance, for three momenta $I_1,I_2,I_3$ incoming to a vertex, momentum conservation indeed fixed $I_3$ to be $I_1 \cup I_2$ minus their intersection; this is what was taken into account in this counting. However, because $|I_j|=p$, not only is $I_3$ fixed, but moreover $I_1$ and $I_2$ are now restricted to satisfy $|I_1 \cap I_2|=p/2$ in order to have a solution. As was mentioned, we will see that this still does not mean that at leading order we can have only vertices of degree two.
What it does mean is that taking into account all the possible momenta conservations, if we cannot express them as setting only sums of two momenta to be zero at a time, we will obtain a further suppression (that is, the suppression will be greater than just fixing $V-1$ momenta) and the result will be subleading. This point will be essential below.

With at least $V-1$ constraints we get at most $E-V+1$ free momenta and thus a suppression of at least
\begin{equation}
\epsilon ^E \frac{1}{\epsilon ^{E-V+1} } =\epsilon ^{V-1} .
\end{equation}
It is simple to check that the circular diagrams of the form of figure \ref{fig:traces3} indeed saturate this bound. This justifies the claim above that this is the leading order connected contribution for $V$ traces.

To arrive at the main claim of this section, let us use the following graph theory terminology. A \emph{path} is a sequence of vertices $v_1-v-\cdots -v_2$ connected by edges such that all vertices (and edges) are distinct. A connected graph is a graph where every two vertices are connected by a path. A graph \emph{cycle} is a sequence of vertices $v_1-v_2-\cdots -v_n-v_1$ connected by edges, such that we do not visit an edge twice, nor a vertex twice except for the first and the last one.

\begin{figure}[h]
\centering
\includegraphics[width=0.5\textwidth]{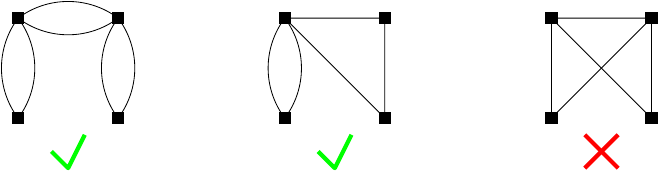}
\caption{Examples of diagrams going as $\epsilon ^3$ and sub-leading ones.}
\label{fig:cactus}
\end{figure}

The graphs saturating the $\epsilon^{V-1} $ leading behavior, and so those which contribute at leading order, are connected 1PI graphs,\footnote{Recall that a graph is 1PI if it is connected and cannot be made disconnected by cutting a single edge.} such that any pair of cycles have no edge in common (equivalently, any two cycles have at most one vertex in common.)
In graph theory, connected graphs in which no pair of cycles share an edge are known as \emph{cactus graphs} (or cactus trees) \cite{weissteinCactus}; for examples, see figure \ref{fig:cactus}.
The claim is then that \emph{the leading order contribution to the connected moments scales like $\epsilon ^{V-1} $, and comes only from the connected 1PI cactus graphs}.

\begin{figure}[h]
\centering
\includegraphics[width=0.3\textwidth]{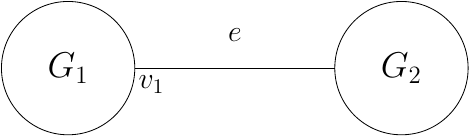}
\caption{A non-1PI graph.}
\label{fig:1PI}
\end{figure}

The restriction to 1PI diagrams in the claim above is easy to understand using the Feynman diagram structure.
To show this, consider a graph that is not 1PI, where $e$ is the edge that connects two connected components that are otherwise not connected to each other (see fig.\ \ref{fig:1PI}). Since $e$ is the only external leg of the connected components, by momentum conservation of each of these components we see that $r(e)=0$, which is a contradiction as $|\{i \,:\, r_i(e)=1\}| = p$. Thus the diagram's contribution must vanish identically.

\begin{figure}[h]
\centering
\includegraphics[width=0.4\textwidth]{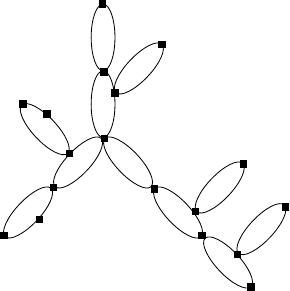}
\caption{ A generic 1PI cactus diagram. It is a tree of bubbles, with the nodes between the bubbles being vertices, and any other vertices of degree 2 can appear on the edges. }
\label{fig:1PI_cactus_dg_g_2}
\end{figure}

Moreover, we can give a simple characterization of these diagrams that give the dominant contribution.
A generic 1PI cactus diagram is just a \emph{tree} of ``bubbles'' attached at vertices. Along the edges, there can be arbitrarily many vertices of degree two (where each bubble contains at least two vertices). A generic example is shown in figure \ref{fig:1PI_cactus_dg_g_2}. This fact, as well as the dominance of 1PI cactus diagrams, are proved in the following two subsections.

\subsection{General structure of 1PI cactus diagrams} \label{sec:gen_1pi}

In order to characterize the general structure of 1PI cactus diagrams, we should concentrate on vertices of degree greater than 2, since as we saw a vertex of degree two only imposes that the two incoming momenta are equal (in $\mathbb{Z}_2$). Therefore, consider a vertex $v$ of degree $>2$ in a 1PI cactus diagram, with edges $e_1$, $e_2,\cdots$.

Consider going from $v$ along $e_1$, denoting the other endpoint by $v_1$ (we denote this by $v\overset{e_1}{-}v_1$). Then necessarily we can complete it to a cycle ending on $v$. Indeed, erase $e_1$ --- since any non-vanishing graph is 1PI --- it is still connected, so we have a path from $v_1$ to $v$ not going through $e_1$; it necessarily completes to a cycle $v \overset{e_1}{-}v_1-\cdots -v_i \overset{e_i}{-}v$ with $e_i \neq e_1$ (it is indeed a cycle). Now start from some other $e_j \neq e_1,e_i$ and do the same, getting another cycle ending on $e_k$, with all $v_1,v_i,v_j,v_k$ being distinct because no two cycles share the same edge. Therefore we see that $v$ looks like a flower, with all its edges forming distinct cycles.
In particular the degree of each vertex is even for 1PI cactus graphs (this is not true for a general cactus graph).

\begin{figure}[h]
\centering
\includegraphics[width=0.3\textwidth]{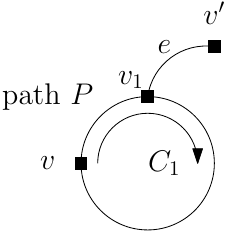}
\caption{A cycle $C_1$ around $v$.}
\label{fig:path_C1}
\end{figure}

\begin{figure}[h]
\centering
\includegraphics[width=0.4\textwidth]{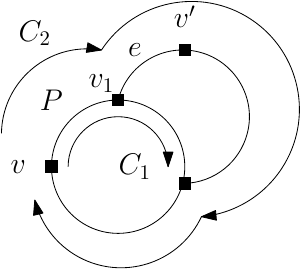}
\caption{The case described in the analysis in the text, where we get two cycles $C_1$ and $C_2$ with a common edge.}
\label{fig:path_C2}
\end{figure}

Every node on each of these cycles can be of degree two, or it can be of higher degree. In the latter case, we saw that it can have further cycles emanating from it. But let us show that they are only of the form we met, i.e., distinct cycles. Consider the first such vertex $v_1$ of higher degree on one of the cycles $C_1$ that we found that encircles $v$; see fig.\ \ref{fig:path_C1}. Let $e$ be an additional edge of $v_1$. Deleting $e$, since again the graph is 1PI, we can find a path $P'$ from $v'$ to $v$. If this path does not share a common edge with $C_1$, then we get the cycles $C_1$ and $P-e-P'$ which have a common edge (where $P$ is the path from $v$ to $v_1$, see the figure), being a contradiction. Otherwise, the path comes to $C_1$. It cannot cross before $v_1$ since $v_1$ is the first higher degree vertex. If it crosses $C_1$ after $v_1$, then we can still get two cycles $C_1$, and $C_2$ with a common edge, as shown in fig.\ \ref{fig:path_C2}.

The only remaining possibility is that the path gets back to $v_1$, meaning it is a cycle. This is possible, since the analogous construction of $C_2$ would not give a cycle.\footnote{In more detail, $C_2$ defined similarly to before will visit the vertex $v_1$ twice and so does not match the definition of a cycle.} Applying the entire same argument with $v$ replaced by any other vertex on these cycles, we get that every such cycle can have more distinct cycles, or `bubbles'', emanating from its vertices (that is, we can attach to any such bubble more bubbles), but these are the only cycles that we have. In particular, the most general non-vanishing cactus graph (which is 1PI and all vertex degrees being at least two) takes the form of the example in fig.\ \ref{fig:1PI_cactus_dg_g_2}. More precisely, as claimed, it is a tree of bubbles attached at vertices, where any additional vertices of degree two can appear on edges. There is nothing more than those bubbles; this is because as we saw, every edge belongs to a cycle necessarily, and we mapped all the possible cycles.\footnote{Note that in this construction, we can attach a new bubble to any other bubble, but we cannot attach a new bubble simultaneously to two existing bubbles, resulting in cycles sharing an edge.}

\subsection{Dominance of cactus diagrams} \label{sec:dominance}

We saw that contributions to the connected moments cannot be larger than the scaling $\epsilon ^{V-1}$. We will now prove the claim from section \ref{sec:cactus} that 
\textit{a connected graph saturates this bound if and only if it is a  1PI cactus graph}. (As we saw, the contribution of non-1PI graphs vanishes.)

In order to prove this, let us recall when the bound $\epsilon ^{V-1}$ was tight. At any vertex, we have momentum conservation stating that the sum of the incoming momenta vanishes. However, the important point is that when taking into account momentum conservation in other vertices as well, we may find out that not only the sum of all momenta vanishes, but also there are subsets of momenta where the sum of momenta vanishes in each subset separately.\footnote{For example, a vertex with four edges naively has $r_1+\cdots +r_4=0$, but taking into account more vertices may result in $r_1+r_2=0$ and $r_3+r_4=0$ separately, as in the case of a diagram made out of two cycles emanating from one vertex.} After taking all the momentum conservation into account, when reducing those subsets to be of the smallest size possible, the question is whether there is at least one subset containing three or more momenta. If that is the case, then we saw that we will get a larger suppression than $\epsilon ^{V-1}$. This is because the momentum conservation not only fixes one momentum (giving the counting that lead to $\epsilon ^{V-1}$) but also constraints the others. For example, in the case of a subset of three momenta corresponding to chords $I_1$,$I_2$,$I_3$, not only that $I_3$ was completely fixed, but $|I_1 \cap I_2|=p/2$ necessarily, which gave a further suppression. If on the other hand, all momentum conservations lead to subsets including only \emph{pairs} of momenta, we will have the $\epsilon ^{V-1}$ leading behavior.

Let us start with the direction $ \Leftarrow$, showing that all cactus diagrams scale as $\epsilon^{V-1}$. Consider a vertex $v$ of degree $>2$ with edges $e_1$, $e_2,\cdots$, as in the previous subsection (if its degree is 2, then necessarily it gives a momentum conservation involving only two momenta). If we show that any such vertex has no further suppression, then we will get the minimal $\epsilon ^{V-1} $.

Given the general structure of cactus diagrams that we found in the previous subsection, this follows immediately. Indeed, because of the bubble structure found, as we saw the momentum on $e_1$ equals that on $e_i$ (in the notation used in the previous subsection), the momentum on $e_j$ is the same as on $e_k$, and so on.
Thus at each vertex we actually get that the sums of pairs of momenta vanish based on these bubbles, and we have the minimal $\epsilon ^{V-1} $ suppression.


In the other direction $ \Rightarrow$, let us show that if we have two cycles with a common edge, we get a further suppression and so no $\epsilon ^{V-1} $ behavior. That is, we must show that there is a minimal subset of momenta satisfying momentum conservation, containing at least three momenta.

\begin{figure}[h]
\centering
\includegraphics[width=0.3\textwidth]{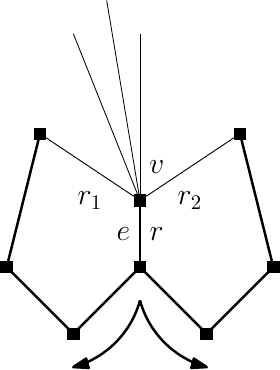}
\caption{Two cycles with a common edge.}
\label{fig:other_direction}
\end{figure}

Consider a pair of two such cycles with a common edge. Since they are different, we can find a vertex $v$ of degree $>2$, having edges with momenta $r,r_1,r_2,\cdots $; see fig.\ \ref{fig:other_direction}.
In order to construct a minimal set of momenta satisfying momentum conservation in a well defined way, let us apply the following painting procedure:
\begin{itemize}
    \item We go from $v$ in the direction of the common edge $e$ with momentum $r$, painting the edge $e$.
    \item In each step there are the vertices that were touched by the edges we painted in the previous step --- in the next step we apply momentum conservation, painting all the rest of the edges of these vertices that are not already painted.
    \item We apply this painting procedure at each vertex we come across other than $v$.\footnote{We never get back to a vertex we visited before, since all of its surrounding is already painted.} Therefore we will stop (necessarily after a finite number of steps) when there is no vertex having both painted and unpainted edges, with the only allowed exception being $v$.
\end{itemize}
We should think about this procedure as an equation, where the sum of the momenta painted in the previous step equals the sum of the newly painted momenta.

\begin{figure}[h]
\centering
\includegraphics[width=0.3\textwidth]{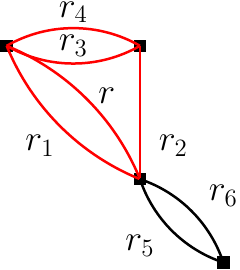}
\caption{Applying momentum conservation for a common edge to two cycles. In this case we get $r=r_1+r_3+r_4=r_1+r_2$. Here the resulting equation includes only $r,r_1,r_2$ without $r_5,r_6$, but in other cases we may get more momenta. This diagram scales as $\epsilon ^4$.}
\label{fig:minimal_momentum_conservation}
\end{figure}

We can do this process in any order we would like with the same result. Therefore, we can just as well go first along the two cycles. We thus get $r_1$ and $r_2$ necessarily. After we complete all vertices other than $v$, we may get some additional edges of $v$.
Thus we found a subset of momenta satisfying $r+r_1+r_2+\cdots =0$. By the procedure we did, there is necessarily no smaller subset containing $r$ that satisfies momentum conservation, while this subset includes at least three momenta. This means that this diagram is suppressed more than $\epsilon ^{V-1}$ as we saw. An example of this construction is shown in figure \ref{fig:minimal_momentum_conservation}. This completes the proof showing the dominance of cactus diagrams.

\subsection{Partition functions and density of states correlations} \label{sec:pf_dos_cor}

With the understanding of which graphs to consider, we move to calculate the leading order contribution to multi-trace moments and thermal partition functions. Let us first consider the contribution of each cactus diagram to the connected moments $M_c(k_1,\cdots ,k_n)$. Recall that from the disorder average, we have the contractions that link the different traces, as shown in the cactus diagrams, and the remaining contractions will form the usual chord diagrams in each trace separately. From now on we restrict to only cactus diagrams. For these, every full multi-trace chord diagram will be multiplied by two factors: (1) a power of $\epsilon $ according to the number of constraints, which as we saw is $\epsilon ^{V-1}$ for a cactus diagram, and (2)
a combinatorial factor that counts how many multi-trace chord diagrams result in the same internal chord diagrams for each trace (one such example appears in figure \ref{fig:double_trace_CD}); this contribution is described precisely below.  As we sum over the chord diagrams inside each trace, we get that $M_c(k_1,\cdots ,k_n)$ is $\prod_i M(k_i)$ times the two factors mentioned above.

All that remains is to describe the combinatorial factor in item (2). For every vertex $i$ of degree $d_i$ in a cactus diagram, each of the $d_i/2$ inner chords will be contracted to other vertices. So first we need to choose which chords participate in the multi-trace contraction, which is $\binom{k_i/2}{d_i/2}$ options. Clearly, there are $(d_i/2)!$ options to choose to which of the chords in each vertex we contract. Then we need to contract individual $J$'s (and not only chords) between the different traces. So what remains is simply to note that there are two ways to contract two chords. Similarly, for a cycle made of $n$ vertices of degree 2, there are $2^n$ options to contract the chords. Therefore we should assign a factor of 2 for every edge in the multi-trace diagrams. The exceptional case $n=2$ is accounted for by recalling that it has a $\mathbb{Z}_2$ symmetry, and we should divide by a symmetry factor $\hat S$ for each diagram. For a single such component of $n=2$ we have a factor of 2 contribution to $\hat S$ giving indeed the correct counting. In these kinds of cactus diagrams where the vertices are labeled, these $\mathbb{Z}_2$ symmetries are actually the only symmetries we have.

To summarize this, we have
\begin{equation} \label{eq:moments_as_cactus_diagrams}
    M_c(k_1,\cdots ,k_V)= \epsilon ^{V-1} \sum _{\text{Cactus diagrams}} \frac{2^E}{\hat S} \prod _{i=1}^V \left\{ \left( \frac{d_i}{2} \right)! \binom{k_i/2}{d_i/2} M(k_i)\right\},
\end{equation}
where $\hat S$ is the symmetry factor of each diagram, as described above.

Now let us consider the connected thermal partition function (remembering from the beginning of this section that all $d_i$ are even)
\begin{equation}
\begin{split}
    \mathcal{Z}_c(\beta _1,\cdots ,\beta_V) &= 
    \sum_{k_1,\cdots ,k_V} \frac{(-\beta_1)^{k_1}}{k_1!} \cdots \frac{(-\beta_V)^{k_V}}{k_V!} M_c(k_1,\ldots,k_V) \\
    & = \epsilon ^{V-1} \sum _{k_1,\cdots ,k_V}  \sum_{\text{Cactus diagrams}} \frac{2^E}{\hat S} \prod_{i=1}^V \left\{ \beta_i^{d_i} \left( \frac{\partial }{\partial \beta_i^2} \right)^{d_i/2} \frac{(-\beta _i)^{k_i}}{k_i!} M(k_i) \right\}.
\end{split}
\end{equation}

The cactus diagrams in the last formula are for fixed vertices associated to chosen $k_1,\cdots ,k_V$ (that is, the vertices are labeled). Therefore there is a large degeneracy in this formula, where we will need to include many different labelings of the same cactus diagram. Instead, it will be more convenient to have only one cactus diagram of each kind in the sum, without including permutations of the vertices. We can take care of this degeneracy by including a factor of $V!$, and then diving by the symmetry factor related to vertex permutations, as usual in Feynman diagrams; this symmetry factor will be included as usual in the symmetry factor $S$ of unlabeled graphs. Therefore we can write this as
\begin{equation} \label{eq:connected_Z_cactus}
    \mathcal{Z}_c(\beta_1,\cdots ,\beta_V)=\epsilon^{V-1} V! \sum_{\text{Cactus diagrams}}\frac{2^E}{S} \prod_{i=1}^V \left\{\beta_i^{d_i} \left( \frac{\partial}{\partial \beta_i^2}\right)^{d_i/2} \mathcal{Z}(\beta_i)\right\}.
\end{equation}
Let us clarify this formula. As mentioned, we want to count each distinct cactus diagram just once.
Then, in this formula, we fix an assignment of the $\beta _i$'s to the vertices of the cactus diagram in an arbitrary way. Of course, at the end, we know that the partition function is symmetric in the $\beta _i$'s, so the obtained expression in this formula is implicitly understood as symmetrized in the $\beta _i$.
As mentioned, the over-counting that can happen in specific diagrams is accounted for as usual by the symmetry factor $S$ (which is now enlarged with respect to the previous case of $\hat S$ since the vertices are not labeled, similarly to the usual internal vertices in a Feynman diagram). We give an example demonstrating how to use this formula below.

Finally, we can translate this to the joint density of states as well\footnote{Note for this equation, that $\rho _0$ is even.}
\begin{equation}\label{eq:connected_density_cactus}
        \rho _c(E_1,\cdots ,E_V) = \epsilon^{V-1} V! \sum_{\text{Cactus diagrams}} \frac{2^E}{S} \prod_{i=1}^V \left( - \frac{\partial }{\partial E_i} \frac{1}{2E_i} \right)^{d_i/2} 
        \left( E_i^{d_i} \rho_0 (E_i)\right). 
\end{equation}

We note that all these results hold to leading order in $N$, or equivalently in $\epsilon$. Furthermore, we again stress that the average density of states $\rho_0(E)$, thermal partition function $\mathcal{Z}(\beta)$, and moments $M(k)$, depend on the scaling and limit one considers the SYK model in; but the result itself is universal and valid in any large $N$ scaling, either the fixed $p$ or the double scaling limit.

The formula \eqref{eq:connected_density_cactus} for the connected density of states can be written as a superposition of integer (positive or vanishing) powers of the dilation operator $ E \frac{\partial }{\partial E}$ for the various energies acting on the leading order density of states $\rho _0$.

\subsubsection*{An example}

\begin{figure}[h]
\centering
\includegraphics[width=0.5\textwidth]{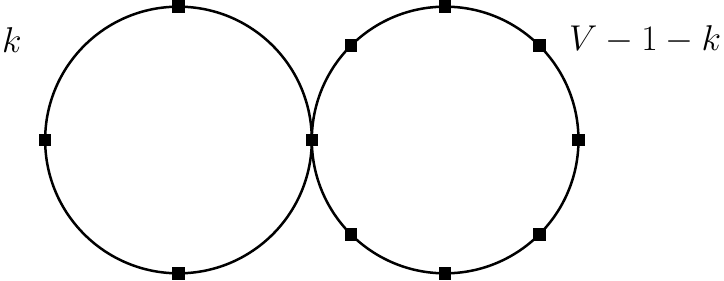}
\caption{The only form of connected diagrams with one vertex of degree 4, and the rest of degree 2. Excluding the vertex of degree 4, there are $k$ vertices on one cycle, and $V-1-k$ on the other.}
\label{fig:single_degree_4}
\end{figure}

As an example consider a diagram with $V$ vertices where one vertex is of degree 4 and the rest have degree 2. It is not hard to see that the only such connected diagrams are of the form shown in fig.\ \ref{fig:single_degree_4} (and they are cactus diagrams). The contribution to $\rho _c$ of such a diagram for a fixed number $k$ of degree 2 vertices on one of the circles  is
\begin{equation}
    \begin{split}
        & \epsilon ^{V-1} V! \frac{2^{V+1}}{4} \cdot \frac{1}{V} \bigg[  D_2(E_1) D_1(E_2) \cdots D_1(E_V)+ \\
        & \qquad \qquad \qquad \qquad +D_1(E_1) D_2(E_2) D_1(E_3) \cdots D_1(E_V)+\cdots \bigg] \prod _i \rho _0(E_i),
    \end{split}
\end{equation}
where the operator $D_k$ acting on $E_i$ is
\begin{equation}
    D_k(E_i) = \left( - \frac{\partial }{\partial E_i} \frac{1}{2E_i}\right) ^k E_i^{2k} .
\end{equation}
The symmetry factor is just 4 because each cycle has a $Z_2$ symmetry.\footnote{Note that when $V$ is odd and $k=(V-1)/2$, there is an additional symmetry factor of $2$ for exchanging the two cycles.}
We wrote the result explicitly, but it is simpler to write it in the symmetrized form as instructed in \eqref{eq:connected_density_cactus}. This means that we can simply forget the superscripts and just write this as
\begin{equation}
    \begin{split}
        & \epsilon ^{V-1} V! \frac{2^{V+1}}{4}D_2 D_1^{V-1} \prod _i \rho _0(E_i).
    \end{split}
\end{equation}
Again, this formula is understood by distributing arbitrarily the $V$ energies in the $V$ factors of $D_k$, and understanding the expression as symmetrized in the energies.

\subsection{Time scales and comparison to RMT fluctuations}\label{sec:TimeScl}

We have written the leading contributions to connected multi-trace correlators using dilatation operators acting on the spectrum. Actually this is the first out of a whole series of transformations on the spectrum (we discuss the next term in section \ref{sec:double_trace_3_lines} and comment on the others in section \ref{sec:GenFluc}). Each of them has an amplitude, and may become important in a different range of energies, or time scales. In this section we make some comments on the time scales associated with the leading operator, and compare to numerical data.\footnote{We would like to thank the authors of \cite{Cotler:2016fpe} for sharing their data with us.}

It is a well known empirical fact that the statistics of the nearest neighbor eigenvalue spacing in the SYK model is the same as that of a random matrix ensemble (RMT). The exact ensemble (GOE, GUE, or GSE) depends on the particle hole symmetry class, leading to a complete classification of the RMT ensemble based on the values of $N \mod 8$ and $p\mod 4$ \cite{You_2017,Garc_a_Garc_a_2016,Kanazawa_2017,Cotler:2016fpe}. This universal level spacing statistics is given by an exponentially suppressed term in the double trace spectral density, $\rho(E,E')$, which dominates at energy separations of the typical eigenvalue separation $E-E' \sim 2^{-N/2}$ (the first term on the RHS of equation \eqref{TwoPtAnstz}). The perturbative moment method we used does not allow us to find this universal term as it is non-perturbative or exponentially suppressed in $N$.\footnote{There may be a way to find this term in the moment expansion by re-summming some of the contributions from a large number of connections between the moments, as was alluded to in \cite{Cotler:2016fpe}, though we have not been able to preform such a re-summation.}

The range of accessible moments implicitly determines the scale of energy separation that we can probe. If we can reliably compute moments up to the $k$'th moment, then we can resolve energies (and energy separations) up to $1/k$ of the energy range. However, the approach above does not necessitate the explicit evaluation (or approximation) of any moment - rather we write the connected part as a dilation operator acting on the moments, for which we can take the exact value at each $k$. So the bottleneck question is up to what order is the operator reliable (in amplitude or in energy range). 

Generally, explicit evaluation and re-summation of moments is reliable for finite $k$ (in the limit of $N\rightarrow\infty$). But since we can act with our operator on the exact partition functions we are not limited by this. We can therefore hope that the leading contributions to the connected double trace spectral density are valid down to perhaps polynomially small energy separation of order $E-E'\sim N^{-\# p}$. Actually there are arguments that in this case the situation is considerably better, and that the connected double trace spectral density computed by the moment method is correct up to exponentially small energy separation (but still much larger than the typical level spacing of $2^{-N/2}$).

 The authors in \cite{Altland_2018} used a sigma model to compute the double trace spectral density and argued that at small separations $\omega = E-E' \sim 2^{-N/2}$ this spectral density consists of a random matrix part plus a one loop correction. The one loop correction can be written as a sum over massive modes which are suppressed by a power law in $N$: 
\begin{equation}\label{TwoPtAnstz}
    \rho_2(E+\omega, E - \omega) = [\rho(E)]^2 \left(\rho^{GUE}_2(\omega) + 2\Delta^2 \Re \sum_{k\geq 0,even} \binom{N}{k}\frac{1}{(i\omega + \epsilon(k))^2} \right),
\end{equation}
where $\rho^{GUE}_2(\omega)$ is the contribution from random matrix theory, $\Delta =  2^{-N/2}$ is the average energy spacing in the bulk of the spectrum, and $\epsilon(k) = T_k^{-1} - 1 $ are the masses of the massive modes with $T_k = \binom{N}{p}^{-1} \sum_{j=1}^p (-1)^j \binom{k}{j}\binom{N-k}{p-j} $. These massive modes become important at level spacing of the order $\omega \sim N^{\#} 2^{-N/2} $, which are still exponentially suppressed but much larger than the RMT scale. Furthermore, a calculation in \cite{verbaarschot2019} showed that the leading term in the massive modes agrees with the leading term moment calculation, while a two loop calculation in the sigma model scales like the next leading term in the moment expansion.
Thus there is some evidence that we should be able to trust the perturbative moment series up to energy separations that are exponentially small in $N$, and not just suppressed in powers of $N$.

(A caveat to this is that we were not been able to precisely match \eqref{TwoPtAnstz} to our formulas. There are two related reasons for this - the first is that \eqref{TwoPtAnstz} smears over the average energy, and the second is that it is more reliable at the center of the distribution, where the $\sigma$-model is defined. If we use our formula there, around the point $E=0$, then the leading correction is just a scaling $\left(1+\frac{1}{2}\epsilon \dd{}{E} E\right)\rho(E) \mid_{E=0} = \left(1+\frac{1}{2}\epsilon \right)\rho(E)\mid_{E=0} $, which matches \cite{verbaarschot2019}.)

We can translate the energy separation to a time scale by considering the spectral form factor \cite{Br_zin_1997,Liu_2018,del_Campo_2017}
\begin{equation} \label{eq:specformOG}
    g(t;\beta) = \frac{\mathcal{Z}(\beta+it,\beta-it)}{[\mathcal{Z}(\beta)]^2}.
\end{equation}
The late time behavior of the spectral form factor contains a ramp and a plateau \cite{Cotler:2016fpe,saad2018}, which is described by the appropriate $\beta$-ensemble. This ramp dominates the spectral form factor at exponential times which was approximated in \cite{Cotler:2016fpe} as $t_{dip}\sim e^{S_0/2}$, where $S_0 \propto N$ is the zero temperature entropy in the large $N$ limit (we will see in \eqref{eq:t_dip} that $t_{dip}$ gets modified when including the effects of the global modes). Thus we expect that the perturbative moment expansion is relevant up to these exponential times.

To be a little more precise, we mentioned before that the leading correction to the multi-trace correlator, which we computed so far, is only the first of an infinite (in the large $N$ limit) set of corrections. When we say that the moment expansion is relevant up to the time scales above, we mean the sum of these terms. However, for the purposes of comparisons at finite $p$ and $N$, which we do next, we use only the leading term computed above.

\begin{figure}[h]
    \centering
    \includegraphics[width = 16 cm]{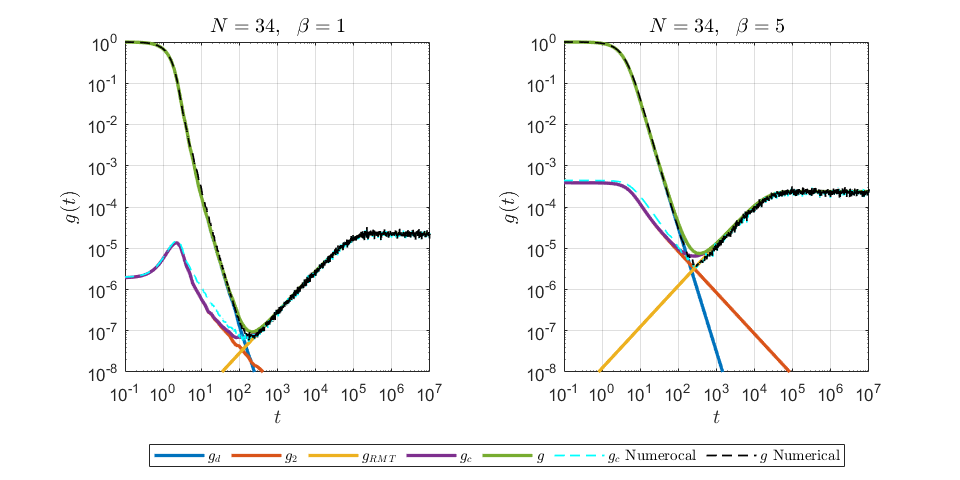}
    \caption{The approximate spectral form factor as a function of time for $N=34$ and $p=4$. $g= g_d+g_2+g_{RMT}$ is the full spectral form factor, $g_d$ is the disconnected portion given by the first term in \eqref{eq:specform}, $g_2$ is the leading connected portion given by the second term in \eqref{eq:specform}, and $g_{RMT}$ is the universal random matrix contribution given by the last term in \eqref{eq:specform}. The connected spectral form factor $g_c = g - g_d$ is also shown. These are matched to the numerical computation of $g$ and $g_c$ from \cite{Cotler:2016fpe}.}
    \label{fig:specform_N34}
\end{figure}

From our double trace moments we can approximate the known contributions to the spectral form factor as 
\begin{equation} \label{eq:specform}
\begin{aligned}
    \mathcal{Z}(\beta+it,\beta-it) \approx |\mathcal{Z}(\beta + it)|^2 & + 
    \frac{\epsilon}{2} \left(\beta^2+t^2\right) \left|\dd{\mathcal{Z}(\beta')}{(\beta')}\right|^2_{\beta' = \beta+it}\\
    & +\int dE~e^{-2\beta E} \min\left\{ N_d^2 \frac{t}{2\pi} ,N_d \rho(E)\right\},
    \end{aligned}
\end{equation}
where the first and third terms are from \cite{Cotler:2016fpe} - the first term is the disconnected contribution and the third term is the RMT contribution given in equation (42) there.\footnote{Note that this term is correct only for the GUE universality class, or when $N\mod 8 = 2,6$. For the GSE and GOE universality classes the ramp behavior is different, and this term is more complicated. See \cite{Liu_2018} for a recent overview of the spectral form factor in each RMT universality class.} The second term is the leading order contribution calculated in \eqref{eq:double_thermal_partition}. $N_d$ is a symmetry factor counting the degeneracy of each energy level (for even $N$ this factor is $N_d = 2$ for $N \mod 8 \neq 0$, and 1 for $N\mod8 = 0$). The normalized spectral form factor $g$ (given in \eqref{eq:specformOG}) can similarly be written as $g \approx g_d + g_2 + g_{RMT}$, where $g_d$ is the disconnected part, $g_2$ is the leading order correction and $g_{RMT}$ is the random matrix theory contribution. In figure \ref{fig:specform_N34} we present a plot of the approximate spectral form factor and its three contributions, which seems to match numerical computations of $g$ and the connected part $g_c$ based on data from \cite{Cotler:2016fpe}. See appendix \ref{app:spectrum} for more details on how this comparison was done, as well as further comparisons for $N=28,30,$ and $32$.

A more careful comparison of our $g_c$ and the numerical $g_c$ shows that they match with a typical deviation of less than 10\%. The maximal deviation is around 35\%. This is also to be expected - jumping ahead, the size of the first subleading correction is given in \eqref{SubLeadCont}. If we plug in $p=4, N=34$, and compute the ratio between \eqref{SubLeadCont} and the leading correction we obtain a deviation of 33\%, so this is the best that we can hope to do for this value of $N$.

\begin{figure} [h]
    \centering
    \includegraphics[width = 16 cm]{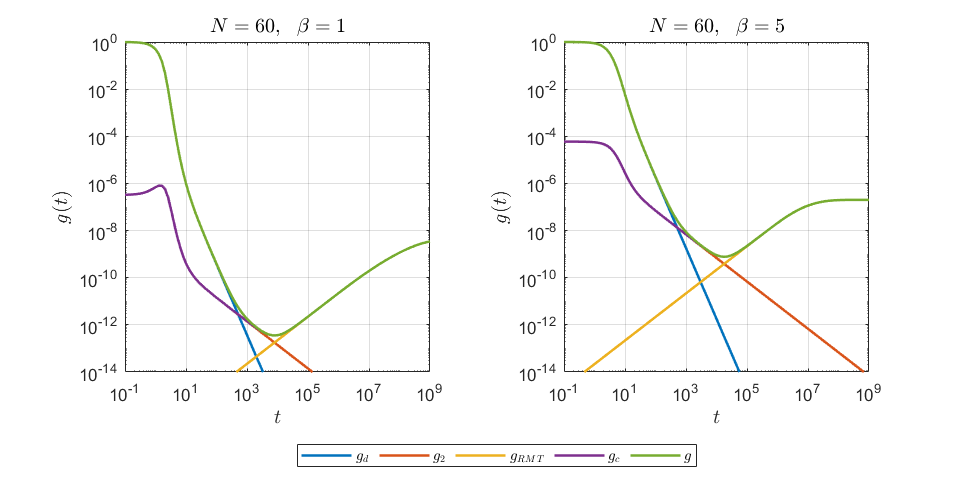}
    \caption{The approximate spectral form factor as a function of time for $N=60$ and $p=4$. There is a clear crossover region around $t \sim 10^3$ to $t \sim 10^4$ where the spectral form factor is dominated by $g_2$.}
    \label{fig:specform_N60}
\end{figure}

In the holographic Schwarzian regime, using \eqref{eq:Z_Schwarzian}, we can approximate the second moment contribution to the spectral form factor as
\begin{equation}
    g_2 \approx \frac{\epsilon E_0^2 \beta^3}{2 \sqrt{\beta^2+t^2}} e^{-\frac{C}{\beta}}.
\end{equation}
This term at late times decays as $g_2 \sim t^{-1}$, which is slower than the disconnected term $g_d\sim t^{-3}$. Thus we expect there to be some intermediate time frame where this term is the leading contribution to the spectral form factor, with a crossover time $t_c \approx \sqrt{2/(\epsilon E_0^2)}$. This crossover region is not seen for $N=34,p=4$, as the crossover time is very similar to the dip time, however for larger $N$ this intermediate region should exist. We present an example of such crossover region in figure \ref{fig:specform_N60}, where we plot the expected spectral form factor for $N=60$. Numerical verification of such crossover region may be within reach with the recent advances in numerical techniques that have been used to calculate certain correlation functions numerically for $N=60$ \cite{kobrin2020manybody}. 

This slower decay of $g_2$ compared to the disconnected spectral form factor also implies that the dip time (defined as the time of transition from the decay to the ramp) will be larger. As mentioned above, the dip time as inferred from the disconnected contribution behaves as $t_{dip}\sim e^{S_0/2}$ \cite{Cotler:2016fpe}. Taking into account the correction from $g_2$ and finding the time where it crosses the ramp behavior, we get
\begin{equation} \label{eq:t_dip}
    t_{dip}^{(2)} \sim \sqrt{\epsilon} E_0 e^{S_0}
\end{equation}
(where we show the scaling, dropping overall constant and $\beta $ factors). This is a much later time compared to $t_{dip}$. In fact, its exponential entropy dependent term is the same as that of the plateau time $t_p \sim e^{S_0+\frac{C}{2\beta }}$ found in \cite{Cotler:2016fpe}. They are still exponentially separated in $N$, in a temperature dependent way, owing to the $C/({2\beta})$ term (as $C$ is linear in $N$).

\section{Dual Vector Model} \label{sec:vector}

In this section we will take the Feynman rules that gave us cactus diagrams as the leading order contribution to the moments of the Hamiltonian, and show that they are equivalent to a zero dimensional vector field theory. We then show that this dual theory is in fact closely related to a theory for the random couplings, which allows us to extend this analysis to non-Gaussian distributions for the random couplings.

To obtain the vector model, we first go over the Feynman rules we found for evaluating a cactus diagram, and extend them to any Feynman diagram such that only cacti remain to leading order. The general rules for cactus diagrams are as follows:
\begin{enumerate}
    \item Each vertex represents a single trace, with $k$ Hamiltonian insertions.
    
    \item Each line connecting vertices is a propagator with the scaling of $\epsilon$.
    
    \item Each closed loop comes with a value of $1/\epsilon$ from the sum over index sets.
\end{enumerate}
Under these rules, a cactus diagram with $n$ vertices is of order $\epsilon^{n-1}$.

These rules are the same as for a large $M = 1/\epsilon$ 0-dimensional vector model. We can think of each trace as a vertex, and associate to each index set $I$ a dual scalar vector field $\phi_I$. Specifically a trace with $k$ insertions becomes a single vertex with a value
\begin{equation}
     M(k) \left(\sum_{I} \phi_I^2\right)^{k/2}.
\end{equation}
Furthermore, we take the propagator for $\phi_I$ to be Gaussian, with
\begin{equation}
    \expt{\phi_I \phi_J} = \epsilon ~ \delta_{IJ}.
\end{equation}
In this large $1/\epsilon$ vector model we see that propagators come with a factor of $\epsilon$, as desired, and closed index loops indicate a summation over all indices and thus give a $1/\epsilon$.

It is simple to check that the leading order term of any number of traces in this theory will be 1, which is achieved by contracting all pairs of identical index sets $\phi_I$ from the same trace with each other. This corresponds to the leading order disconnected moment, and indeed it has the right value of 1 (times the disconnected contribution). The leading order connected piece is indeed cactus diagrams, as before, and the combinatorial factor it gives is identical to the one above as the counting of the number of diagrams is exactly the same. 

\begin{figure}
    \centering
    \includegraphics[width = 8 cm]{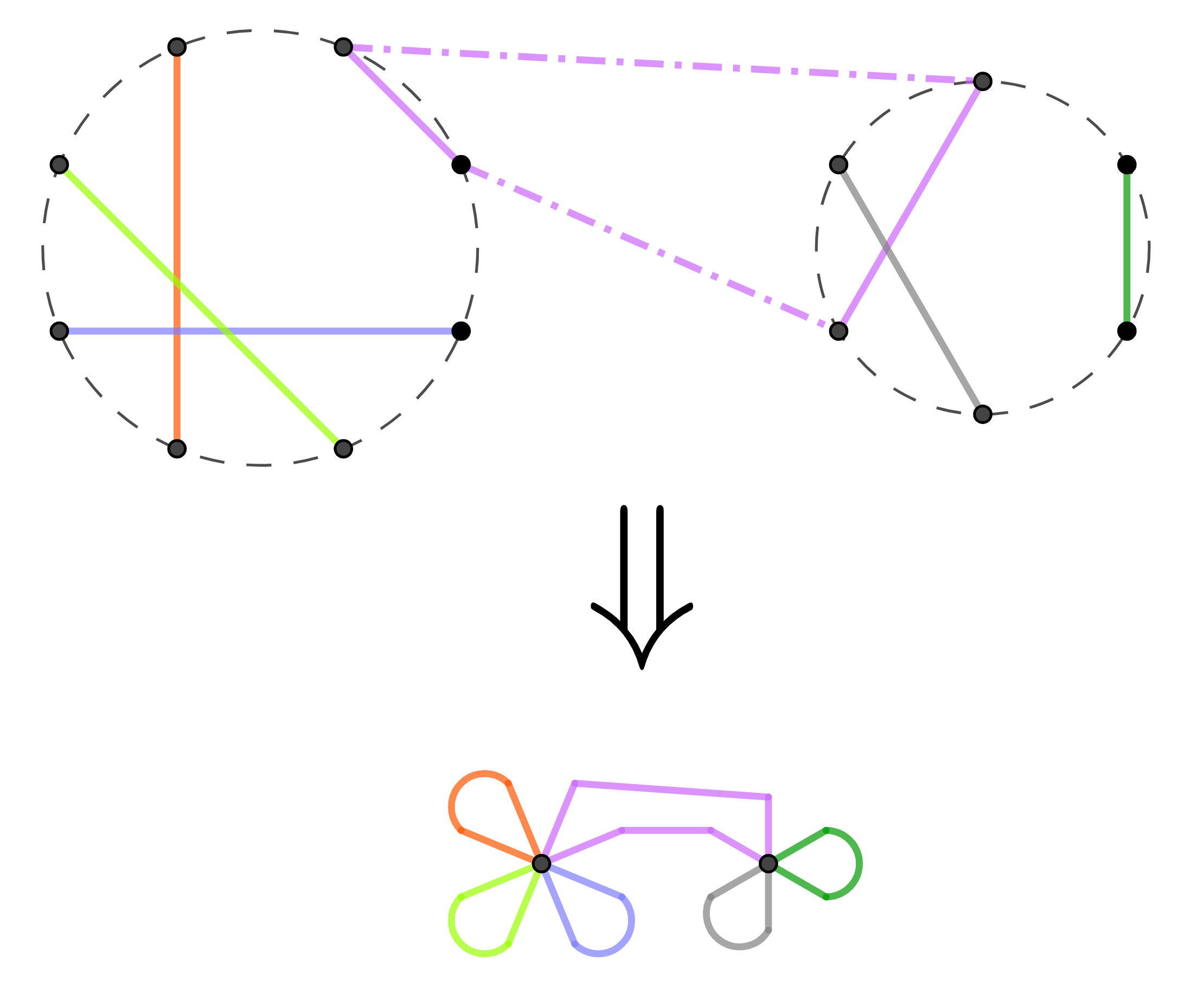}
    \caption{An example of a pair of chord diagrams translated into the vector model. Edges in both graphs correspond to index sets, with different index sets colored differently. Each trace became a vertex with $k/2$ pairs of index sets coming out, some of which are contracted to each other.}
    \label{fig:trace_to_vec}
\end{figure}

We can also translate the multi-trace chord diagrams to this large $1/\epsilon$ vector model, by simply thinking of each chord as a $\phi$ propagator and each trace as a vertex. An example of this procedure can be seen in figure \ref{fig:trace_to_vec}.

The complete joint moments (to leading orders) can be calculated using this dual model as
\begin{equation} \label{eq:moments_dual}
    M(k_1,\ldots , k_n) = C^{-1} \int d\phi_I~ e^{-\sum_{I}\phi_I^2 / (2 \epsilon)}
    \prod_{i=1}^n \left\{ M(k_i) \left(\sum_{I} \phi_I^2 \right)^{k_i/2} \right\}
\end{equation}
where $C$ is the normalization $C \equiv \int d\phi_I~ e^{-\sum_{I}\phi_I^2 / (2 \epsilon)}$.
We will denote the expectation in this vector model by
\begin{equation} 
\expt{\mathcal{O}(\phi)}_{\phi} = C^{-1} \int d\phi_I~ e^{-\sum_{I}\phi_I^2 / (2 \epsilon)} \mathcal{O}(\phi),
\end{equation}
so that $\expt{1}_\phi = 1$.
Then the connected expectation is defined recursively as in \eqref{eq:def_connected_moments}, only now with respect to the expectation over the dual fields $\phi_I$.
This dual vector model gives the correct leading order behavior of the connected moments.

We can also exponentiate the moments \eqref{eq:moments_dual} to get the joint thermal partition function as an expectation in this vector model
\begin{equation} \label{eq:thermal_from_vector}
    {\cal Z}(\beta_1,\ldots , \beta_n) = \expt{
    \prod_{i=1}^n \mathcal{Z}\left(\beta_i \sqrt{\sum_I \phi_I^2}\right)}_{\phi} .
\end{equation}
To calculate the leading order connected contribution we consider only the connected parts, or joint cumulants, on both sides of \eqref{eq:thermal_from_vector} with respect to the expectation over the $\phi_I$'s.

\subsection{The vector model as a theory for the couplings} \label{sec:vec_as_coup}

Already in the dual model results for the moments and the thermal partition function, equations \eqref{eq:moments_dual} and \eqref{eq:thermal_from_vector}, it is apparent that the expectation over $\phi_I$'s is similar to the expectation over the random couplings $J_I$. Not only do they correspond to the same index sets $I$, but they also have the same Gaussian distribution (once we re-scale the $\epsilon^{1/2}$ factor in the Hamiltonian). In this subsection we will make this equivalence exact. As a consequence of this, we can extend our results to non-Gaussian random couplings.

Throughout this subsection we will assume that the random couplings are independent identically distributed (i.i.d) with zero mean, unit variance (by their definition, with $\mathcal{J}$ extracted), and bounded moments (but not necessarily Gaussian).
We can also easily extend to other cases as we comment below.
Under these assumptions the SYK model is self averaging and independent of the exact distribution of the random couplings in the large $N$ limit \cite{Erdos14,feng2018spectrum}.\footnote{Single trace expectations are independent of the exact distribution of the couplings, while the multi-trace expectations factorize to single trace expectations at leading order.} As we will see, multi-trace connected expectations do however depend on the exact distribution of the couplings \cite{feng2018spectrum2}, and so we should consider different distributions for the couplings when computing connected multi-trace expectations. We will further assume in what follows that the distribution is even to make the computations simpler, though this is not strictly required.

The idea is that at the leading orders we can perform the trace first, before the expectation over the random couplings, and what remains is a theory for the couplings. We can do it perturbatively around a Gaussian model.

This is done as follows. Consider a particular contraction of $J_I$'s in the averaged multi-trace. The argument here is valid only for the contractions that are leading in $N$ for a given connected component with an even number of insertions. As derived above, at leading order in $N$ all the $J_I$'s \emph{in a given trace} will eventually come in pairs. As a result we can first evaluate each trace as a chord diagram, giving us a factor of the disconnected moment. 
There are various different contractions of the couplings that will give the same internal chord diagrams. At leading order in $N$, the same set of contractions is obtained by considering the expectation value of product of terms of the form $J_{I_1} J_{I_2} \cdots J_{I_1} \cdots J_{I_2} \cdots $ for every trace, with ordering according to the chord diagram in this trace. (That is, we impose the ordering of the chord diagrams using large $N$.) But this means the coefficients coming from the $J$'s of all chord diagrams are the same, and are just $\langle (\sum _I J_I^2)^{k_1/2} \cdots (\sum _I J_I^2)^{k_V/2}\rangle _J$. Note that in the traces over the fermions there are constraints on the indices, which just give the overall suppression by a power of $
\epsilon$, as explained before.

For clarity, let us give a simple example where we show this explicitly. We can do the following sequence of equalities for the particular contraction shown here, with an implied summation over index sets:
\begin{equation}
    \begin{split}
        &  \tr \Big( 
        \contraction[3ex]{}{J_{I_1}}{\Psi_{I_1} J_{I_2} \Psi_{I_2} J_{I_3} \Psi_{I_3} J_{I_4} \Psi_{I_4}\Big) \tr \Big(}{J_{I'_1}}
        \contraction[2ex]{J_{I_1} \Psi_{I_1} J_{I_2} \Psi_{I_2} }{J_{I_3}}{\Psi_{I_3} J_{I_4} \Psi_{I_4}\Big) \tr \Big( J_{I'_1} \Psi_{I'_1} J_{I'_2} \Psi_{I'_2}}{J_{I'_3}}
        \contraction{J_{I_1} \Psi_{I_1}}{J_{I_2}}{\Psi_{I_2} J_{I_3} \Psi_{I_3}}{J_{I_4}}
        \contraction{J_{I_1} \Psi_{I_1} J_{I_2} \Psi_{I_2} J_{I_3} \Psi_{I_3} J_{I_4} \Psi_{I_4}\Big) \tr \Big( J_{I'_1} \Psi_{I'_1}}{J_{I'_2}}{\Psi_{I'_2} J_{I'_3} \Psi_{I'_3}}{J_{I'_4}}
        J_{I_1} \Psi_{I_1} J_{I_2} \Psi_{I_2} J_{I_3} \Psi_{I_3} J_{I_4} \Psi_{I_4}\Big) \tr \Big( J_{I'_1} \Psi_{I'_1} J_{I'_2} \Psi_{I'_2} J_{I'_3} \Psi_{I'_3} J_{I'_4} \Psi_{I'_4}\Big)=\\
        & =
         \tr \Big( 
        \contraction[3ex]{}{J_{I_1}}{\Psi_{I_1} J_{I_2} \Psi_{I_2} J_{I_1} \Psi_{I_1} J_{I_2} \Psi_{I_2}\Big) \tr \Big(}{J_{I'_1}}
        \contraction[2ex]{J_{I_1} \Psi_{I_1} J_{I_2} \Psi_{I_2} }{J_{I_1}}{\Psi_{I_1} J_{I_2} \Psi_{I_2}\Big) \tr \Big( J_{I'_1} \Psi_{I'_1} J_{I'_2} \Psi_{I'_2}}{J_{I'_3}}
        \contraction{J_{I_1} \Psi_{I_1}}{J_{I_2}}{\Psi_{I_2} J_{I_1} \Psi_{I_3}}{J_{I_2}}
        \contraction{J_{I_1} \Psi_{I_1} J_{I_2} \Psi_{I_2} J_{I_1} \Psi_{I_1} J_{I_2} \Psi_{I_2}\Big) \tr \Big( J_{I'_1} \Psi_{I'_1}}{J_{I'_2}}{\Psi_{I'_2} J_{I'_1} \Psi_{I'_1}}{J_{I'_2}}
        J_{I_1} \Psi_{I_1} J_{I_2} \Psi_{I_2} J_{I_1} \Psi_{I_1} J_{I_2} \Psi_{I_2}\Big) \tr \Big( J_{I'_1} \Psi_{I'_1} J_{I'_2} \Psi_{I'_2} J_{I'_1} \Psi_{I'_1} J_{I'_2} \Psi_{I'_2}\Big)=\\
        & =
         \Big(
        \contraction[2em]{}{J_{I_1}}{ J_{I_2} J_{I_1} J_{I_2} \Big) \Big(
        }{J_{I'_1}}
        \contraction{J_{I_1}}{J_{I_2}}{ J_{I_1}}{ J_{I_2}}
        \contraction[1.5em]{J_{I_1} J_{I_2} }{J_{I_1}}{ J_{I_2} \Big) \Big(
        J_{I'_1} J_{I'_2} }{J_{I'_1}}
        \contraction{J_{I_1} J_{I_2} J_{I_1} J_{I_2} \Big) \Big(
        J_{I'_1}}{ J_{I'_2}}{ J_{I'_1}}{ J_{I'_2}}
        J_{I_1} J_{I_2} J_{I_1} J_{I_2} \Big) \Big(
        J_{I'_1} J_{I'_2} J_{I'_1} J_{I'_2}
        \Big)  \cdot
        \epsilon^3 \cdot \tr \Big( \Psi_{K_1} \Psi_{K_2} \Psi_{K_1} \Psi_{K_2}\Big)
        \tr \Big( \Psi_{K_1} \Psi_{L_2} \Psi_{K_1} \Psi_{L_2}\Big)=\\
        &=  \Big(
        \contraction[2em]{}{J_{I_1}}{ J_{I_2} J_{I_1} J_{I_2} \Big) \Big(
        }{J_{I'_1}}
        \contraction{J_{I_1}}{J_{I_2}}{ J_{I_1}}{ J_{I_2}}
        \contraction[1.5em]{J_{I_1} J_{I_2} }{J_{I_1}}{ J_{I_2} \Big) \Big(
        J_{I'_1} J_{I'_2} }{J_{I'_1}}
        \contraction{J_{I_1} J_{I_2} J_{I_1} J_{I_2} \Big) \Big(
        J_{I'_1}}{ J_{I'_2}}{ J_{I'_1}}{ J_{I'_2}}
        J_{I_1} J_{I_2} J_{I_1} J_{I_2} \Big) \Big(
        J_{I'_1} J_{I'_2} J_{I'_1} J_{I'_2}
        \Big)
        \cdot (\text{Chord diagram 1}) \cdot (\text{Chord diagram 2}).
    \end{split}
\end{equation}
Let us repeat the steps. In the first equality, we used the fact that in the leading order in $N$ contractions, the $J_I$'s come in pairs. This is the crucial step from which the argument already follows. But to be explicit about the chord diagrams we can continue with the evaluation. In the second equality we separated the indices of the $J_I$'s and the fermions $\Psi_I$'s, and it results in the explicit binomial coefficient (the variances from the contractions are the same in the second and third lines). In the third equality we used once again the argument in \eqref{eq:double_tr_ind_indices} by which we can make the indices in the traces independent, with the appropriate binomial suppression.

Therefore, we see that
\begin{equation} \label{eq:C_theory_moments}
    \expt{ \tr H^{k_1} \cdots \tr H^{k_n}}_J = \expt{ \prod_{i=1}^n \left\{ \left( \epsilon \sum_I J_I^2 \right)^{k_i/2} M(k_i) \right\} }_J .
\end{equation}
We note that \eqref{eq:C_theory_moments} only holds at leading order in $N$ for each connected component.
As \eqref{eq:C_theory_moments} matches the expression for the moments from the dual vector model, \eqref{eq:moments_dual}, it follows that the $\phi_I$'s of the vector model are really the random couplings $J_I$, up to a normalization of $\epsilon^{-1/2}$. 

The distribution of the couplings $J_I$ could include higher moments (independent of $\epsilon$) and the perturbative analysis goes through, with the result \eqref{eq:C_theory_moments}, where the average over $J$ given with the corresponding distribution. We analyze a simple example below.
We could just as well consider couplings that are not independent, and the result remains the same. The difference is that the scaling with $\epsilon$ of the higher moments should be taken appropriately, and it can be chosen so as not to affect the single trace moments.

\subsubsection{An example of non-Gaussian couplings}

As a simple example of such a computation with a non Gaussian distribution, consider adding a 4-point interaction term $\sum_I \frac{u}{4!} J_I^4$ to the distribution of the couplings. This adds a new vertex with 4 identical $\phi_I$ fields to the vector model. Notice that all the moments of $J_I$ for fixed $I$ are independent of $N$ and uniformly bounded (in the large $N$ limit), so we have universality of the single trace quantities, in the sense that they are independent of $u$. For example, $Z(\beta)$ does not depend on $u$ in the large $N$ limit but $Z_c(\beta_1,\beta_2)$ does.

\begin{figure}[h]
\centering
\includegraphics[width=0.5\textwidth]{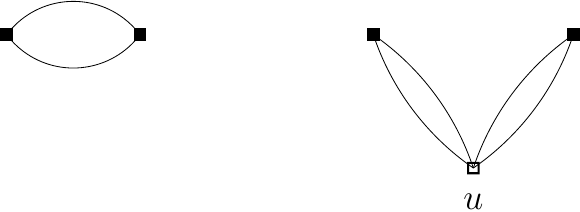}
\caption{Leading contributions to the double-trace in a non-Gaussian distribution.}
\label{fig:trace2_pert}
\end{figure}

Consider the double-trace moments, and let us consider them as a perturbation theory in $u$ at leading order in $\epsilon$ (or at each order in $\epsilon$). At leading order, we have the diagrams shown in fig.\ \ref{fig:trace2_pert}, where the notation is as before, and in addition by non-filled squares we denote the perturbations (recall that the filled ones stand for the traces). The first diagram is the only one we had before, and gave us something of the order $\epsilon $. The second diagram goes as $\epsilon ^2\frac{1}{\epsilon} u$ and so is also proportional to $\epsilon $.\footnote{The first factor of $\epsilon^2$ comes from the insertions, and the $1/\epsilon$ is from the sum over the free index $I$. Note that we do not allow self contractions of the interaction vertex.} We see it affects the (connected) double trace at the leading order in $\epsilon$. Its effect on the disconnected part is subleading. Thus the double-trace moments up to order $u$ (in the convention above) is
\begin{equation}
    M_c(k_1,k_2) = M(k_1) M(k_2) \cdot \epsilon \frac{k_1 k_2}{2}\left( 1+\frac{u}{2} + \cdots \right) .
\end{equation}

We can continue in perturbation theory in the standard way to find higher order corrections in $u$, or add additional interaction vertices for the couplings which can also be taken into account in a similar manner. Thus the vector model is a powerful tool to find the leading order multi-trace correlates for arbitrary distributions of the coupling.\footnote{So long as the distribution satisfies the standard requirements of being bounded with a normalized variance, so that the single trace expectations are universal and self averaging.}

We would like to emphasize the point that $M(k)$ does not depend on $u$, but $M(k_1,k_2)$ does. I.e., single trace quantities are universal, while connected multi-trace quantities are not, and even the leading corrections change as the distribution changes. This means that at the level of a single trace there is a single gravity description (independent of the distribution of the couplings) but the gravity description of the theory with several boundaries is not universal at the time scales that we are interested in. 

\subsection{The $h_2$ fluctuation parameter, and the gravitational description of the leading global mode}\label{H2DualAct}


We would like to interpret the results above in the gravitational dual theory (going back to the Gaussian distribution for the couplings). In equation \eqref{eq:moments_dual} we wrote (to leading order) the connected multi-trace correlator as a simple integral, where in the integrand we have the single trace quantities after the ensemble average. Therefore, to obtain the gravitational interpretation we just need to rewrite each trace in terms of the gravity dual, interpret the result, and we are done.

The only twist comes in the interpretation stage. The integral that we carry out is over $\binom{N}{p}$ quantities, and these do not necessarily have a meaning in gravity, which is the effective theory of the averaged quantities (in the SYK context). We would therefore like to replace this integral by a ``compressed'' version. In other words, we would like to find a set of ``minimal'' modifications to the gravity action that produces the full cactus expansion for any number of traces. We therefore begin by ``minimizing'' the vector model to the smallest possible set of degrees of freedom, which is just the fluctuation of $h_2=\sqrt{\sum \phi_I^2}$. This is similar to a single instance of the average over $\alpha$ states discussed in \cite{Coleman:1988cy,marolf_maxfield}.

This is straightforward. Starting from equation \eqref{eq:thermal_from_vector}, we can go to radial coordinates for the $\phi_I$'s with $h_2 \equiv \sqrt{\sum_I \phi_I^2}$, and integrate the $\binom{N}{p}-1$ dimensional sphere. Then we see that the joint thermal partition function is simply
\begin{equation} \label{eq:thermal_from_r}
    {\cal Z}(\beta_1,\ldots , \beta_n) = A^{-1}  \int_0^\infty dh_2~ h_2^{1/\epsilon - 1}e^{-h_2^2/(2 \epsilon)}
    \prod_{i=1}^n \mathcal{Z}\left(h_2 \beta_i \right)=\int_0^\infty dh_2 P_{h_2}(h_2)
    \prod_{i=1}^n \mathcal{Z}\left(h_2 \beta_i \right),
\end{equation}
where $A \equiv \int_0^\infty dh_2 ~ h_2^{1/\epsilon - 1}e^{-h_2^2/(2 \epsilon)}$, and $P_{h_2}(h_2) dh_2$ (implicitly defined by the equality) is a probability measure on the fluctuation parameter $h_2$.

From here we can do a simple saddle point calculation of this integral. The large $1/\epsilon$ action for $h_2$ has a saddle point at $h_2=1$
and expanding around this saddle point gives the cactus diagrams we saw before. For example, the leading order connected two trace thermal partition function is obtained by setting $h_2 = 1+\sqrt{\epsilon}\rho$, and expanding the integral for small $\epsilon$. 

The important point is that  ${\cal Z}$ {\it is already the ensemble averaged single trace partition function}. That is,
\begin{equation}\label{PartFunA}
 \mathcal{Z}\left(h_2 \beta_i \right)=\langle \tr(e^{-h_2 \beta_i H(J)}) \rangle_{J,\langle J^2\rangle=1}=\langle \tr(e^{-\beta_i H(J)}) \rangle_{J,\langle J^2\rangle=h_2^2}
\end{equation}
and the non-trivial connected multi-trace correlator is induced only by the fluctuation parameter $h_2$. In the 2nd equality we emphasize that it is also the partition function in an ensemble where the average size of the couplings is rescaled by $h_2$. 


The main point now is that if the ensemble averaged theory has a gravitational description, then we can replace each ${\cal Z}$ in \eqref{eq:thermal_from_r}  by the gravity expression for the quantity, i.e., 
\begin{equation} \label{eq:multi_Z_grav}
    {\cal Z}(\beta_1,\ldots , \beta_n) =  \int dh_2~ P_{h_2}(h_2)
    \prod_{i=1}^n \mathcal{Z}_{\text{grav}}\left(h_2 \beta_i\right)\ .
\end{equation} 
In the SYK model that we are discussing, this can be concretely one of the following
\begin{itemize}\label{GraAlphas}
    \item $H_{\text{grav}}=H_{\text{Schwarzian}}$ in the low energy limit. For example we can take\footnote{with an additional constraint on the initial and final state} $H_{\text{Schwarzian}}=H_{\text{Liouville}}$  (as in \cite{Bagrets_2016,Bagrets_2017}).
    \item The full large $p$ SYK partition function \cite{MaldecenaStanford,Kitaev_talk}. 
    \item $H_{\text{grav}}=T_{\text{double scaled}}$ where the latter is the transfer matrix of the double-scaled SYK model (as in \cite{Berkooz_2019}).  
\end{itemize}
In any case, if we think about \eqref{eq:multi_Z_grav} as coming from some Euclidean path integral over some geometry, then each space is multiplied by a fluctuating parameter which rescales the entire action.

Another way of phrasing this is to define an effective Hamiltonian and an effective partition function
\begin{equation}
    H_{\text{eff}} = h_2H, \qquad \qquad
    \mathcal{Z}_{\text{eff}} = \expt{\tr \left(e^{-\beta H_{\text{eff}}} \right)}_J.
\end{equation}

Equation \eqref{eq:multi_Z_grav} is perhaps familiar from the discussion of wormholes \cite{Saad:2019lba,stanford2019jt,marolf_maxfield,penington2019,Almheiri_2020,Pollack_2020,marolf2020observations,Bousso2020,Coleman:1988cy,Giddings:1988cx,Giddings:1988wv}. There too, wormholes are equivalent to fluctuations in parameters in the theory. There are, however, some differences. 

The first is that there are no wormholes. In fact, this effect is considerably larger than anything that can be obtained by a smooth 2D surface with multiple boundaries. Smooth surfaces will introduce a connected component which scales like 
\begin{equation}
    \text{connected from wormhole}\sim \exp(-\kappa N)
\end{equation}
where $\kappa$ is a finite number which depends on the topology of the surface (and $N$ is the number of fermions). Here the size of the effect is 
\begin{equation}
    \text{Connected}\sim\epsilon= {N \choose p}^{-1} \sim N^{-p},\ \text{ for finite }~ p;
\end{equation}
or 
\begin{equation}
    \text{Connected}\sim \epsilon = {\binom{N}{p}}^{-1} \sim \left(e \sqrt{\frac{N}{\lambda}} \right)^{-\sqrt{\lambda N}} \left(\lambda N \right)^{1/4} e^{-\frac{\lambda}{2}}, ~~~~\lambda=p^2/N,
\end{equation}
for the double scaled SYK model. In particular, it is a perturbative effect for finite $p$, which is perhaps an irrationally large value. This is of course just the power of $N$ in which we will expect to see the effects of replicas.  

The other difference is that ${\cal Z}$ in equation \eqref{PartFunA} is still an ensemble averaged quantity, and not the partition function for a fixed value of the parameters. 
This, however, can be fixed at a small cost. We can consider a different averaged partition function, which is the ensemble average conditioned on the value of $\sum J_I^2$
\begin{equation}
    m_{k| h_2} = { \int d^{\cal N}J_I e^{-\sum_I J_I^2/2} \delta(\epsilon \sum J_I^2 - h_2^2  ) \tr(H^k)  \over  \int d^{\cal N}J_I e^{-\sum_I J_I^2/2} \delta(\epsilon \sum_I J_I^2 - h_2^2) }
\end{equation}
where $h_2$ is slightly off $1$. We can then use the fact that
\begin{equation}\label{CondMmnt}
    (h_2^2)^{k/2} m_k = m_{k|h_2}(1+{\cal O}(\epsilon k^2)) .
\end{equation}
So to leading order we can replace each ${\cal Z}$ by the conditioned partition function. 

\medskip
{\it Life in a given realization:} Consider now taking \eqref{CondMmnt}, and resumming it to the partition function, to obtain
\begin{equation} \label{eq:Cond_multi_Z}
    {\cal Z}(\beta_1,\ldots , \beta_n) =  \int dh_2~ P_{h_2}(h_2)
    \prod_{i=1}^n \mathcal{Z}_{\text{grav}}\left(\beta_i| \epsilon\sum J_I^2=h_2^2\right)\ 
\end{equation} 
then we can discuss the dual of a given realization. In a given realization of the theory has a specific $h_2^2$ value which slightly differs from 1. This is the leading piece of information about the specific realization. It seems that, to the order that we are discussing, correlation function in this specific realization are the same as correlation functions in a universe obtained by integrating over all the couplings with the conditional distribution (i.e. constrained by $\epsilon\sum J_I^2=h_2^2$). So on the one hand we are provided with the leading information about the distribution, and on the other hand we are still averaging on almost the same number of random coupling, giving rise to a gravitational dual (as in the standard coupling-averaged SYK duality). 

This is at the leading level of precision. More and more detailed information about the specific realization appears in higher and higher order corrections, and we turn to these in the next section.

\section{New Fluctuation Parameters}\label{sec:NewFluc}

The multi-trace diagrams that are not of the cactus type give higher order corrections in $N$. However, if one of the $k$'s in $M_c(k_1,\cdots ,k_V)$ is odd, there are no contributing cactus diagrams and a non-zero answer comes from these higher order terms. Higher order contributions of this type are therefore both easy to isolate, and provide a useful example of how generic higher order corrections work in general.

We will start in section \ref{sec:double_trace_3_lines} by analyzing in detail the case of two traces $M_c(k_1,k_2)$ for $k_1, k_2$ odd, and find its leading behavior. Then in section \ref{sec:subleading_theory_J} we will generalize this to other multi-trace moments, and in section \ref{sec:h3_efctv} we show how to change the effective Hamiltonian to generate such connected terms. In particular we will show that if we require locality of the effective Hamiltonian, then not only would there be new fluctuations parameters, but we necessarily have to introduce new fields into the theory with specific couplings. We will occasionally refer to them as {\it fluctuation fields}.

\subsection{The odd moments of the double trace} \label{sec:double_trace_3_lines}

\begin{figure}[h]
\centering
\includegraphics[width=0.8\textwidth]{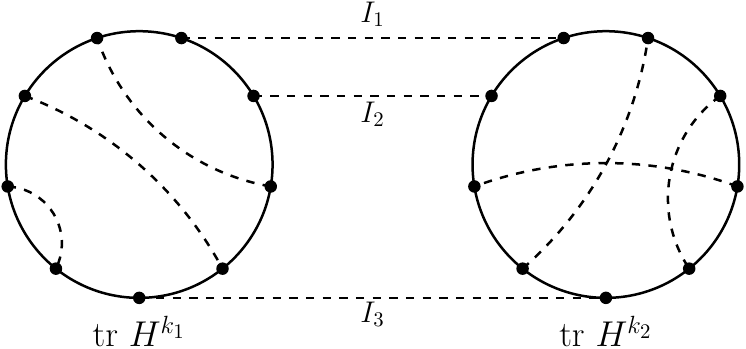}
\caption{An example of a diagram contributing to the odd double trace with $k_1=k_2=9$. Here we show the full multi-trace chord diagram, including the internal chords within each trace.}
\label{fig:odd_double_trace_CD}
\end{figure}

As we found before, the leading contributions to the double trace partition function come from the minimal number of couplings connecting the two traces. Thus the next order correction to the double trace partition function comes from connecting three couplings between the traces,\footnote{Recall that when we connect a single coupling between the two traces we get zero in each trace.} which contributes to the connected odd moments $M_c(k_1,k_2)$.
An example is shown in figure \ref{fig:odd_double_trace_CD}.

We shall start by computing the lowest of these moments, $M_c(3,3)$, and then generalize to other $k_1,k_2$. We must fully connect the Hamiltonians between the two traces for this contribution to be non-vanishing, and additionally all fermionic indices must appear exactly twice within each trace. As mentioned, this immediately tells us that every one of the three $\Psi_{I_j}$'s in each trace must share half of their indices with the other two $\Psi_{I}$'s.\footnote{We have such a single solution for the intersections size only in this case. This is another distinguishing feature of this first subleading correction. In section \ref{sec:GenFluc} we will discuss the situation where four or more index sets are correlated, in which case there is a large number of possibilities for partial overlaps.} 


$M_c(3,3)$ is now straightforward to compute. Having balanced the fermions in each trace as was just mentioned, ensemble average over the couplings implies that we have the same index sets in the two traces. The only thing that can change between the two traces is the ordering of the fermions inside them. Since the trace is cyclic we only have two possible such configurations, and since we are summing over all configurations we are bound to receive contributions from both. This gives us
\begin{equation} \label{eq:M(3,3)Contributions}
\begin{split}
    M_c(3,3) =  3 {\binom{N}{p}}^{-3}~~~
    \smashoperator[l]{ \sum_{\substack{|I_1|=|I_2|=|I_3|=p/2 \\ I_i \cap I_j = \emptyset, ~i\neq j}}}& \bigg( \tr(\Psi_{I_1}\Psi_{I_2}\Psi_{I_2}\Psi_{I_3}\Psi_{I_3}\Psi_{I_1}) \tr(\Psi_{I_1}\Psi_{I_2}\Psi_{I_2}\Psi_{I_3}\Psi_{I_3}\Psi_{I_1}) \\
    & + \tr(\Psi_{I_1}\Psi_{I_2}\Psi_{I_2}\Psi_{I_3}\Psi_{I_3}\Psi_{I_1}) \tr(\Psi_{I_1}\Psi_{I_2}\Psi_{I_3}\Psi_{I_1}\Psi_{I_2}\Psi_{I_3}) \bigg)
\end{split}
\end{equation}
where we have implicitly rearranged the ordering of the indices in each index set relative to the previous convention. This, however, does not introduce any additional signs since we rearrange the index set in the two traces simultaneously. We see that in the first line the traces have the same ordering, and in the second line we have switched $(\Psi_{I_2}\Psi_{I_3})\leftrightarrow (\Psi_{I_3}\Psi_{I_1})$. Since we know that $\Psi_{I_a}\cap\Psi_{I_b}=0$ for $a\neq b$, we have $\Psi_{I_a}\Psi_{I_b}=(-1)^{p/2}\Psi_{I_b}\Psi_{I_a}$,
so that if $p/2$ is odd the two contributions in (\ref{eq:M(3,3)Contributions})  cancel each other, while if $p/2$ is even they add up. We can take the sum over all possible index sets explicitly in this case, and find that
\begin{equation}\label{SubLeadCont}
    M_c(3,3) = \left\{\begin{array}{cc}
         6 {\binom{N}{p}}^{-3} \binom{N}{3p/2} \binom{3p/2}{p} \binom{p}{p/2}, \qquad \qquad & 4 \mid p, \\
         0, & 4 \nmid p.  
    \end{array} \right.
\end{equation}
For the rest of the section we will assume $p$ is divisible by 4 and so this contribution does not vanish.

Now let us consider arbitrary $k_1,k_2$.
We should go over all possibilities of choosing, in each trace, which three Hamiltonian insertions participate in the triple contraction as above. When we draw the chord diagrams it is convenient to cut the diagram (or start the trace) at one of those three insertions, which we can think of as the ``first'' one. Since the trace is cyclic, we need to divide each trace by 3, as any of them will be counted once as the ``first'' one. The distances between those three insertions can be arbitrary, hence we should sum over them. Each trace therefore contributes 
\begin{equation}\label{fluc33B}
    \begin{split}
    W_{k} & \equiv {\binom{N}{p/2}}^{-3}
    \sum_{\substack{|I_1|=|I_2|=|I_3|=p/2 \\ k_1+k_2+k_3 = k}}
    \expt{\tr\left(\Psi_{I_1}\Psi_{I_2}H^{k_1}
    \Psi_{I_2}\Psi_{I_3}H^{k_2}
    \Psi_{I_3}\Psi_{I_1}H^{k_3}\right)}_J =\\
    & = {\binom{N}{p/2}}^{-3}
    \sum_{\substack{|I_1|=|I_2|=|I_3|=p/2 \\ k_1+k_2+k_3 = k}}
    \expt{\tr\left(\Psi_{I_2}H^{k_1}
    \Psi_{I_2}\Psi_{I_3}H^{k_2}
    \Psi_{I_3}\Psi_{I_1}H^{k_3}\Psi_{I_1}\right)}_J.
    \end{split}
\end{equation}
Note that we dropped the constraint that $I_1,I_2,I_3$ are distinct, which is justified at large $N$.\footnote{When viewed as chord diagrams, we do not assign a value to the intersections of the corresponding chords, but rather only to an intersection of a Hamiltonian chord with one of the $I_j$ chords. Since at large $N$ we have no elements in common to three such sets of size of order $p$, we can drop the empty intersection condition. The calculation can also be done without this assumption with the same large $N$ result.} The normalization factor is chosen in correspondence which will be useful when handling $W_k$ as a 6-pt function later on.

With this definition, the full double-trace connected moment for odd $k_1,k_2$ is given by
\begin{equation}\label{fluc33A}
    M_c(k_1,k_2) = \frac{k_1 k_2}{9} M_c(3,3)W_{k_1-3}W_{k_2-3} + \text{higher order terms}
\end{equation}
(as mentioned, we choose for each trace the first element, giving a factor of $k$, and divide by 3 for overcounting). 
  
We see that the computation reduces to a calculation of a certain (ordered) 6-point function. In fact, up to a numerical value, we could consider a generic 6-point function (where every pair of operators are next to each other), and allow all possible contractions of those operators, except for neighboring ones (as in normal ordering). For $p/2$ even, this is just what we have here. We will see such a computation explicitly in subsection \ref{sec:h3_efctv}. 

Computing this 6-point function depends on the precise large $N$ limit considered, and can be done in the scaling in which $p$ is independent of $N$ using \cite{Gross_2017}. Here we will perform this calculation explicitly in the double-scaled limit. But it is important to emphasize that for any scaling the result is a product of two 6-point functions, one in each space. 

\begin{figure}[h]\label{Fig6Pt}
\centering
\includegraphics[width=0.8\textwidth]{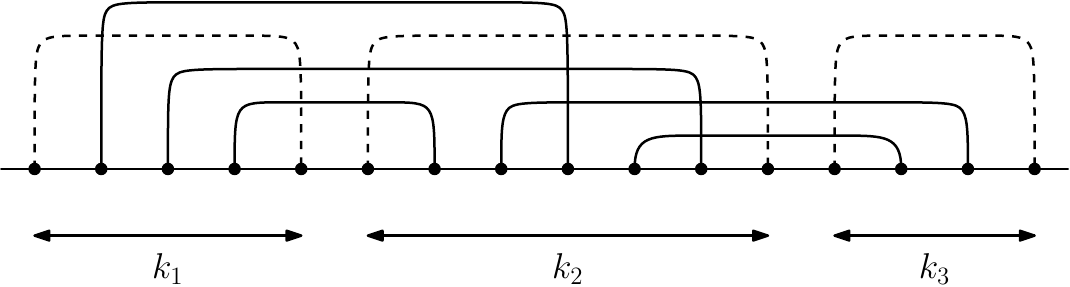}
\caption{A demonstration of the 6-point function that is needed for the calculation of odd moments.}
\label{fig:6_point_func}
\end{figure}

The 6-point function of interest, \eqref{fluc33B}, is demonstrated in figure \ref{fig:6_point_func}. It can be calculated similarly to what is done in \cite{Berkooz_2019}, which we closely follow. The idea is that the intersections with the first and third dashed chords are easily accounted for, since any open chord that we have in the regions between the dashed chords, necessary crosses them. This is accounted for by the operator $S$ that gives a factor of $q^{1/2}$ for every open chord (since here the dashed chords correspond to operators of size $p/2$). All that remains is to account for intersections with the dashed chord in the middle. Propagation along such a contracted pair is explained in section 3.1 of \cite{Berkooz_2019}. Denoting an operator of size $p/2$ by $\hat M$, and using notations as in \cite{Berkooz_2019},\footnote{$T$ is the transfer matrix that represents the Hamiltonian in chord space. $U$ ($D$) are matrices with 1's one diagonal above (below) the main diagonal, and $P_i^{(m)}$ is defined there, with $\tilde q = \sqrt{q}$. $\hat N$ counts the number of chords at the point where this operator appears. The state $|n \rangle$ stands for the state with $n$ open chords. $(a;q)_n$ are the Pochhammer symbols and $H_n(x|q)$ are the $q$-Hermite polynomials.} $W_k$ is given by
\begin{equation}\label{Exact33}
\begin{split}
& W_k = \sum_{k_1+k_2+k_3 = k} \innn{0}{T^{k_1} q^{\hat{N}/2}  \hat M  T^{k_2} \hat M q^{\hat{N}/2} T^{k_3}}{0} \\
&= \sum_{k_1+k_2+k_3 = k} \sum_{n,m = 0}^\infty \sum_{i = 0}^m P_i^{(m)}
 \innn{0}{T^{k_1}}{n}q^{n/2} \innn{n}{D^i ST^{k_2} S U^i}{m}  q^{m/2} \innn{m}{T^{k_3}}{0} \\
 &= \sum_{k_1+k_2+k_3 = k} \sum_{n,m,i = 0}^\infty \frac{q^{n+m+i}}{(q;q)_n (q,q)_m}
 \int_0^\pi \prod_{j=1}^3 \left\{ \frac{d\theta_j}{2\pi} \left(q,e^{\pm2\theta_j};q\right)_{\infty}
 \left( \frac{2\cos \theta_j}{\sqrt{1-q}}\right)^{k_j} \right\} \\
&~~~~~~~~~~~~~~~\times  H_{n+i}(\cos \theta_1|q)H_{n}(\cos \theta_2|q)H_{m}(\cos \theta_2|q)H_{m+i}(\cos \theta_3|q) .
\end{split}
\end{equation}
This can be simplified, written in terms of Al Salam-Chihara polynomials $Q_i$ (defined in equation (B.14) of \cite{Berkooz_2019})
\begin{equation}
    \begin{split}
        & W_k = \sum_{k_1+k_2+k_3 = k} \sum_{i = 0}^\infty q^i
 \int_0^\pi \prod_{j=1}^3 \left\{ \frac{d\theta_j}{2\pi} \left(q,e^{\pm2\theta_j};q\right)_{\infty}
 \left( \frac{2\cos \theta_j}{\sqrt{1-q}}\right)^{k_j} \right\} \\
&~~\times \frac{\left(q^2;q\right)_{\infty}^2}{\left(q e^{i(\pm\theta_1\pm\theta_2)};q\right)_{\infty}
\left(q e^{i(\pm\theta_3\pm\theta_2)};q\right)_{\infty}
\left(q^2;q\right)_i^2} Q_i\left(\cos \theta_1 |q e^{\mp i \theta_2};q\right) Q_i\left(\cos \theta_3 |q e^{\mp i \theta_2};q\right) .
    \end{split}
\end{equation}
This expression for $W_k$ completes the result for the double-trace connected odd moments.

\subsection{Multi-trace correlators} \label{sec:subleading_theory_J}

\begin{figure}[h]
\centering
\includegraphics[width=0.6\textwidth]{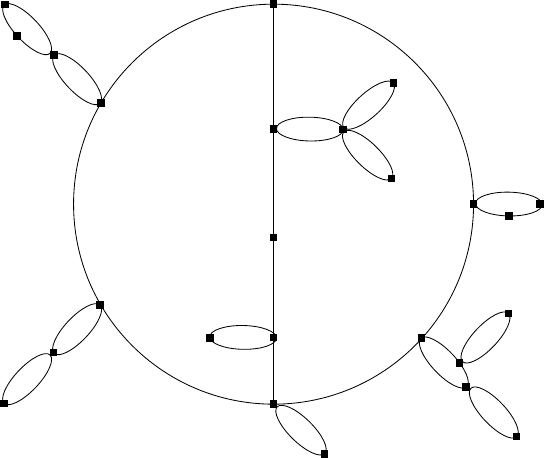}
\caption{A demonstration of the general diagrams contributing to leading order in the case of two vertices of odd degree, which are made of cactus diagrams built on top of the basic diagram.}
\label{fig:2_odd_vertices_cactus}
\end{figure}

In the previous subsection we discussed the case of two traces, with odd powers of $H$. It is easy to generalize it to the case of a general number of vertices $V$, where two of them are of odd degree. In this case, the general diagrams that contribute to the connected moment are cactus diagrams built on top of the basic diagram of the previous subsection. A generic diagram of this sort is demonstrated in figure \ref{fig:2_odd_vertices_cactus}.

In order to represent these contributions in the theory for the couplings $J$ (as in section \ref{sec:vector}), we use, as before, that whenever three index sets are paired the trace forces those three indices $I_1,I_2,I_3$ to satisfy $|I_1 \cap I_2|=p/2$ and then $I_3=I_1 \oplus I_2$ (where $\oplus$ stands here for the XOR operation). Then, the vertices with odd degree $k$ correspond to the operator
\begin{equation} \label{eq:operator_odd_degree}
    O_k=\frac{k}{3} W_{k-3} \left(\epsilon ^{3/2} \sum _{|I_1 \cap I_2|=p/2} J_{I_1} J_{I_2} J_{I_1 \oplus I_2}\right) \left(\epsilon \sum _{I} J_{I}^2\right)^{(k-3)/2} ,\qquad \text{($k$ odd)}.
\end{equation}
This allows for the leading order case of having a single constraint involving three paired chords, with the rest being of the form of cactus diagrams.

Just as before, the propagator is still normalized to $1$ as it simply comes from the value assigned to each chord (that is, a contraction of two $J$'s) 
\begin{equation}
    \langle J_{I_1} J_{I_2} \rangle = \delta_{I_1,I_2} .
\end{equation}

The traces of even degree are just as before
\begin{equation} \label{eq:operator_even_degree}
    O_k = m_k \left(\epsilon \sum _{I} J_{I}^2\right)^{k/2}, \qquad \text{($k$ even)}.
\end{equation}

The generalization to further higher orders should now be clear. Any vertex corresponds to a sum of terms of the form
\begin{equation}
    J_{I_1} \cdots J_{I_k}
\end{equation}
such that every SYK site appears an even number of times among the $I_j$'s, or, more concisely, $\sum_{j=1}^k r_j=0$ where $r_j$ is the set $I_j$ represented in $\mathbb{Z}_2^N$, multiplied by the appropriate amplitude. This is simply the trace constraint. For every given structure of indices, the trace translates to a certain form of a chord diagram (that is not just a contraction of pairs), and the value of the diagram (with the appropriate combinatorial factor) is the amplitude.
The expressions above, equations \eqref{eq:operator_odd_degree} and \eqref{eq:operator_even_degree}, give the leading contribution in this sum, for odd and even degrees respectively.

\subsubsection{More than two vertices of odd degree}

\begin{figure}[h]
\centering
\includegraphics[width=0.6\textwidth]{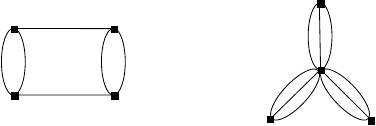}
\caption{Two different diagrams contributing to four traces with odd degree. Note that additionally there is a diagram looking like a tetrahedron, contributing at the same order as the diagram shown on the LHS.}
\label{fig:three_vertices_examples}
\end{figure}

We saw what happens exactly at leading order when we have two vertices of odd degree. Now we would like to discuss the general case. To be concrete, suppose that there are non-zero diagrams contributing when we use \eqref{eq:operator_odd_degree} for the odd degree vertices, and \eqref{eq:operator_even_degree} for the even degree vertices. Are there other possibilities that may give a result that is of lower order (i.e., more dominant)?
As we will show here, the short answer to this question is ``no'', namely that a non-vanishing result with the vertices \eqref{eq:operator_odd_degree} and \eqref{eq:operator_even_degree} will be the leading order.
For example consider the case of four traces with all nodes having an odd degree. Two of the diagrams contributing to this case are shown in figure \ref{fig:three_vertices_examples}. While all vertices are of the form \eqref{eq:operator_odd_degree} in the diagram on the left, there is one vertex not of this form for the diagram on the right (requiring three invariants, each consisting of three $J$'s). In these cases it can be checked that the order of the first diagram is $\epsilon ^4$, while the second one is $\epsilon ^{9/2}$ which is subleading.

In order to show the statement above, first note that if we use basic invariants of the $J$'s that are higher than the $J_{I_1}J_{I_2}J_{I_1\oplus I_2}$ and $J_I^2$ appearing in \eqref{eq:operator_odd_degree} and \eqref{eq:operator_even_degree}, then clearly the result will be of higher order. What is not clear is what happens when the vertices are made of one or more of these dominant quadratic and cubic invariants (as we have in the example on the right hand side of figure \ref{fig:three_vertices_examples})\footnote{Which scales as a disconnected diagram.}.
In all these cases, we can represent them by the familiar double line notation. Namely, we split all index sets of size $p$, and consider now index sets of size $p/2$. The random couplings are then of the form $J_{I_1 I_2}$ and the cubic invariant is $J_{I_1 I_2} J_{I_2 I_3} J_{I_1 I_3}$.

Suppose then that we have a Feynman diagram with $V_n$ vertices of degree $n$ (for $n=2,3,\cdots $), $E$ (double line) propagators and $L$ loops (of single lines). Then as we saw each propagator gives $\epsilon $, while each loop gives an order of $\epsilon ^{-1/2} $ (since the sets are of size $p/2$). Then each diagram is of the order (using $\chi =V+L-E$ and $E= \sum _n \frac{nV_n}{2}$)
\begin{equation}
    \epsilon ^{E-L/2}=\epsilon ^{-\frac{\chi}{2}+ \sum _n \left( \frac{1}{2}+\frac{n}{4} \right)V_n }.
\end{equation}
As expected, increasing the degree of vertices can only lead to a smaller result. Indeed, for the first diagram in figure \ref{fig:three_vertices_examples} we have $\chi =2$, $V_3=4$ giving $\epsilon ^4$. In the second diagram, we are connecting three times the basic diagram (of two vertices connected by three edges) each one having $V_3=2$ (and $\chi =2$) giving in total $(\epsilon ^{3/2})^3$. Each such addition of a diagram can lower the degree by $\epsilon ^{-\chi /2}$, that is at most by $\epsilon ^{-1}$, however, if we compare diagrams with the same number of vertices, it has an extra vertex giving $\epsilon ^{\frac{1}{2}+\frac{n}{4}}$, which suppresses by at least $\epsilon ^{5/4}$. Therefore, it can only lead to higher order results. Note that for $n=2$ we have $\epsilon ^{\frac{1}{2}+\frac{n}{4}}=\epsilon $, and this is why we found many diagrams of the same order in the even degree case, which are the cactus diagrams.

\subsection{The effective Hamiltonian of the $h_3$ fluctuation}\label{sec:h3_efctv}

In subsection \ref{sec:double_trace_3_lines} we wrote the connected odd correlation function in equation \eqref{fluc33A}. In the double scaling limit, an exact expression for $W_k$ is given in \eqref{Exact33} but an inspection of figure \ref{fig:6_point_func} reveals that it is related to a 6-point function for any value of $p$. In subsection \ref{sec:subleading_theory_J} we showed how to incorporate this in the theory for the couplings. However, in the latter, we had to explicitly use $W_k$ as a new object in the vertex. Here we show another way to represent the correction to the double-trace of odd moments that automatically generates this $W_k$ correction via a relatively simple modification of the single trace Hamiltonian. 

The ideology is similar to the one we used before in section \ref{H2DualAct}. There we rewrote the connected part of the multi-trace correlator, induced by fluctuations of $h_2$, as
\begin{equation}
    \mathcal{Z}_{c}(\beta_1,\cdots,\beta_n)=  \int_{conn} dh_2 P(h_2) Z(h_2\beta_1) \cdots Z(h_2\beta_n)\
\end{equation}
where the $Z$'s on the RHS are the single trace partition functions after averaging over the couplings or, equivalently but more interesting for us, the gravitational partition functions. The fluctuation parameter $h_2$ is just $h_2^2=\epsilon \sum_I J_I^2$.   

Here, we would like to write the odd double-trace contribution in a similar form. The fluctuation parameter that we need to include is $h_3^3=\epsilon^{3/2} \sum_{|I_1\cap I_2| = p/2 } J_{I_1} J_{I_2} J_{I_1 \oplus I_2}$. But for now let us not commit to using $h_3$ and denote it as a general parameter $\alpha$
\begin{equation}\label{h3FrmWrk}
 \mathcal{Z}_{c}(\beta_1,\cdots,\beta_n) \sim \int_{conn} d\alpha dh_2 P(\alpha,h_2) Z(\beta_1,h_2,\alpha)\cdots Z(\beta_n,h_2,\alpha)
\end{equation}
where $h_2$ is the random variable which encodes the $h_2$ fluctuation, and $\alpha$ is a new random variable (or a set of a few variables) which reproduces the $h_3$ fluctuation parameter. $P(h_2,\alpha)$ is their joint probability measure.

The reason for using $\alpha$ is that we will present two possible options of implementing \eqref{h3FrmWrk}. The first uses a general $\alpha$ not related to $h_3$, but we currently have it working only for the double trace. The second uses $h_3$, and it works more generally, but it entails another peculiar ingredient to be discussed below. 


The $Z$'s are modified single trace ``partition functions'' computed separately for each trace, for which we can find a gravitational dual (for every relevant value of $\alpha$). These new ``partition functions'' are approximately the familiar single trace partition functions (after ensemble average, or equivalently written in the gravitational language), modified by small $\alpha$ dependent terms. 

The key requirement that goes into constructing the $Z$'s is the requirement that the $\alpha$-dependent modifications will be ``minimal'' when interpreted gravitationally. By this we mean, for example, that the modification will be local and will include a minimal set of particles and interactions, and with a ``reasonable'' (in some loose sense) gravity interpretation.


In the first realization, the form of the modification will be the following. At step one, we consider a random operator\footnote{As a reminder, a random operator $O$ of length $p'$ is of the form 
\begin{equation} \label{eq:RandomOp}
    O= i^{p'/2}\binom{N}{p'}^{-1/2} \sum_{L} \tilde J_{L} \Psi_{L} 
\end{equation}
        where $L$ are index sets of length $p'$, and the $\tilde J$ are Gaussian independent random variables normalized such that $\langle \tilde J_L \tilde J_{L'} \rangle =\delta _{L,L'}$.} $O$ of length $p/2$ (or a corresponding field in the bulk), and change the Hamiltonian into
\begin{equation}\label{HamRep}
    {\hat H} = H + \alpha :O^2: \ ,
\end{equation}
and then
\begin{equation}\label{ModPrtA}
    Z(\beta,\alpha)=\langle e^{-\beta {\hat H}} \rangle_{J,O} .
\end{equation}
Here $H$ is the original Hamiltonian and we are averaging over the random coefficients in $H$ and in $O$. By the normal ordering of $O^2$ we meant that we do not allow self contractions. 

In step 2, we replace the RHS of \eqref{ModPrtA} by the suitable gravitational dual. Each such $Z$ has such a dual gravitational interpretation, since we are doing a small double-trace (in the sense that we deform the theory by $O^2$) deformation of a background which has a good gravitational dual  \cite{Aharony:2001pa,Witten:2001ua,Berkooz:2002ug}. The net result will be a deformation of the original gravitational action  by a small double-trace deformation with a random coefficient (which is correlated across different universes).  

Basically, we will simply verify explicitly that
\eqref{HamRep}, with an appropriate $P(\alpha)$, generates the correct fluctuation for two traces. Still, it is instructive to first argue why we modify the Hamiltonian by an $O^2$ term. The original expression that we are interested in can be written as
\begin{equation} \label{eq:two_traces_exp}
\begin{split} 
    & \langle \tr(e^{-\beta_1 H}) \tr (e^{-\beta_2 H})\rangle_J = \\
    & \qquad=\left\langle 
    \tr\biggl[ \left( 1-{\beta_1\over L_1} H\right) \cdots \left( 1-{\beta_1\over L_1} H\right)  \biggr]
     \tr\biggl[ \left( 1-{\beta_2\over L_2} H\right) \cdots \left( 1-{\beta_2\over L_2} H\right)  \biggr]
     \right\rangle_J
\end{split}
\end{equation}
where there are $L_1$ ($L_2$) products in the first (second) trace and $L_1,L_2\rightarrow\infty$. To get the desired odd moments, we go over all possibilities of choosing three Hamiltonian terms out of the first trace in \eqref{eq:two_traces_exp}, as well as three Hamiltonian terms in the second one. We would like to correlate them in such a way as to generate \eqref{fluc33A}, or more precisely figure \ref{fig:6_point_func}. The diagram in that figure corresponds to a 6-point function, where each Hamiltonian is replaced by a a pair of operators. Effectively we replace the 3 Hamiltonian insertions (that we chose in each trace) by
\begin{equation}
    \begin{split}
       & \left( 1-\frac{\beta _1}{L_1} H\right)^{A_1}   \left( 1-\frac{\beta _1}{L_1} H-\alpha \frac{\beta _1}{L_1} O_1 O_2\right)  
         \left( 1-\frac{\beta _1}{L_1} H\right)^{A_2}  
        \left( 1-\frac{\beta _1}{L_1} H-\alpha \frac{\beta _1}{L_1} O_2O_3\right) \times\\
       & \qquad \times\left( 1-\frac{\beta _1}{L_1} H\right)^{A_3} \left( 1-\frac{\beta _1}{L_1} H-\alpha \frac{\beta _1}{L_1} O_3 O_1\right)  \left( 1-\frac{\beta _1}{L_1}H\right)^{A_4}, 
    \end{split}
\end{equation}
where $A_1+A_2+A_3+A_4=L_1-3$, and $\alpha$ is a small parameter proportional to $M_c(3,3)^{1/3}$. It will shortly be promoted to a random variable. The subindex in $O_j$ indicates the identification of the operators. This reproduces both the original $\alpha^0$ partition function and the $\alpha^3$ $W_{L_1-3}$ terms. The net result seems to involve adding $O_iO_j$ terms to the Hamiltonian with a coefficient proportional to $\alpha$.

Changing the Hamiltonian into $H+\alpha \sum_{i\not=j, i,j=1,2,3} O_iO_j$ is still cumbersome, since they are all operators with the same dimension (1/2) and the same propagators. In fact, we can do with a modification of the form  
\begin{equation}
    \left[ 1-\frac{\beta_1}{L_1}\left( H+\alpha  :O^2: \right) \right] ^{L_1} 
\end{equation}
if we focus on the $\alpha^3$ term in each trace.
So choosing 3 $H$'s in the trace and replacing them by the 6-point function is correctly captured just by changing the effective Hamiltonian.

In this replacement we of course generate additional terms that we do not want.
So there are several consistency conditions that we should impose on the Hamiltonian modification:
\begin{itemize}
\item It should not modify, at leading order in $N$, the single trace partition function.
\item Since we can bring down also 2 interaction terms in each trace, it might give us cactus-like contributions --- we need to verify that this is not the case, at least in leading order.
\item It should reproduce the odd double-trace leading order result.
\end{itemize}

Let us verify that \eqref{HamRep} satisfies these constraints, with an appropriate $P(\alpha)$. The last condition is given in terms of moments as the requirement that
\begin{equation} \label{eq:effective_odd_double_trace}
    M_c(k_1,k_2) = \langle M_{\text{eff}}(k_1) M_{\text{eff}}(k_2)\rangle _{conn,\alpha }
\end{equation}
where
\begin{equation}
    M_{\text{eff}}(k) = \langle \tr \hat H^k \rangle _{J,O}
\end{equation}
is the effective moment after tracing over the microscopic operators and couplings.

As above, $\alpha $ is a small parameter in which we expand. Explicitly, it should go as $\binom{N}{p}^{-1/4}$ in order to give the $N$ dependence of the odd double trace. (This is because we will see that the leading contribution will come from six insertions of $\alpha$, and we saw that the odd double trace scales like $\binom{N}{p}^{-3/2}$.)  We should specify a probability distribution for $\alpha $ that reproduces the 6-point function contribution that we saw above. Let us take
\begin{equation}
     P(\alpha ) = \frac{1}{2} \left( \delta (\alpha -{\kappa} )+\delta (\alpha +{\kappa} ) \right);
\end{equation}
we will specify $\kappa $ in a moment. Note that this distribution satisfies
\begin{equation}\label{sec:disth3}
 \langle \alpha ^{n_1} \alpha ^{n_2} \rangle -\langle \alpha ^{n_1}\rangle \langle \alpha ^{n_2}\rangle  =
\begin{cases}
\kappa ^{n_1+n_2} & \text{if }n_1,n_2 \text{ are odd} \\
0 & \text{otherwise}
\end{cases} .
\end{equation}
Since $:O^2:$ is normal ordered, the effective trace vanishes at first order in $\alpha $. Working up to sixth order in $\alpha $, the only contribution to the connected double trace is
\begin{equation}
\begin{split}
     \langle M_{\text{eff}}(k_1) M_{\text{eff}}(k_2)\rangle _{conn,\alpha } \approx ~
     &\kappa^6 \frac{k_1}{3} \sum_{i_1+i_2+i_3=k_1-3} \langle \tr :O^2: H^{i_1} :O^2: H^{i_2} :O^2: H^{i_3}\rangle \\ &\times \frac{k_2}{3} \sum_{j_1+j_2+j_3=k_2-3} \langle \tr :O^2: H^{j_1} :O^2: H^{j_2} :O^2: H^{j_3}\rangle ,
\end{split}
\end{equation}
where we have used cyclicity to start the trace before one of the operator insertions, just as before. We have not included various lower powers of $\kappa$ which correct the disconnected moments (i.e., correct each moment separately), since we consider the connected moment.
Each averaged trace is what we have in the definition of $W_k$.\footnote{To determine the proportionality constant, note that each averaged trace contains 8 different contractions of the $O$'s, where each one is just the trace in the definition of $W_k$, times some power of $q^{1/4}$ from the intersections of the $O$ chords \cite{Berkooz_2019}. So each of the two traces equals $W_k$ times a simple constant factor of $(1+q^{1/4})^3$ (since every pair in $O^2$ can be exchanged).}
These are non-vanishing only for $k_1,k_2$ odd.
So, choosing
\begin{equation}
    \kappa = \frac{1}{1+q^{1/4}}  M_c(3,3)^{1/6}
\end{equation}
we get indeed $M_c(k_1,k_2)$ as in \eqref{fluc33A}, establishing \eqref{eq:effective_odd_double_trace}.

This effective Hamiltonian has been constructed in order to reproduce the connected piece in the two trace correlator. One can ask whether, for example, it reproduces correctly all the diagrams in figure \ref{fig:2_odd_vertices_cactus}. The answer is no --- it 
actually does generate all the correct diagrams for a multi-trace correlator with odd vertices, but it does not give the correct combinatorial factors. So one can either try and complicate the model, or re-start in a more systematic way (which will entail some additional twists).

Perhaps the most ``natural'' choice is to replace $\alpha $ in \eqref{HamRep} by
\begin{equation}
h_3 = \left(\epsilon^{3/2} \sum_{|I_1\cap I_2| = p/2 } J_{I_1} J_{I_2} J_{I_1 \oplus I_2}\right)^{1/3} ,
\end{equation}
 and then to consider the vector model of the couplings derived in section \ref{sec:subleading_theory_J}. This is in analogy to the definition of $h_2$ as $h_2 = \left(\epsilon \sum_I J_I^2\right)^{1/2}$. This choice also has a ``natural'' probability distribution $P(h_2,h_3)$ derived directly from the probability distribution of the couplings (or dual vector model.)

Replacing $\alpha $ by $h_3$ has the advantage of automatically satisfying the first and last conditions from the list above, however it does modify the even terms. Additionally there is some ambiguity as to which root to pick out of the possible three (if we allow phases). Unlike previously where taking $h_2$ to be the positive or negative square-root gives the same result, in this case the different roots do not give the same answer. Indeed the different choices of $h_3$ ``interfere'' with each other, and in particular if we sum over all three possibilities only the desired contributions with three $h_3$ terms remain. 

This observation tells us how to modify the Hamiltonian in a consistent way: simply sum over all different choices of the cube-root in each trace separately. This is done by correcting the $: O^2:$ term using a new ``random'' variable $\chi$ which is to be drawn from the discrete measure over the complex plane with equal support at the points $1,e^{2i\pi/3}$, and $e^{-2i\pi/3}$. This new effective Hamiltonian now also satisfies the second condition, and allows us to write the effective action in a compact form
\begin{equation}\label{HamRepNew}
     H_{\text{eff}} = h_2 H + h_3 \chi :O^2: \ ,
\end{equation}
and 
\begin{equation}\label{ModPrtANew}
    Z_{\text{eff}}(\beta,h_2,h_3)=\langle \tr\left( e^{-\beta H_{\text{eff}}}\right) \rangle_{J,O,\chi} .
\end{equation}

The probability measure $P(h_2,h_3)$ for Gaussian couplings is given by integrating over the dual vector field with the defining constraints:
\begin{equation}
    P(h_2,h_3) = \int D J_I e^{-\frac{1}{2} \sum_I J_I^2} 
    \delta\left(h_2 - \sqrt{\epsilon \sum_I J_I^2} \right)
    \delta\left(h_3 - \left(\epsilon^{3/2}\sum_{|I_1\cap I_2| = p/2 } J_{I_1} J_{I_2} J_{I_1 \oplus I_2}\right)^{1/3} \right) .
\end{equation}

It is worth noting several points. The first is that the $\chi$ variable is local to each universe. I.e. in each $Z$ we sum over its own $\chi$ variable independently of the other universes. The only information which carries between universes is $h_2$ and $h_3$ as expected. 

Averaging over $\chi$ for each universe, in and of itself, is not a problem. It just means that the theory splits into sectors, labelled by $\chi$, and the full theory is the sum over these sectors. In each sector, the Hamiltonian is slightly different. 

There is one fly in the ointment which is that the effective Hamiltonian \eqref{HamRepNew} is no longer Hermitian as $\chi$ is complex, though the final result in the multi-trace correlator is always real due to the sum over complex conjugate values of $\chi$. We are not sure how to interpret this from a gravitational standpoint, and if the loss of Hermiticity has any gravitational consequences, especially since the underlying microscopic Hamiltonian is always Hermitian.

\section{Fluctuation Parameters, Fixed Realization and Wormholes} \label{sec:gravitystuff}

\subsection{General fluctuation parameters}\label{sec:GenFluc}

We discussed until now the connected contribution of multi-trace correlators induced by fluctuations of $h_2=\left(\epsilon\sum J_I^2\right)^{1/2}$ and $h_3=\left(\epsilon^{3/2}\sum_{I_1+ I_2+ I_3=0} J_{I_1}J_{I_2}J_{I_3}\right)^{1/3}$. These terms were easier to handle because they were the leading contribution to appropriate moments (they are also unique in that they are ``isolated'' in a sense that we will discuss below). These, however, are just the first out of an infinite (in the large $N$ limit) number of fluctuation parameters. We will focus on the latter and briefly discuss their contribution to the connected two-trace correlator.

One estimate for the size of contributions from higher fluctuation parameters is given in   \cite{Cotler:2016fpe,verbaarschot2019}. For the case of $\langle \text{tr}(H^k) \text{tr}(H^k)\rangle$, when the random coefficients are all paired between the traces in their order along the circle, then
\begin{equation}\label{KKpairings}
    \left. M_{c, \text{ladder}}(k,k) \right|_{\text{k pairings}}
    = {\binom{N}{p}}^{-k} 2^{-N} \sum_{m=0}^N \binom{N}{m} \left[ \sum_{j=0}^p (-1)^j \binom{m}{j} \binom{N-m}{p-j} \right]^k.
\end{equation}
This is of the order of ${\binom{N}{p}}^{-k/2}$ for small $k$, and of the order $2^{-N}$ at large $k$.

This clumps many fluctuation parameters together, but we can be more specific about the strength of each one separately. Consider the case that we want to contract $n$ Hamiltonians between the two traces\footnote{The two traces can have $k_1,k_2\ge n$ insertions in total.}. The requirement that each Majorana fermion appears twice means that we need to split the $n$ multi-fermions index sets into smaller index sets, such that each of the latter appears in two of the $n$ Hamiltonians. More specifically, we split the multi-index sets of the Majoranas in groups as
\begin{align}\label{eq:SpltFlcPrm}
    \begin{split}
        &\Psi_1 = \Psi_{1,1}\cdots \Psi_{1,n}
        \\
        &\Psi_2 = \Psi_{2,1}\cdots \Psi_{2,n}
        \\
        &\qquad\qquad \vdots
        \\
        & \Psi_n = \Psi_{n,1}\cdots\Psi_{n,n} \ ,
    \end{split}
\end{align}
with the constraint that 1) $\Psi_{i,j}=\Psi_{j,i}$, 2) $\Psi_{i,i}=\emptyset$, 3) the intersection of $\Psi_{i,j_1}$ and $\Psi_{i,j_2}$ is empty for all $1\leq i\leq n$, $j_1 \neq j_2$, and 4) that the total length in each row is $p$.  Notice that these four conditions are equivalent to contracting the $\Psi_I$'s in an $O(N)$ invariant way, and hence we are really just enumerating the possible $O(N)$ invariant combinations of $n$ couplings.

If we denote $|\Psi_{i,j}|=n_{ij}$, then the $n_{ij}$'s satisfy
\begin{align}\label{eq:Condnij}
n_{ij}=n_{ji}, \qquad \sum_{j}n_{ij}=p, \qquad n_{ii}=0, \qquad n_{ij}\in\mathbb{N}\cup\{0\}.    
\end{align}
The associated chord diagram is the same as in figure \ref{fig:odd_double_trace_CD} except with $n$ insertions correlated between the two traces.

Each fluctuation parameter corresponds to a different set of $n_{ij}$'s that satisfy the constraints \eqref{eq:Condnij}.
For a fixed set of $n_{ij}$'s the contribution of such a fluctuation parameter is determined by the number of random parameters $J$ such that their index sets satisfy the four constraints listed above. In the limit $N\gg p,n_{ij}$, this fluctuation parameter contributes to the double trace correlator with strength
\begin{equation}\label{eq:FlucParamSup}
    {N^{pn/2}/\prod_{i<j} n_{ij}!\over N^{np}/(p!)^n}= N^{-np/2}(p!)^{n/2} \prod_i { p \choose  n_{i1}\ ..\ n_{in}} ^{1/2}
\end{equation}
where the numerator comes from the number of indices which satisfy the decomposition \eqref{eq:SpltFlcPrm}, and the denominators comes from the normalization of the $J$'s and the fact that we have $2n$ of them participating in this term ($n$ in each trace)\footnote{The corrections to this formula go like $1/N$ in the fixed $p$ limit, or a function of $\lambda=p^2/N$ in the double scaled limit.}. In the double scaling limit, we rearrange the expression above to be $(\sqrt{\lambda}/e)^{pn} \prod_{i<j} {1/n_{ij}!} $.

We can carry out a saddle point estimate of the strength of this contribution in the large $N, p, n_{ij}$ limit. We define 
\begin{equation}
A = \left(2\pi\right)^{n\left(n-1\right)/4}\left(2\pi p\right)^{n/2}\left(\frac{\lambda}{e}\right)^{np/2},
\end{equation}
and $n_{ij}=px_{ij}$. Then we can evaluate the following integral
\begin{equation}
   Ap^{n\left(n-1\right)/2-n}\int_{0}^{1}d^{n\left(n-1\right)/2}x_{ij}e^{-p\left[\sum_{i<j}\left(x_{ij}+\frac{1}{2p}\right)\ln\left(px_{ij}\right)\right]}\prod_{i=1}^{n}\delta\left(\sum_{j}x_{ij}-1\right),
\end{equation}
for which there is a saddle point at $x_{ij}=1/(n-1)$. This is of course just the range where each of the $\Psi_i$ overlaps with the others to a similar extent. Evaluating the width around the saddle point, we obtain that the contribution for $(N,p,n)$ is
\begin{equation}
    \left(2\pi\right)^{n(n-1)/2}p^{-n(n+2p-3)/4}\sqrt{\frac{1}{2(n-1)(n-2)^{n-1}}}\left(\frac{\lambda}{e}\right)^{np/2}\left(\frac{n-1}{2}(2p-n+1)\right)^{-n(n-3)/4},
\end{equation}
for $\lambda = p^2/N$. 

In the limit $p\gg n \gg 1$ this becomes
\begin{align}
    2^{n^2/4}p^{-n^2-np/2}n^{-n^2/4-n/2}\left(\frac{\lambda}{e}\right)^{np/2}.
\end{align}
This regime is relevant for the double scaled limit, when we take the fluctuation parameters to be smaller than $p$ (which is taken to infinity). We see that there is no divergence associated with the increasing $n$.

In the case of finite $p$ and $n$ large, most of the $n_{ij}$'s are zero and the analysis above needs to be modified. We will not do this analysis here.

\subsection{Fluctuation parameters and the dual of a single realization}\label{sec:SnglRlz}

We would like here to briefly suggest a possible solution to the question of what is the dual of a given realization of the couplings $J_I$, and how it differs from the ensemble average. In the discussion, we will focus on lessons that can be drawn from the two-trace correlator. The phrasing of the question will actually be different from ``what is the dual of...'' but rather it would be: suppose we are provided information about the couplings $J$ up to a certain precision, what is the gravitational dual which captures all the observables up to that set precision? 

The advantage of this phrasing is that it is naturally compatible with the grading in the strength of the fluctuation parameters, and it is also more operational in the sense that it sets a benchmark precision. Its main advantage though is that, at any finite order in precision, we still do an average over a very large set of random couplings, making a gravitational dual perhaps more reasonable. Of course, the precision can always be further increased. 

We have really discussed it before in sections \ref{H2DualAct}, \ref{sec:h3_efctv} and  \ref{sec:GenFluc}, but here we summarize the structure. There were several ingredients in the computation of the connected two-trace correlator, which we summarize right below. We have argued for the first four in general, and have shown the last item for the $h_2$ and $h_3$ fluctuation parameters. The sequence of ingredients leads to a natural suggestion to what is effectively the dual of a given realization.
\begin{itemize}
    \item In each trace, correlate sets of $H$'s according to \eqref{eq:SpltFlcPrm} in general. Each such splitting is associated with a fluctuation parameter, $h_2$ and $h_3$ being the simplest ones.
    \item The expectation values of the fluctuation parameters is zero (except $h_2$). However, their variance contributes to the two-trace correlator. 
    \item In each trace the contribution is similar to a  $n(n-1)$ functions, made out (in the generic case) of $n(n-1)/2$ insertions of pairs of the same operator. Each operator is made out of $n_{ij}=n_{ji}$ fundamental fermions as in the previous subsection. We will refer to these operators as $O_{ij}$ ($O_{ij}=O_{ji},\ O_{ii}=1)$.
    \item We label by $h_{\{ n_{ij}\}}$ the fluctuation parameter that corresponds to the set $\{n_{ij},\ i<j\}$ as in the previous subsection. Then these $n(n-1)$ functions are among the ones generated by the substitution
    \begin{equation}\label{eq:FlucSubGen}
        H_{\text{eff}} = H + h_{\{ n_{ij}\}} \sum_{\hat i} \prod_{\hat j} O_{{\hat i}{\hat j}} \ .
    \end{equation}
    We emphasize that this generates the required contribution, but in general it generates much more, and we have not explored whether they can be cancelled in general by a suitable choice of the $h$'s joint distribution. To do this one might have to extend the set of fluctuation parameters. 
    \item If we are able to cancel all the unwanted contributions that originate from equation \eqref{eq:FlucSubGen} (by a suitable probability measure of the fluctuation parameter) as we have seen for $h_2$ and $h_3$, then the multi-trace connected correlation function is given by a substitution
    \begin{equation}
        \int d^Lh \ P(\{ h\}) \prod_i\ \tr \left[\exp(-\beta_i  H_{\text{eff}}(\{ h\})) \right]
    \end{equation}
    where $\{ h\}$ is the set of $L$ leading fluctuation parameters, and $L$ is fixed by the degree of precision that we want to obtain. $P$ is the probability measure on the space of fluctuation parameters. 
\end{itemize}

Our proposal is therefore the following. If we have the theory in a given realization, we still need to specify the precision to which we are working. This amounts to specifying a limited set of fluctuation parameters which take the values prescribed by the given realization. We can then take the ensemble average, conditional on the values of these small number of fluctuation parameters. The dual of this is described by adding the additional operators (field in the bulk) $\{ O_n \}$ and using the Hamiltonian $ H_{\text{eff}}$. 

For a slightly more ``operational'' perspective, consider the case that we have some higher dimensional theory, which we will call the UV theory, which flows in the IR to some $AdS_2\times M$ near horizon. Suppose also that the latter (the IR theory) is effectively described by some SYK like theory, or a set of weakly coupled SYK models, for which the rules above apply. We would like to consider the experience of an observer outside the IR SYK-like region. 

This outside observer probes the near horizon IR region and the black-hole with the set of fields at her/his disposal. The set of allowed fields is really determined by the theory outside the black-hole - actually all the way to the boundary of space. As they are defined using the UV degrees of freedom, they are not necessarily easily defined in terms of the IR degrees of freedom. An operator in the UV theory flows to some operator in the IR SYK model, but they do not need to span or coincide with any set of preconceived operators in the latter. Since the Hamiltonian, in terms of the UV degrees of freedom, flows to a random operator in terms of the IR SYK degrees of freedom, we will assume that this is true for all UV operators. So we are probing the black-hole with random operators of the class that we discussed before. The outside observer throws quanta of the corresponding fields, and measures the response of the black-hole by measuring the outgoing quanta, i.e. by measuring a correlation function of random operators in the IR SYK model.

Let us go back to the Hamiltonian. Given a specific UV theory we flow to a specific realization of the SYK Hamiltonian in the IR. Under the suggestion above we are instructed to add the set of random operators with all values of $n<p$. In the conformal limit it means that there is a spectrum of low mass virtual particles (that correspond to dimensions smaller than 1, and a continuum of such particles in the double scaled limit). Time evolution using the Hamiltonian ${\hat H}$ now means that quanta of these fields  generate, bubble and decay as part of the time evolution. The details of the realization are encoded in their couplings. 

It's important to emphasize that none of these fields need to exist in the UV theory. They are suggested only for the internal consistency of the near horizon, when trying to discuss the theory in a given realization. It does mean, however, that this consistency implies a very large amount of sort of ``quantum hair'', confined to the near horizon area.

\subsection{The relation to geometric wormholes} \label{sec:wormholes}

Since we are discussing connected multi-trace correlators, a natural question is whether there is any relation to the wormholes studied in \cite{Saad:2019lba} for JT gravity, or more generally in  \cite{Coleman:1988cy,Giddings:1988cx,marolf_maxfield,marolf2020observations,penington2019,Almheiri_2020}. It seems that the answer is no. This can be seen from several points of view:

1. {\it The strength of correlation that we are discussing is much larger: } In this work we have seen that for the SYK model, the leading connected contribution to the spectral form factor $\left<Z(\beta_1)Z(\beta_2)\right>_{\text{c,SYK}}\propto {N \choose p}^{-1} \xrightarrow{p \text{ is finite}}N^{-p}$. Note that this is of the same strength as a perturbative process (in the sense that it is a power of $N^{-1}$). In the double scaling limit ($\lambda=p^2/N$),  the spectral form factor scales like $\exp(-{1\over 2}\sqrt{\lambda N}\log(N))$. 

In contrast, following \cite{Saad:2019lba}, we know that the first connected contribution to the spectral form factor in JT comes from a connected topology, which is reduced by a factor of $e^{-\kappa N}$, where there $N$ represents the ground state entropy of JT theory (and $\kappa$ is some coefficient of order 1).   This is a non-perturbative result, and the two corrections obviously do not match. This is a consequence of the fact that JT itself, which is a radical truncation of the full SYK model, is dual to an RMT model with a $\sinh(\sqrt{E})$ envelope, but otherwise has standard $\beta$-ensemble level statistics, unlike the full SYK model.  
 
We might ask if there's some local deformation we can apply to JT gravity, namely - some $W(\phi)$ term we can add to the Lagrangian that will reproduce the results we see in this work. In \cite{Witten:2020wvy} it was shown that any deformation of this type still results in a matrix model. As such, the first connected component will scale like $e^{-\kappa N}$, and will not match the one computed here.

2 {\it Relevant time scales:} The standard $e^{-\kappa N}$ JT wormholes build the ramp and the plateau \cite{saad2018}. These are very late time phenomena. In contrast, the corrections we find here are early time corrections (as evident in figure \ref{fig:specform_all}), and tend to zero at late times.

3. {\it Bulk vs.\ boundary corrections:}
Finally, we would like to highlight again one point, which is that the gravitational implementation of the connected correlator (with the exception of the $h_2$ contribution) is done via an analogue of a multi-trace deformation of the theory \cite{Aharony:2001pa,Witten:2001ua,Berkooz:2002ug}. After all, we are (schematically) deforming the Lagrangian by a local term ${\cal L}\rightarrow {\cal L}+ \sum C_{1 \cdots k} O_{1} \cdots O_{k}$ where the $O$'s are operators whose dimension sums to 1, and $C$ are fluctuation parameters. This is rather different from the wormholes of \cite{Saad:2019lba,penington2019,Almheiri_2020}, in which wormholes connect the two spaces via the interior\footnote{Of course, it could be that a more refined study will identify a modification of the action also in the bulk.}.

Note, however, that the necessity of a large number of fluctuating fields is certainly a statement about the bulk and may affect its internal dynamics (for example, when discussing the stability of bulk configurations). Furthermore, double trace deformations can drive a change in the bulk by their effect on the quantization of the fields as in \cite{Gubser:2002vv,Gubser:2002zh}.

\section{Multi-Trace Operator Correlations} \label{sec:operators}

The simplest generalization of the multi-trace thermal partition function is to consider multi-trace correlation functions of random operators. We will look at random operators of the form 
\begin{equation} \label{eq:def_operator}
    A = i^{\tilde{p}/2} {\binom{N}{\tilde{p}}}^{-1/2}\sum_{|I| = \tilde{p}} \tilde{J}_I \Psi_I ,
\end{equation}
where the couplings $\tilde{J}_I$ are independent Gaussian variables with zero mean and unit variance, as already defined in (\ref{eq:RandomOp}). We will further take the disorder average over these couplings. These are the same type of operators considered in \cite{Micha2018,Berkooz_2019}. If we are working in the double-scaled limit then we require these operators to have a well defined double-scaled limit, ${\tilde p}^2/N$ fixed, as in \cite{Berkooz_2019}; though our results also hold in the usual large $N$ limit.

There are several arguments as to why this class of operators is interesting to consider, and we refer the reader to \cite{Micha2018,berkooz2020complex} for further details. The only argument that we will repeat is in the spirit of section \ref{sec:SnglRlz}: if we are given some UV theory which flows to an IR of the form $AdS_2\times M$, which is approximately described by an SYK model (or an array of such models), then the probes at our disposal are determined by the theory away from the near horizon region, which knows little about the SYK regime. The only expectation that we can have about such operators is that they are statistically similar to the local energy-momentum tensor, which is one of these probes. 

We can also motivate this choice by noting that these correlation functions reduce to the standard SYK correlation functions after taking the disorder average over $\tilde{J}_I$. For example the 2-point function of $A$ is really 
\begin{equation} \label{eq:random_operator}
    \expt{\tr(A(t) A(0))}_{J,\tilde{J}} = 
    i^{\tilde{p}}{\binom{N}{\tilde{p}}}^{-1} \sum_{|I| = \tilde{p}} 
    \expt{\Psi_I(t) \Psi_I(0)}_J,
\end{equation}
which is the class of 2-point functions that are computed in the large $N$ limit of the SYK model, say in \cite{MaldecenaStanford,Polchinski_2016}. The higher point functions after disorder averaging are also identical to the regular correlation functions computed, say in \cite{Gross_2017}. As such, they seem like a reasonable choice of operators to consider. These operators are also reminiscent of end-of-the-world branes \cite{penington2019} as they have some internal degrees of freedom, $\tilde{p}$, and they can be correlated between different traces as the Hamiltonian. 

\subsection{Multi-trace 2-point functions}

We will start by computing the connected two point functions where we insert two operators on each side of the trace, and subtract the disconnected part. We will denote the regular thermal 2-point function by
\begin{equation}
    G^{(2)}(\beta_1,\beta_2;\tilde{p}) \equiv \expt{ \tr\left(e^{-\beta_1 H} A e^{-\beta_2 H} A\right) }_{J,\tilde{J}}.
\end{equation}
The exact expression of these correlation functions is not important for what follows, though we note that such functions were computed in \cite{Micha2018,Berkooz_2019} in the double scaled limit, and in \cite{MaldecenaStanford,Polchinski_2016,Gross_2017} for the large $N$ finite $p$ limit. Furthermore, one can take $\beta_{1} = \beta + \tau$ and $\beta_2 = -\tau$ to get the thermal Euclidean 2-point function, though we found the $\beta_{1,2}$ variables to be more convenient for this specific computation.

\begin{figure}[h]
\centering
\includegraphics[width = 8 cm]{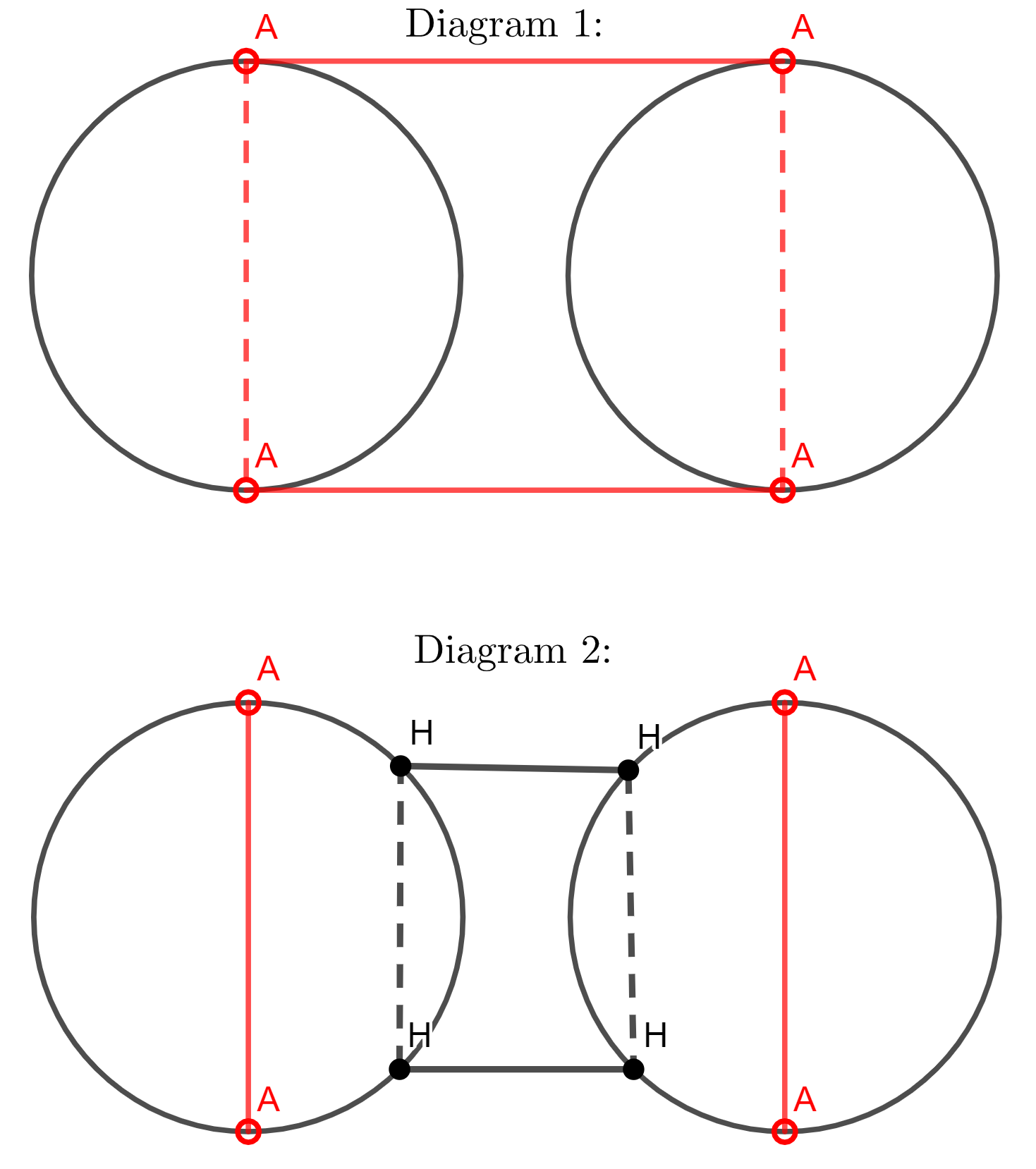}
\caption{The two different contractions that may contribute to the connected 2-point function $G^{(2,2)}$. Connection via $A$ chords, as in the top diagram, is dominant when $\tilde{p}<p$. Connection via $H$ chords is dominant when $p<\tilde{p}$.}
\label{fig:4ptfun}
\end{figure}

We will be interested in the connected two point functions
\begin{equation}
    \begin{split}
        G_c^{(2,2)}(\beta_1,\beta_2,\beta_1',\beta_2';\tilde{p}) \equiv &\expt{ \tr\left(e^{-\beta_1 H} A e^{-\beta_2 H} A\right)\tr\left(e^{-\beta_1' H} A e^{-\beta_2' H} A\right) }_{J,\tilde{J}} \\
    & \qquad \qquad - G^{(2)}(\beta_1,\beta_2;\tilde{p}) G^{(2)}(\beta_1',\beta_2';\tilde{p}) .
    \end{split}
\end{equation}
The leading order contribution to these functions will again be minimally connected, as was the case in the thermal partition function in \eqref{eq:double_thermal_partition}. In this case, however, there are two different diagrams that may dominate, depending on the ratio of $p,\tilde{p}$, and they are given in figure \ref{fig:4ptfun}.

The first diagram in figure \ref{fig:4ptfun} is simply contracting the four $A$ operators. There are two ways to do this, and each is suppressed by a factor of $ {\binom{N}{\tilde{p}}}^{-1}$, which together give a contribution that we denote by $I_1$, with
\begin{equation} \label{eq:I_1_2-point}
    I_1 = 2{\binom{N}{\tilde{p}}}^{-1} G^{(2)}(\beta_1,\beta_1';\tilde{p}) G^{(2)}(\beta_2,\beta_2';\tilde{p}).
\end{equation}

The second diagram in figure \ref{fig:4ptfun} can be computed in the same manner as was done for the double trace moments. Its moments are identical to those of \eqref{eq:two_pairings}, only with $k_i =  \tilde{k}_1+\tilde{k}_2$, where $\tilde{k}_i$'s are the moments in the expansion of $e^{-\beta_i H}$. We can immediately switch $\tilde{k}_i$ with derivatives $\beta_i \partial_{\beta_i}$, resulting in a value of 
\begin{equation}
    I_2 = \frac{1}{2} {\binom{N}{p}}^{-1}\left(\beta_1 \partial_{\beta_1} +\beta_2 \partial_{\beta_2}\right) G^{(2)}(\beta_1,\beta_2;\tilde{p})
    \left(\beta_1' \partial_{\beta_1'} +\beta_2' \partial_{\beta_2'}\right) G^{(2)}(\beta_1',\beta_2';\tilde{p}).
\end{equation}
The full connected correlation function is the sum of $I_1$ and $I_2$.
\\

The leading order behavior of $G_c^{(2,2)}$ is
\begin{equation}
    G_c^{(2,2)}(\beta_1,\beta_2,\beta_1',\beta_2';\tilde{p})\sim \left\{ \begin{array}{cc}
        {\binom{N}{\tilde{p}}}^{-1}  & \tilde{p} < p, \\
        {\binom{N}{p}}^{-1}  & \tilde{p} > p .
    \end{array} \right.
\end{equation}
If we had $n$ such traces the leading order behavior would be
\begin{equation} \label{eq:G_n_correlation}
    G_c^{(2,\ldots,2)}(\beta_{1}^{(1)},\beta_{2}^{(1)},\ldots,\beta_{1}^{(n)},\beta_{2}^{(n)};\tilde{p})\sim \left\{ \begin{array}{cc}
        {\binom{N}{\tilde{p}}}^{n-1}  & \tilde{p} < p, \\
        {\binom{N}{p}}^{n-1}  & \tilde{p} > p .
    \end{array} \right.
\end{equation}
If $\tilde{p} > p$ then the exact result of the $n$ trace function will be a sum of cactus diagrams with derivative operators, as in \eqref{eq:connected_Z_cactus}. If $\tilde{p}< p $ then the leading order behavior will just be a constant multiplying the disconnected result, similar to \eqref{eq:I_1_2-point}.

Of course, the correlation induced by $A$ in diagram I in figure \ref{fig:4ptfun} is a new effect on top of figure II, and as such we can trust it (there is also a clear regime where it is larger than subleading fluctuation parameters associated with $H$). It is associated with a fluctuation parameter derived from $A$: $\sum_I A_I^2$. Correspondingly, the vector model we find for $A$ is in the same spirit as the one we found for $h_2$.





\subsection{The associated vector model for operators}

In the case where $\tilde{p} \neq p$ we can construct a dual vector model for each random operator which captures the leading order behavior, similar to the one we found for the Hamiltonian in section \ref{sec:vector}. This time the 0-dimensional vector field represents the random couplings of the operator in addition to the Hamiltonian.

Starting from the operator $A$ defined in \eqref{eq:def_operator}, we consider the zero dimensional fields $\phi^{(A)}_I$ taking an index $I$ from $1$ to $\epsilon_A^{-1} \equiv \binom{N}{p_A}$, the number of index sets, or random couplings, in the operator $A$. Similar to the Hamiltonian case, the joint generating functional of $A$ is \footnote{We take the couplings of $A$ to be Gaussian. If the couplings of $A$ obeyed a different distribution, then that distribution should be used instead, as was in the Hamiltonian case.}
\begin{equation} \label{eq:action_of_A}
\begin{split}
&\expt{\prod_{i=1}^n\tr\left[\exp\left(-\beta_i H + \int d\tau S_i(\tau) A(\tau) \right)\right]}_{J,\tilde{J}} 
    = \\
     &\qquad \qquad \quad  = \int D\phi_I~ D\phi_I^{(A)} ~ e^{-\sum_{|I|=\tilde{p}} (\phi^{(A)}_I)^2/(2 \epsilon_A) - \sum_{|I|=p}\phi_I^2/(2 \epsilon)} \\
    & \qquad \qquad \qquad
    \times \prod_{i=1}^n \expt{\tr\left[\exp\left(-\sqrt{\sum_{|I|=p}\phi_I^2}~\beta_iH + \sqrt{\sum_{|I|=\tilde{p}} (\phi_I^{(A)})^2} \int d\tau S_i(\tau) A(\tau) \right)\right]}_{J,\tilde{J}} .
\end{split}
\end{equation}
As before, we can take only the connected part of both sides if we so desire. Additionally we can then go to the variables $\phi_I^2$ and $(\phi_I^{(A)})^2$ and reduce the integration to these two fluctuation parameters. Since the integration on these two variables is after we do the average over $J,{\tilde J}$, then we can replace $\langle . \rangle_{J,{\tilde J}}$ by their GR expression and obtain the GR description of these two fluctuation parameters.  

We stress that this vector model is only accurate at leading order in $\epsilon_A$ or $\epsilon$, and for all values of $p,\tilde{p}$, as long as $p\neq\tilde{p}$. This means that any correlation functions whose leading order scaling is smaller cannot be computed using this dual vector model. In particular the correlations functions involving an odd number of $A$ insertions in some traces is more suppressed in $\epsilon_A$, similarly to an odd number of Hamiltonian insertions considered in section \ref{sec:NewFluc}, and so cannot be computed using this vector model. To compute these odd correlation functions we must add additional terms to the effective action, similarly to what was done for the Hamiltonian in section \ref{sec:h3_efctv}. Another interesting correlation function that cannot be computed using this vector model is the correlations of single operator insertions. Both these cases will be discussed in the next subsection.

\subsection{1-point functions and an odd number of operator insertions}

A slightly more complicated observable is the fluctuation (across the ensemble) of the 1-point function of the operator $A$. This will teach us about the expectation value of the operator $A$ in a specific realization. Thus we want to find the leading order behavior of the connected thermal 1-point function 
\begin{equation}
    G^{(1,1)}(\beta_1 ,\beta_2;\tilde{p}) \equiv \left\langle \tr\left(e^{-\beta_1 H} A \right)
    \tr\left(e^{-\beta_2 H} A \right)\right\rangle_{J,\tilde{J}} ,
\end{equation}
or its moments
\begin{equation}
    m_{\tilde{p}}(k_1,k_2) \equiv \left\langle \tr\left(H^{k_1} A \right)
    \tr\left(H^{k_2} A \right)\right\rangle_{J,\tilde{J}}.
\end{equation}

For the expectation value to not vanish, we must have that the two $A$'s are contracted, and that each fermion in the $A$ chain must have a pair in one of the $H$ operators that are contracted between the traces. As before, only minimally connected diagrams contribute at leading order, so the number of $H$'s contracted between the sides will be $n = \lceil \tilde{p}/p \rceil$, except for $\tilde{p} < p$ when $n = 2$. In the following we will discuss a couple of simpler cases:

\subsubsection*{Case 1: $\tilde{p} = p$}

\begin{figure}[h]
\centering
\includegraphics[width=0.6\textwidth]{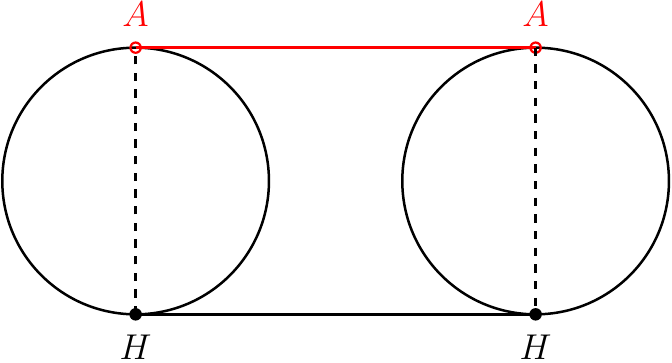}
\caption{Double-trace of one-point functions of operators with Hamiltonian size.}
\label{fig:correlators_A_case1}
\end{figure}

If $\tilde{p} = p$ then we can have all the fermions in $A$ be in a single $H$ which is contracted between the traces. All the other $H$'s on each side will be paired, so the leading order moments will be odd, and their contribution will be similar to the double trace moments
\begin{equation}
    m_{p}(2k_1-1,2k_2 - 1) = {N \choose p}^{-1} m_{2k_1} m_{2k_2}.
\end{equation}
This is one way of writing the result. Another way is arrange, in each trace, the insertions of $H$ along a circle, along with one insertion of $A$. The trace is non-zero if the $\Psi_I$ from one of the H is the same as one of the $\Psi_{I'}$ from one of the Hamiltonians, and each of the $2k-1$ Hamiltonians can each participate in this overlap. Then write $m_{2k_1-1,2k_2-1}$ as 2-pt function of an operator of dimension 1 in each of the traces. 
This case is shown in figure \ref{fig:correlators_A_case1}.

The thermal two point function in this case will be
\begin{align}\label{AAConnect}
\begin{split}
    G^{(1,1)}(\beta_1,\beta_2;p) &=  {N \choose p}^{-1}\sum_{k_1,k_2\geq 1} \frac{(-\beta_1)^{2k_1-1}(-\beta_2)^{2k_2-1}}{(2k_1-1)!(2k_2-1)!}  m_{2k_1} m_{2k_2} \\
    &= {N \choose p}^{-1} \frac{d\mathcal{Z}(\beta_1)}{d\beta_1} \frac{d\mathcal{Z}(\beta_2)}{d\beta_2} .
\end{split}
\end{align}


In the same spirit as in section \ref{sec:h3_efctv} and section \ref{sec:GenFluc} we would like to find a ``minimal'' description of this connected correlator, after we do the ensemble average. This can be implemented using the substitution
\begin{equation}
    \tilde{H}= H + \alpha A
\end{equation}
where $A$ is the operator above and $\alpha$ is a random variable with 
\begin{equation}
    \left\langle\alpha\right\rangle=0,\qquad \langle \alpha^2 \rangle = {N \choose p}^{-1}.
\end{equation}
In gravity, the interpretation is the following. A marginal operator corresponds to a massless field, and adding $\alpha A$ to the Hamiltonian corresponds to shifting the VEV of the field in the bulk. In the theory in which we carry out an ensemble average over the coefficients of $A$, $\alpha$ is averaged over. In the language above, $\alpha$ encodes the fluctuation parameter $\sum_I J_I {\tilde J}_I$.     We interpret the case of a fixed realization as having a fixed $\alpha$.

The ansatz is arranged such that we reproduce the correlator \eqref{AAConnect}. At the level of a single trace correlator it (1) does not introduce a 1-pt function for $A$ since we take $\langle \alpha \rangle=0$, and (2) introduces a correction to the partition function at order $\epsilon$, which is small compared to terms we are neglecting. At the level of connected multi-trace correlation function, it introduces corrections which are subleading to the fluctuation parameter $h_2$ (which is the fluctuating parameter in front of $H$). For example, in the connected part of 2 traces it will introduce a term proportional to $\langle \alpha^4 \rangle - \langle \alpha^2\rangle^2$, but this is of order $\epsilon^2$, whereas the cactus diagrams that we discussed in section \ref{sec:cactus} give a contribution of order $\epsilon$.

Finally we would like to point out that we could also think of $A$ not as a random operator but rather as a fixed operator, and this same computation would yield the variance of the thermal expectation value of $A$. For example we can take $A = \Psi_I$ for a fixed index set $I$ and using this chord diagram formalism calculate the probability distribution of the thermal expectation value of $\Psi_I$.\footnote{Of course the thermal expectation value of $\Psi_I$ must vanish on average, but it's variance is non-zero.} As this computation shows, the operators with the largest expectation value, on average, would be operators of the same length as the Hamiltonian. Thus living in a given realization of the couplings, the first operators whose expectations deviate from the average value are the ones of length $p$, and their typical deviation is of order $\epsilon^{1/2}\sim N^{-p/2}$.\footnote{Note that $G^{(1,1)}$ measures the variance, which is the square of the standard deviation.} This also measures the extent of the breaking of the $O(N)$ symmetry: at large $N$ it looks like the theory has an $O(N)$ symmetry, just like averaged model has, but really the exact realization of the couplings breaks this symmetry by a small amount of order $\epsilon^{1/2}$.

\subsubsection*{Case 2: $p \neq \tilde{p}\leq 2p$}\label{GenPsml}

In this case we must pair the fermions in $A$ with those in at least two insertions of $H$ in each trace. The random coefficients of the $H$'s are then contracted between the two sides. We must give half of the fermions in $A$ to one of the $H$'s, and the other half to the other $H$. The rest of the fermions in the two $H$'s are paired. This contribution is similar to the one discussed on section \ref{sec:double_trace_3_lines}. For example we can think of figure \ref{fig:odd_double_trace_CD}, where now two of the nodes that connect the different traces are $H$ nodes, and one is an $A$ node. The moments will once again factorize in a similar way to the odd double trace moments (see section \ref{sec:double_trace_3_lines}) 
\begin{equation} \label{eq:OperatorMoment}
    m_{\tilde{p}}(k_1,k_2) = m_{\tilde{p}}(2,2) W_{\tilde{p}}(k_1-2)W_{\tilde{p}}(k_2-2).
\end{equation}
The combinatorial factor is
\begin{equation}
    m_{\tilde{p}}(2,2) = 2{N \choose p}^{-2} {N \choose p - \tilde{p}/2} {\tilde{p} \choose \tilde{p}/2} ,
\end{equation}
so the order of this will be $\epsilon^2\leq m_{\tilde{p}}(2,2)<\epsilon$. We note that this contribution only happens if $\tilde{p} \mod 4 \equiv 0$, as otherwise the two possible contractions anti-commute and thus cancel each other, similar to the argument as to why the contributions to the odd moments vanish if $p\mod 4 \equiv 2$ in section \ref{sec:double_trace_3_lines}.

The function $W_{\tilde{p}}(k)$ is also very similar to $W(k)$ from equation \eqref{fluc33B} - only the size of the index set of the three chords is no longer $p/2$, but rather two of them have length $\tilde{p}/2$, and the third has length $p-\tilde{p}/2$. Explicitly, in the large $N$ limit, it is given by the three point function
\begin{equation} {\label{eq:W Tilde}}
    W_{\tilde{p}}(k)  \equiv {\binom{N}{\tilde{p}/2}}^{-2}
    {\binom{N}{p - \tilde{p}/2}}^{-1}
    \sum_{\substack{|I_1|=|I_2|=\tilde{p}/2 \\ |I_3| = p-\tilde{p}/2\\ k_1+k_2+k_3 = k}}
    \expt{\tr\left(\Psi_{I_1}\Psi_{I_2}H^{k_1}
    \Psi_{I_2}\Psi_{I_3}H^{k_2}
    \Psi_{I_3}\Psi_{I_1}H^{k_3}\right)}_J . 
\end{equation}

We can write these contributions as a correction to the vector model coming from a matrix model type interaction, similar to section \ref{sec:subleading_theory_J}. The idea is to add an additional term to the effective Hamiltonian, and correct the operator $A$ accordingly, in order to allow for an index contraction of $(H,H,A)$. In order to not change the leading order result of an even number of insertions of $A,H$, we need to make sure that these new terms are subdominant with respect to the previous corrections. The Hamiltonian is then
\begin{equation}
    H_{\text{eff}} = h_2H + \beta:O_{\tilde{p}/2} O_{p-\tilde{p}/2}:,
\end{equation}
where the distribution of $h_2$ is given in (\ref{eq:thermal_from_r}). The operator $A$ is corrected by the random operator
\begin{equation}
    A_{\text{eff}} = h_AA+\alpha:O_{\tilde{p}/2} O_{\tilde{p}/2}:,
\end{equation}
where by normal ordering we mean that we don't allow self contractions for the random operator (if the two operators are of the same length). We have introduced the Gaussian variables $\alpha,\beta$, with zero mean and variances to be specified later, and the variable $h_A$ with mean $1$ and variance ${N \choose \tilde p}^{-1}$. Now we can compute the operator moment
\begin{align}
    M_{\tilde{p},c}(k_1,k_2) = \left<M_{\tilde{p},\text{eff}}(k_1)M_{\tilde{p},\text{eff}}(k_2)\right>_{\alpha,\beta,h_2,h_A,conn}, \qquad M_{\tilde{p},\text{eff}}(k) = \left<\tr H_{\text{eff}}^k A_{\text{eff}}\right>_{J,\tilde{J},O}.
\end{align}
To pick up the relevant contribution, we go to the $\beta^2$ contribution from $H_{\text{eff}}$ in each trace. The effective moment then gives
\begin{align}
\begin{split}
    M_{\tilde{p},\text{eff}}(k) &= h_2^{k-2}\alpha\beta^2 \sum_{i_1+i_2+i_3=k-2}\left<\tr H^{i_1}:O_{\tilde{p}/2}O_{p-\tilde{p}/2}:H^{i_2}:O_{\tilde{p}/2}O_{p-\tilde{p}/2}:H^{i_3}:O_{\tilde{p}/2}^2:\right>
    \\
    &=h_2^{k-2}\alpha\beta^2 (1+q^{\tilde{p}^2/(4p^2)}){N \choose \tilde{p}/2}^2{N \choose p-\tilde{p}/2}W_{\tilde{p}}(k-2),
\end{split}
\end{align}
where $q = e^{-2p^2/N}$, and the factor $(1+q^{\tilde{p}^2/(4p^2)})$ comes from different possible contractions of the fluctuation fields.
This means that
\begin{align}
    M_{\tilde{p},c}(k_1,k_2) = \left<h_2^{k_1+k_2-4}\right>\left<\alpha^2\right>\left<\beta^4\right>\left(1+q^{\tilde{p}^2/(4p^2)}\right)^2{N \choose \tilde{p}/2}^4{N \choose p-\tilde{p}/2}^2W_{\tilde{p}}(k_1-2)W_{\tilde{p}}(k_2-2).
\end{align}
This matches (\ref{eq:OperatorMoment}) if we require
\begin{align} \label{eq:OperatorRandomVarValues}
\begin{split}
    \left<\alpha^2\right>\left<\beta^4\right> & = \frac{m_{\tilde{p}}(2,2)}{{N \choose \tilde{p}/2}^4{N \choose p-\tilde{p}/2}^2(1+q^{\tilde{p}^2/(4p^2)})^2} 
    \\
    & = 2{\tilde{p} \choose \tilde{p}/2}{N \choose p}^{-2}{N \choose \tilde{p}/2}^{-4}{N \choose p-\tilde{p}/2}^{-1}(1+q^{\tilde{p}^2/(4p^2)})^{-2}.
\end{split}
\end{align}

Next we need to verify that we have not introduced any large corrections to observables that include an even number of both $A,H$ insertions, in all traces. This motivates us to require that both the $\alpha$ and $\beta$ contractions need to be subleading w.r.t the $A$ and $H$ contractions. That is $\left<\alpha^4\right>\ll\left< h_A^2 \right> = {N \choose \tilde{p}}^{-1}$,
 and $\left<\beta^4\right> \ll \left<h_2^2\right>={N \choose p}^{-2}$, which we can easily satisfy, if we define 
\begin{align}
\begin{split}
    \left<\alpha^2\right> = 2{\tilde{p} \choose \tilde{p}/2}{N \choose \tilde{p}/2}^{-4}, \qquad \left<\beta^4\right> = {N \choose p}^{-2}{N \choose p-\tilde{p}/2}^{-1}(1+q^{\tilde{p}^2/(4p^2)})^{-2}.
\end{split}
\end{align}

As before, we interpret the operators $O_{{\tilde p}/2}$ and $O_{p-{\tilde p}/2}$ as new operators in the theory (or new fields in the bulk), whose existence is required in order to explain, in gravity, this two-trace correlator. Before we  saw this happening for the Hamiltonian, but it is actually a feature associated with any operator observable in the theory.

\subsubsection*{Case 3: $\tilde{p} > 2p$}
These will look like the deformations of high order corrections to the double trace moments. In general they will involve higher point functions, a sum over all the possible ways to distribute $\tilde{p}$ among the various $H$'s, and a sum over all the ways to distribute the remaining indices in the $H$'s among themselves. This function will have a discontinuity every time $\tilde{p} = np$ for some integer $n$, as the leading order moments will change parity, from even moments to odd moments, or vice versa. We have not carried out a full analysis of these cases.



\section{Multi-Trace Correlations from the Replica Path Integral} \label{sec:path_integral}

Computations of SYK model are typically done in the path integral formulation of the theory, as in \cite{MaldecenaStanford,Polchinski_2016,Rosenhaus_2019}. As the resulting action is a large $N$ action, the theory can be solved in the large $N$ limit by finding the saddle points and looking at the fluctuations around them. 

Similarly, multi-trace expectations have previously been computed in the path integral formulation using replicas \cite{Cotler:2016fpe,saad2018,Aref_eva_2019,Wang_2019,khramtsov2020spectral,penington2019}. The main object of interest in these computations is the spectral form factor, which is an analytic continuation of the double trace thermal partition function. The main focus of these computations is finding non-diagonal saddle points of the two replica path integral that generate the RMT contribution to the multi-trace expectations \cite{saad2018}. These additional saddle points are not the source of the global fluctuations, as they are related to connected topologies in the gravity dual. Rather, we will demonstrate how to calculate the leading order contributions to the connected thermal expectation function as a perturbative expansion around the disconnected saddle.

We shall start out by looking at the fermionic path integral and show how the global modes arise in this description. Then we use this intuition to derive a formulation of the two replica action in terms of the bi-local fields $G$ and $\Sigma$. This action is found via an expansion followed by a resummation of the interacting part, and has a completely different form (and saddle point equations) than the naive action derived in \cite{saad2018}. Finally we show how the global modes naturally arise from a perturbative expansion around the disconnected saddle of this action.

\subsection{The path integral derivation}
The thermal partition function of the SYK model is given by the path integral\footnote{This is the partition function before disorder averaging.}
\begin{equation}
    Z(\beta) = \int D\psi_i e^{-S[\psi]},
\end{equation}
with the Euclidean action given by 
\begin{equation}
    S[\psi] = \int_0^{1} d\tau\left\{ \sum_{i=1}^N \psi_i \partial_\tau \psi_i + \beta i^{p/2} { \binom{N}{p} }^{-1/2} \sum_{|I| = p} J_{I}\Psi_{I} \right\},
\end{equation} 
where $\tau$ is the normalized Euclidean time and we set $\mathcal{J}=1$. The expectation value of the thermal partition function is computed through the path integral
\begin{equation}
    \mathcal{Z}(\beta) =\expt{ \int \prod_{i=1}^N 
    D\psi_i~ e^{-S[\psi]} }_J .
\end{equation}
Taking the integral over $J_I$'s leaves us with the following expression for the thermal partition function:
\begin{equation}
    \mathcal{Z}(\beta) = \int \prod_{i=1}^N 
    D\psi_i~ \exp\left\{ -\sum_{i=1}^N \int d\tau ~\psi_i(\tau) \partial_\tau \psi_i(\tau) + \frac{1}{2} (-1)^{p/2}\beta^2 \epsilon \sum_{|I| = p} \int d\tau~ d\tau'~ \Psi_{I}(\tau)\Psi_{I}(\tau') \right\} ,
\end{equation}
with $\epsilon = {\binom{N}{p}}^{-1}$ as before. We will call the action in the path integral above $S_0[\psi]$, and it is nothing but the standard single trace action.

We can now compute the leading connected double trace thermal expectation value using this same path integral formulation of the model. Notice that
\begin{equation}
\begin{split}
    \mathcal{Z}(\beta_1,\beta_2) &= \expt{ \int \prod_{i=1}^N 
    D\psi_i^{(1)}~D\psi_i^{(2)}~ e^{-S[\psi^{(1)}](\beta_1) - S[\psi^{(2)}](\beta_2)} }_J\\
    &= \int \prod_{i=1}^N D\psi_i^{(1)}~D\psi_i^{(2)}~ \exp\left\{ -S_0[\psi^{(1)}] -S_0[\psi^{(2)}] + S_{\text{int}}[\psi^{(1)},\psi^{(2)}] \right\} ,
\end{split}
\end{equation}
with the interaction term
\begin{equation}
    S_{\text{int}}[\psi^{(1)},\psi^{(2)}] = 
        (-1)^{p/2} \epsilon \beta_1 \beta_2 \sum_{I} \int d\tau ~d\tau' \Psi^{(1)}_I(\tau) \Psi^{(2)}_I(\tau').
\end{equation}

We can treat this extra term as a small interaction term, and expand the exponent in a Taylor series, as is typically done in perturbation theory. The zeroth order term in the series is just the disconnected component. The first term in the series vanishes as each free action $S_0[\psi]$ has a $\ZZ_2$ symmetry of $\psi^{(i)} \ra -\psi^{(i)}$, but the interaction term is odd under this symmetry. Thus the leading order of the connected component will be the second order term, and will give the contribution
\begin{equation}
    \begin{split}
        \mathcal{Z}_c(\beta_1,\beta_2) &= \frac{1}{2}\epsilon^2 \beta_1^2 \beta_2^2 \int \prod_{i=1}^N D\psi_i^{(1)}~D\psi_i^{(2)}~
    \int \prod_{j=1}^4d\tau_j\sum_{|I| = |J| = p} \Psi^{(1)}_J(\tau_1)\Psi^{(2)}_J(\tau_2) \Psi^{(1)}_I(\tau_3)\Psi^{(2)}_I(\tau_4)\\
    & \qquad \qquad \qquad \qquad \qquad \times
    e^{-S_0[\psi^{(1)}] - S_0[\psi^{(2)}]} .
    \end{split}
\end{equation}

The path integral will vanish because of the aforementioned $\ZZ_2 \times \ZZ_2$ symmetry if $J \neq I$, in which case we can write the connected thermal partition function as
\begin{equation} \label{eq:double_fromPIFull}
    \begin{split}
    \mathcal{Z}_c(\beta_1,\beta_2) &=\frac{1}{2}\epsilon^2 \beta_1^2 \beta_2^2 \int \prod_{i=1}^N D\psi_i^{(1)}~D\psi_i^{(2)}~
    \int \prod_{j=1}^4d\tau_j\sum_{|I| = p} \Psi^{(1)}_I(\tau_1)\Psi^{(2)}_I(\tau_2) \Psi^{(1)}_I(\tau_3)\Psi^{(2)}_I(\tau_4)\\
    & \qquad \qquad \qquad \qquad \qquad \times
    e^{-S_0[\psi^{(1)}] - S_0[\psi^{(2)}]} \\
    &= 2\epsilon \beta_1^2 \beta_2^2 \frac{d}{d(\beta_1^2)}\mathcal{Z}(\beta_1)~
    \frac{d}{d(\beta_2^2)}\mathcal{Z}(\beta_2) ,
    \end{split}
\end{equation}
which is identical to \eqref{eq:double_thermal_partition} computed using the moment method. 

If we want to calculate $n$ trace thermal partition functions then the computation is similar. The ``free'' part of the action will consist of $n$ averaged actions, one for each of the $n$ replicas (or traces). Additionally we will have an interaction term of the form
\begin{equation}
    S_{\text{int}} = (-1)^{p/2} \epsilon \sum_{i\neq j = 1}^n \beta_i \beta_j \int d\tau d\tau' \sum_{|I| = p} \Psi^{(i)}_I(\tau) \Psi^{(j)}_I(\tau') .
\end{equation}
Expanding the interaction term order by order in $\epsilon$ will again give us only cactus diagrams as the leading order contribution to the connected thermal expectation function.

\subsection{Path Integral in terms of $G$ and $\Sigma$}

Up to now we have worked with the action for the original fermions to compute the connected thermal expectation functions. However the path integral formulation is much more elegant as we can express it in terms of bi-local fields $G$ and $\Sigma$ rather than the original $N$ fermions. Furthermore, in these variables (with the correct scaling) the action becomes a large $N$ action, and thus a saddle point approximation is valid (see for example \cite{Rosenhaus_2019} for details). We would like to reproduce the above computation when considering these bilocal fields, both for completeness and to be able to relate this work to others, such as \cite{saad2018}.

We will define the bi-local fields as
\begin{align} {\label{eq:G,Sigma defs}}
    G_{i}(\tau_1,\tau_2) \equiv \frac{1}{N}\sum_{j=1}^N \psi^{(i)}_j(\tau_1)\psi^{(i)}_j(\tau_2),&
    &&G_{ii'}(\tau_1,\tau_2) \equiv \frac{1}{\sqrt{N}} \sum_{j=1}^N \psi^{(i)}_j(\tau_1)\psi^{(i')}_j(\tau_2).
\end{align}
We use a slightly different normalization for the $G_{i i'}$ field, the reason for which will become apparent in the analysis. Furthermore we shall concentrate on the double trace thermal partition function for brevity, though similar calculations can be preformed for high trace expectations.

We introduce Lagrange multipliers $\Sigma_i$ and $\Sigma_{12}$ to enforce the definitions of $G$, as is typically done in the replica path integral (see for example \cite{saad2018}). After substituting in the definitions for $G$, and taking the integral over the couplings, we arrive at the following expression for the thermal partition function
\begin{equation} {\label{eq:Connected_Z_before_exp}}
    \begin{split}
        Z(\beta_1,\beta_2) = \int & \prod_{j=[1,2],i=[1,N]} D\psi_i^{(j)} ~DG_1 ~DG_2 ~DG_{12} ~D\Sigma_1 ~D\Sigma_2 ~D\Sigma_{12}~ \\
    &\exp\Bigg\{-\int d\tau \sum_{j=[1,2],i=[1,N]} \psi_i^{(j)} d_\tau \psi_i^{(j)} -N\int d\tau_1 d\tau_2 \bigg[ \Sigma_1 (\tau_1,\tau_2) G_1(\tau_1,\tau_2) \\
    &+ \Sigma_2 (\tau_1,\tau_2) G_2(\tau_1,\tau_2)  - \frac{\sqrt{\epsilon}  \beta_1 \beta_2}{N \sqrt{p!}} G^p_{12}(\tau_1,\tau_2) - \frac{\beta_1^2}{2N} G^p_1(\tau_1,\tau_2) \\
    &  - \frac{\beta_2^2}{2N} G^p_2(\tau_1,\tau_2)
     -\frac{1}{N}\sum_{j=[1,2],i=[1,N]} \Sigma_j(\tau_1,\tau_2)  \psi_i^{(j)}(\tau_1)\psi_i^{(j)}(\tau_2) \bigg] \\
     &-\int d\tau_1 d\tau_2 \Sigma_{12} (\tau_1,\tau_2) \left(G_{12}(\tau_1,\tau_2) -\frac{1}{\sqrt{N}}\sum_{i=[1,N]} \psi_i^{(1)}(\tau_1)\psi_i^{(2)}(\tau_2) \right)   
     \Bigg\} .
    \end{split}
\end{equation}

Typically at this stage the fermions are immediately integrated out, leaving the well known $G$--$\Sigma$ action \cite{saad2018}, however we will take another route. We will expand the exponential of the last term (the one involving fermions) in the above path integral action and consider the path integral over the fermions order by order. As the fermions are Grassmann variables, the integral over them will vanish unless they exactly come in pairs. This already tells us that all the odd terms in this expansion will vanish. The zero term is just a one, so the first nontrivial term is the second term which will be
\begin{equation}
    I_2 = \int \prod_{k=1}^4 d\tau_k \frac{1}{2N}\sum_{i,j}  \Sigma_{12}(\tau_1,\tau_2) \Sigma_{12}(\tau_3,\tau_4)  \psi_i^{(1)}(\tau_1)\psi_i^{(2)}(\tau_2) \psi_j^{(1)}(\tau_3)\psi_j^{(2)}(\tau_4),
\end{equation}
which will vanish for all $i \neq j$, by Grassmann integration arguments. This leaves us with the term
\begin{equation}
\begin{split}
    I_2 &= -\int \prod_{k=1}^4 d\tau_k \frac{1}{2N}\sum_{i}  \Sigma_{12}(\tau_1,\tau_2) \Sigma_{12}(\tau_3,\tau_4)  \psi_i^{(1)}(\tau_1)\psi_i^{(1)}(\tau_3) \psi_i^{(2)}(\tau_2) \psi_i^{(2)}(\tau_4) \\
    & = -\frac{1}{2} \int \prod_{k=1}^4 d\tau_k ~ \Sigma_{12}(\tau_1,\tau_2) \Sigma_{12}(\tau_3,\tau_4)  G_1(\tau_1,\tau_3) G_2(\tau_2,\tau_4) ,
\end{split}
\end{equation}
where in the last line we substituted the definitions of the two point functions.

 In general, all higher expansion will reduce to index pairings of the fermions from Grassmann integration arguments, allowing us to substitute in the two point functions. This leads to Wick contractions of $\Sigma_{12}$'s. We note that we neglect corrections arising from the case where more than two fermions share the same index as these are suppressed by higher order powers in $1/N$.\footnote{Similar corrections have already been neglected in the path integral derivation when we substituted $G^p(\tau,\tau')$ for $\sum_I \Psi_I(\tau) \Psi_I(\tau')$.}

 The numerical factor multiplying the Wick contraction of $k$ contraction is $(-1)^{k}$ from anti-commuting the fermions times $1/(2k)!$ from expanding the exponent times $(2k-1)!!$ which is the number of Wick contractions. Overall this gives a factor of $(-1/2)^k/k!$, allowing us to re-sum the exponent, and rewrite the partition function as 
\begin{equation}
    \begin{split}
        Z(\beta_1,\beta_2) = \int & \prod_{j=[1,2],i=[1,N]} D\psi_i^{(j)} ~DG_1 ~DG_2 ~DG_{12} ~D\Sigma_1 ~D\Sigma_2 ~D\Sigma_{12}~ \\
    &\exp\Bigg\{-\int d\tau \sum_{j=[1,2],i=[1,N]} \psi_i^{(j)} d_\tau \psi_i^{(j)} -N\int d\tau_1 d\tau_2 \bigg[ \Sigma_1 (\tau_1,\tau_2) G_1(\tau_1,\tau_2) \\
    &+ \Sigma_2 (\tau_1,\tau_2) G_2(\tau_1,\tau_2) + \frac{1}{N}\Sigma_{12} (\tau_1,\tau_2) G_{12}(\tau_1,\tau_2)  - \frac{\sqrt{\epsilon} \beta_1 \beta_2}{N \sqrt{p!}} G^p_{12}(\tau_1,\tau_2) \\
    &- \frac{\beta_1^2}{2N} G^p_1(\tau_1,\tau_2) - \frac{\beta_2^2}{2N} G^p_2(\tau_1,\tau_2)
     -\frac{1}{N}\sum_{j=[1,2],i=[1,N]} \Sigma_j(\tau_1,\tau_2)  \psi_i^{(j)}(\tau_1)\psi_i^{(j)}(\tau_2) \Bigg] \\
     & - \frac{1}{2} \int \prod_{k=1}^4 d\tau_k~\Sigma_{12}(\tau_1,\tau_2) \Sigma_{12}(\tau_3,\tau_4) G_1(\tau_1,\tau_3) G_2(\tau_2,\tau_4) \Bigg\} .
    \end{split}
\end{equation}
    
Now we are ready to integrate out the fermions, leaving us with the large $N$ action
\begin{equation} \label{eq:action_G_Sigma}
    \begin{split}
        S = N\sum_{i=1,2}&\left\{-\frac{1}{2}\log\left[\det\left(\Sigma_i - \partial_\tau\right)\right] 
        + \int d\tau~d\tau'\left(\Sigma_i(\tau,\tau')G_i(\tau,\tau') 
    - \frac{\beta_i^2}{2N}G_i^p(\tau,\tau')\right) \right\} \\
    & +\int d\tau~d\tau'\left[ \Sigma_{12}(\tau,\tau') G_{12}(\tau,\tau') 
     - \frac{ \sqrt{\epsilon} \beta_1 \beta_2}{\sqrt{p!} } G_{12}^p(\tau,\tau')\right] \\
     &+ \frac{1}{2}\int \prod_{k=1}^4 d\tau_k~ \Sigma_{12}(\tau_1,\tau_2) \Sigma_{12}(\tau_3,\tau_4) G_1(\tau_1,\tau_3) G_2(\tau_2,\tau_4) .
    \end{split}
\end{equation}
The first term in the action \eqref{eq:action_G_Sigma} is the sum of the original actions for the replica diagonal fields. To this we add an action for the replica off-diagonal fields, along with a quadratic interaction term between the diagonal and off-diagonal fields.

We can see that the contributions order by order in $\epsilon$ match up with the previous results if we integrate out $\Sigma_{12}$, which involves completing the square.\footnote{Note that we normalized $\int d\tau = 1$, so there are no divergences and integrating out $\Sigma_{12}$ is formally exact. Furthermore note that we are integrating $\Sigma_{12}$ along the imaginary axis.} This gives a quadratic piece to $G_{12}$, and the final action
\begin{equation} \label{eq:action_G_Sigma_final}
    \begin{split}
        S = N \sum_{i=1,2}&\left\{- \frac{1}{2}\log\left[\det\left(\Sigma_i - \partial_\tau\right)\right] 
        + \int d\tau~d\tau'\left(\Sigma_i(\tau,\tau')G_i(\tau,\tau') 
    - \frac{\beta_i^2}{2N}G_i^p(\tau,\tau')\right) \right\} \\
    & -\int d\tau~d\tau' \frac{ \sqrt{\epsilon} \beta_1 \beta_2}{\sqrt{p!}} G_{12}^p(\tau,\tau')
    - \frac{1}{2}\int \prod_{k=1}^4 d\tau_k~ \frac{G_{12}(\tau_1,\tau_2) 
     G_{12}(\tau_3,\tau_4)}{G_1(\tau_1,\tau_3) G_2(\tau_2,\tau_4)} .
    \end{split}
\end{equation}
The final step is expanding this action order by order perturbatively in $\sqrt{\epsilon}$, after which the quadratic action for $G_{12}$ leads to Wick contractions of the form 
\begin{equation} \label{eq:G_wicks}
    \contraction{}{G_{12}}{(\tau_1,\tau_2)}{G_{12}}
    G_{12}(\tau_1,\tau_2)G_{12}(\tau_3,\tau_4)
    =  -G_{1}(\tau_1,\tau_3)G_{2}(\tau_2,\tau_4) .
\end{equation}
This transforms the calculation of the two replica thermal partition function to a sum of correlation functions of $G_1$ and $G_2$ in the original theory.

For concreteness let us analyze the first few terms in this expansion, and compare them to the previous results. It is clear that the zeroth order term gives the disconnected contribution, while the term of order $\sqrt{\epsilon}$ vanishes as self contractions vanish due to the anti-symmetry of $G_i$ (namely $G_i(t,t) = G_i(0) = 0$ by construction.)

The first non-trivial term is the one of order $\epsilon$, which is
\begin{equation}
  \epsilon \frac{1}{2p!}\beta_1^2 \beta_2^2
  \int d\tau_1 \ldots d\tau_4 G^{p}(\tau_{12})G^{p}(\tau_{34}).
\end{equation}
There is only one way to do the Wick contractions for this term,  and the resulting contribution is
\begin{equation}
    \epsilon \frac{1}{2}\beta_1^2 \beta_2^2 \expt{\int d\tau_1 d\tau_3~ G_1^p(\tau_{13})}\expt{\int d\tau_2 d\tau_4~G_2^p(\tau_{24}) } = \frac{\epsilon}{2} \beta_1 \dd{}{\beta_1}{\cal Z}_0(\beta_1)
    \beta_2 \dd{}{\beta_2} {\cal Z}_0(\beta_1),
\end{equation}
where the expectations are with respect to the standard SYK action for each replica.\footnote{Note that the $p!$ cancels as there are $p!$ equivalent ways of doing the Wick contraction.} This leading order connected contribution is identical to the previously obtained results in equations \eqref{eq:double_thermal_partition} and \eqref{eq:double_fromPIFull}, as should be expected.

The next correction comes from taking down three interaction terms, and it reads
\begin{equation}
    \epsilon^{3/2} \frac{1}{6(p!)^{3/2}}\beta_1^3 \beta_2^3
  \int d\tau_1 \ldots d\tau_6 G^{p}(\tau_{12})G^{p}(\tau_{34})G^{p}(\tau_{56})
\end{equation}
Here too there is only a single way to do the Wick contractions (as we do not allow self contractions,) and that is to pair half of each $G_{12}^p$ with half of a different $G_{12}^p$. All the different choices for these contractions give the same contribution, so we only need to count how many different choices there are. There are $\binom{p}{p/2}^3 [(p/2)!]^3 = [p!/(p/2)!]^3$ different Wick contraction (that all give the same result,) the $\binom{p}{p/2}$ factor has to do with choosing which $G_{12}$'s in each group contract with a different $G_{12}$, while the $(p/2)!$ counts the different ways to contract the $G_{12}$'s after choosing the overall pairings. At the end this contribution reduces to the product of two six point functions
\begin{equation}
    \left(\epsilon \binom{p}{p/2} \right)^{3/2}
    \frac{\beta_1^3 \beta_2^3}{6} \expt{\int d\tau_1 d\tau_2 d\tau_3~G_1^{p/2}(\tau_{12})G_1^{p/2}(\tau_{23})G_1^{p/2}(\tau_{13})}^2 .
\end{equation}

This resulting term is very similar to the next leading order term, which was shown to be a 6-point function in section \ref{sec:double_trace_3_lines}. Not only does it have the correct power of $\epsilon$, the combinatorics of $p$ agree as well because at finite $p$ the factor in \eqref{SubLeadCont} reduces to 
\begin{equation}
         \begin{split}
             {\binom{N}{p}}^{-3} \binom{N}{3p/2} \binom{3p/2}{p} \binom{p}{p/2}
         &\approx \frac{N^{-3p} N^{3p/2} (3p/2)! p!
         }{(p!)^{-3} (3p/2)! p! [(p/2)!]^{3}}
          =  \left(\epsilon \binom{p}{p/2} \right)^{3/2} .
         \end{split}
\end{equation}
The factorization form and the lengths of the operators in the correlation functions is also the same as in section \ref{sec:double_trace_3_lines}. Indeed this is the same correlation function as in \eqref{fluc33A} only from the path integral approach.

Higher order terms in this pertubrative expansion have the same form as the ones studied in section \ref{sec:GenFluc}. This is because the counting of all possible Wick contraction at order $n$ is equivalent to counting how many ways there are to split $n$ multi-fermions into groups such that each group appears twice, and so the suppression factors in this expansion will be identical to the ones computed in section \ref{sec:GenFluc}.

Finally, we note that this calculation is a perturbative expansion around the disconnected saddle of \eqref{eq:action_G_Sigma_final}, namely the solution where $G_{12} = 0$. One can also consider connected saddle points of \eqref{eq:action_G_Sigma_final}, similar to \cite{saad2018}, which are related to a connected geometry in the bulk; though that is beyond the scope of this paper.

\section{Discussion} \label{sec:discussion}

In this paper we analyzed the leading order contributions to connected multi-trace expectation values in the SYK model. We showed that at early times the contributions can be organized into a well defined perturbative expansion in powers of $\epsilon = \binom{N}{p}^{-1} \sim N^{-p}$. The lowest order contributions arise from connected cactus diagrams, as derived in section \ref{sec:multi_trace}, or equivalently using a large $M = \binom{N}{p}$ vector model for the random couplings which was described in section \ref{sec:vector}. They are associated with a specific fluctuation parameter $\epsilon\sum_I J_I^2$.

This is actually the first of an infinite set of fluctuation parameters, which give higher order corrections to the moments, with decreasing strength. 
The next term in this expansion, which is the first correction of the odd moments, was systematically studied in section \ref{sec:NewFluc}; while the general structure of this expansion was further examined in section \ref{sec:GenFluc}.

The puzzling aspect of this expansion is its gravitational interpretation. Our analysis is valid at any energy range (captured by the moment method), and if we are studying the SYK model in a low temperature regime, albeit in early times, we expect there to be a clear gravitational dual. As these are connected contributions, one might expect them to arise from connected geometries between two disconnected boundaries in the sum over geometries implied by the gravitational path integral. However such contributions are topologically different and so are suppressed by factors of $e^{-S_0} \sim e^{-N}$ \cite{Saad:2019lba}, rather than the polynomial suppression in $N$ that we observe. Thus the gravitational path integral of the dual theory must contain an additional non-geometric way to couple disconnected boundaries. Based on the leading order contributions studied in section \ref{H2DualAct}, it seems like these contributions correspond to a global mode that collectively re-scales the AdS radius of the dual theory. The rest of the perturbative expansion can also be seen as additional fluctuation fields connecting the different copies of the boundary theory via the coefficient of their interaction terms without resorting to wormholes. Actually, the form of the interaction suggests that the between-universe correlations are only given by boundary terms. However, once we have identified the need for more fluctuation fields, then the rules of the AdS/CFT correspondence tell us that they propagate in the bulk.

It would be interesting to precisely specify the gravitational theory where such non-geometric fluctuations naturally arise from the path integral, perhaps by some deformation of JT gravity like \cite{Witten:2020wvy,maxfield2020path}, with the additional light fields that we discussed here. These non-geometric wormholes may also exist in recent stringy constructions of the SYK model \cite{goel2021string}, and so may be related to certain constructions in string theory. A more ambitious goal would be to understand under what more general circumstances these non-geometric global fluctuations arise in a theory of quantum gravity. For example, we might expect them to be there in any black holes with an $AdS_2$ near horizon limit. 

One may speculate that perhaps these global fluctuations, and the corresponding bulk fields, are a generic feature of a UV complete theory of quantum gravity, at least when the background is associated with high entropy configurations. The only piece of evidence that we can provide is to consider the case in which there are no global modes at all, in the sense above. This is just a standard RMT model for the Hamiltonian and the various observables in the theory. I.e. $H$, or any other operator, is drawn from an ensemble $dH \exp( -{\cal N} \text{Tr}(V(H)) )$ where ${\cal N}$ is the dimension of the Hilbert space. This is not the case for the SYK model, but JT gravity alone  falls into this class \cite{Saad:2019lba}. The problem is that it is difficult to have weakly coupled fields, i.e, familiar matter, propagating in any bulk dual. Consider two such fields, corresponding to operators ${\cal O}_{1,2}$ in the field theory. If the statistics of these operators is RMT, i.e., independent of each other and of the Hamiltonian, then the correlator $\langle O_1(t_1)O_2(t_2)O_1(t_3)O_2(t_4)\rangle$ will be exponentially suppressed (i.e. a suppression of the order $1/{\cal N} \sim e^{-S_0}$),\footnote{This scaling follows directly from the eigenvalue thermalization hypothesis \cite{Deutsch1991,Srednicki1994,Pollack_2020}.} which does not correspond to weakly coupled gravity. But if there are strong correlations between matrix elements, these imply global modes (although perhaps not as simple as the ones we discussed here).

Generally, it is unclear if the correct gravitational picture is that of a single dual theory to the whole SYK ensemble, as suggested by \cite{marolf_maxfield,Bousso2020}, or perhaps that there is a gravitational dual to each realization of random couplings. In the second case the effective gravitational picture arises from measuring coarse-grained observables which only weakly depend on the exact values of the couplings, similar to \cite{saad2019late,Pollack_2020,Engelhardt_2021}. In both cases, it is not quite clear what is the dual of a single realization if no coarse-graining is done. In this note, we are adding the option that there is actually a range of intermediate possibilities. We actually need to specify first the precision in which we plan to carry out our experiment, i.e., how much information about the specific couplings we want to obtain. If we increase the precision, we need to use a gravitational description which includes more fields, more couplings and more fluctuation parameters. This is not unreasonable since GR is an effective theory and we may have to use more features to describe effects with higher and higher precision. We can perhaps think about all these fields and couplings as already existing in the background, but averaged over if we carry out an experiment with low precision. They are still low  mass fields, not to be integrated out by a Wilsonian argument, but we can ignore them at low precision experiments. This is tantamount to saying that coarse grained observable is given by GR. As we increase the level of precision we need to add more of these fields and couplings. When we reach high precision measurements - as we try and nail down what is the precise realization - we need to include many such fields, which might change the behaviour of the background altogether. For example, it might lead to disconnecting universes, and, simultaneously, introducing many new objects in the bulk, beyond GR. Clearly, more work is needed to elaborate this point of view. 


Another interesting endeavor would be to analyze the global fluctuations in the charged SYK model \cite{Sachdev_2019,gu2019notes}, or the supersymmetric SYK models \cite{SusySYK,Li_2017}. This can be done using the same chord diagram methodology which has been previously used to study these models \cite{Berkooz_2020,berkooz2020complex}. Especially interesting would be to analyze how the existence of a large number of exact ground states in the $\mathcal{N} = 2$ theory, observed in \cite{Kanazawa_2017,Berkooz_2020}, affects these global fluctuations. 

Finally we note that two replica action derived in the path integral analysis of the global fluctuations (in section \ref{sec:path_integral}) is different than the standard action for the two-replica bi-local fields $G$ and $\Sigma$ derived in \cite{saad2018}.\footnote{Though they are mathematically equivalent.} It may be simpler to study the connected saddle points of this new action, perhaps even finding the exact contributions that lead to the universal RMT structure of the ramp and plateau in the spectral form factor.

\acknowledgments

It is a pleasure to thank O. Aharony, A. Altland, D. Bagrets, M. Isachenkov, A. Kamenev, P. Narayan, M. Rangamani, M. Rozali, S. Shenker, J. Sonner, R. Speicher, D. Stanford, J. Verbaarschot and H. Verlinde for useful discussions. We would like to thank the authors of \cite{Cotler:2016fpe} for sharing their data with us.

The work of MB and NB is supported by an ISF center of excellence 2289/18. MB is the incumbent of Charles and David Wolfson Professorial Chair of Theoretical Physics. The work of AR was supported, in part, by a grant from the Simons Foundation (Grant 651440, AK).

\appendix

\section{The spectrum at finite $N$ and $p$, and numerical comparisons} \label{app:spectrum}

The spectrum of the SYK model at finite $N$ and $p$ is extremely well approximated by a $q$--Gaussian distribution \cite{Garc_a_Garc_a_2017,Garc_a_Garc_a_2018_2}, and this is exact in the double scaled limit \cite{Erdos14,feng2018spectrum,Micha2018}. This spectrum is given by
\begin{equation}
    \rho(\theta) = \frac{1}{2\pi}\left(q,e^{\pm2i\theta};q \right)_{\infty},
\end{equation}
where $(a;q)_{n}$ is the $q$-Pochammer symbol
\begin{equation}
    (a;q)_{n} = \prod_{k=0}^{n-1}\left(1-aq^k \right), 
\end{equation}
with the shorthand notation $(a,b;q)_n \equiv (a;q)_n (b;q)_n$. Here the energies are given by 
\begin{equation}
    E(\theta) = \frac{2 {\cal{J}} \cos(\theta) }{\sqrt{1-q}},
\end{equation}
and in the double scaled limit $q$ is simply $q = e^{-2p^2/N}$. At finite $N$ and $p$ we can find an exact expression for $q$ given by (see \cite{Garc_a_Garc_a_2017} for details)
\begin{equation}
    q = {\binom{N}{p}}^{-1} \sum_{k=0}^p (-1)^k \binom{p}{k} \binom{N-p}{p-k}. 
\end{equation}

\begin{figure}[h]
    \centering
    \includegraphics[width = 15 cm]{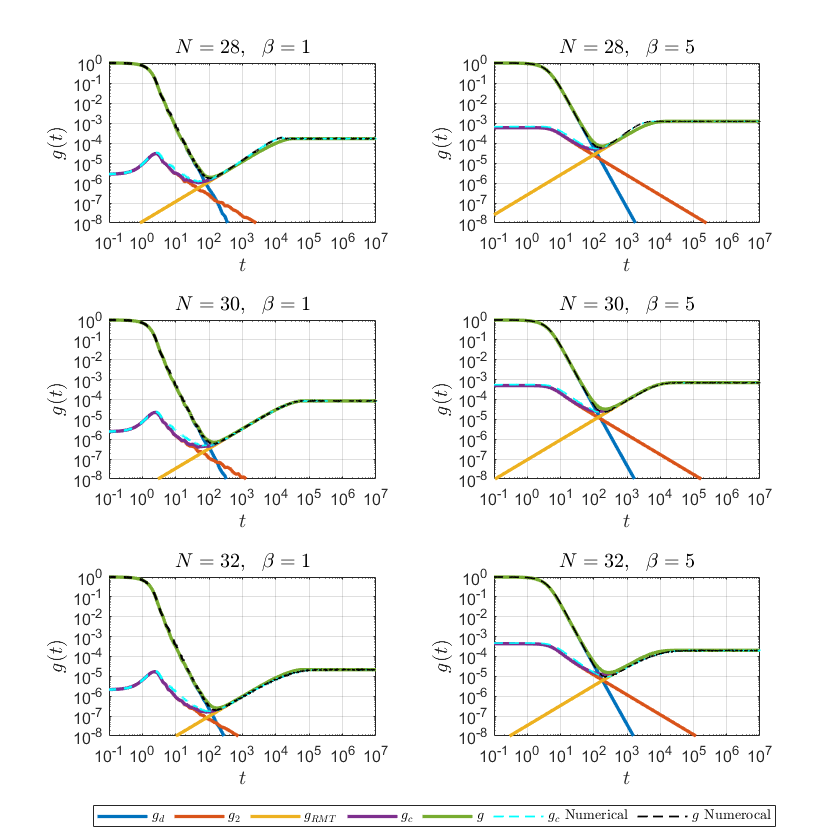}
    \caption{A comparison of the leading order correction to the spectral form factor for $N=28,30,32$ to the numerical results from \cite{Cotler:2016fpe}.}
    \label{fig:specform_all}
\end{figure}

Using this spectral density we can calculate $\cal{Z}(\beta)$ and the first contribution to the connected 2-point thermal partition function (given by \eqref{eq:double_thermal_partition}) via simple numerical integration. However the scaling we choose for the normalization of $H$ is different than the choice in \cite{Cotler:2016fpe}, namely we choose to normalize $2^{-N/2}\expt{\tr(H^2)} = 1$, while they choose to have it proportional to $N$ to get a well defined large $N$ action. Thus to compare to their numerical results we need to set (see \cite{verbaarschot2019} for details)
\begin{equation}
    {\cal J}^2 = \binom{N}{p} \frac{(p-1)!}{2^p N^{p-1}} .
\end{equation}

Then we can compare these results to the numerical results in \cite{Cotler:2016fpe} using simple numerical integration of the spectral density. Apart from the comparison in the text for $N=34$, we also compared the leading term to the numerical results from \cite{Cotler:2016fpe} for $N=28,30,32$, which are given in figure \ref{fig:specform_all}. We note that the slope-plateau transition is not exactly matched as we use the GUE approximation from \eqref{eq:specform} rather than the GOE or GSE forms of the ramp, which are the correct universality classes for $N=32$ and $N=28$ respectively.

\bibliography{TRTR_SYK}
\bibliographystyle{JHEP}

\end{document}